\documentclass[letterpaper]{iopart}
\usepackage{iopams,setstack,amsopn} 

\expandafter\let\csname equation*\endcsname\relax
\expandafter\let\csname endequation*\endcsname\relax
\usepackage{amsmath}

\usepackage{cite}

\usepackage[varg]{txfonts}
\usepackage{graphicx}
\usepackage{color}
\usepackage[colorlinks=true,citecolor=blue]{hyperref}
\usepackage[margin=1in]{geometry}
\usepackage[caption=false]{subfig}
\usepackage{booktabs}

\usepackage[stable]{footmisc}

\usepackage{multirow}

\graphicspath{{figures/}}

\newcommand{\tphisn}{ {T_{\varphi}^{{\rm SN}}} }
\newcommand{\tphiflux}{ {T_{\varphi}^{{\Phi_{\rm ext}}}} }
\newcommand{\tphicc}{ {T_{\varphi}^{I_{c}}} }
\newcommand{\tphing}{ {T_{\varphi}^{n_{g}^{\theta}}} }

\newcommand{\tphi}{ {T_{\varphi} } }
\newcommand{\tone}{ {T_{1} } }

\newcommand{\tonepurcell}{ {T_{1}^{\rm Purcell}} }

\newcommand{\dl}{{\delta\mkern-1.0mu\lambda} }
\newcommand{\la}{\lambda}

\newcommand{\abs}[1]{\left|#1\right|}
\newcommand{\bra}[1]{\langle \, #1 \,|}
\newcommand{\ket}[1]{|\, #1 \, \rangle}
\newcommand{\bket}[2]{\langle \, #1 \,|\, #2 \, \rangle}
\newcommand{\ketb}[2]{| \, #1 \,\rangle\langle\, #2 \, |}
\newcommand{\boket}[3]{\langle\, #1 \,|\, #2 \,|\, #3 \,\rangle}

\newcommand{\cj}{C_{\rm J}}
\newcommand{\cjp}{C_{\rm J}'}

\newcommand{\cs}{C_\Sigma}
\newcommand{\csp}{C_\Sigma'}

\newcommand{\cg}{C_{g}}

\newcommand{\cp}{C'}

\newcommand{\ec}{E_{\rm C}}
\newcommand{\ecj}{E_{{\rm CJ}}}
\newcommand{\ecs}{E_{{\rm C}\Sigma}}

\newcommand{\ej}{E_{\rm J}}

\newcommand{\el}{E_{\rm L}}
\newcommand{\eq}{E^{\rm q}}

\newcommand{\ecp}{E_{\rm C}'}
\newcommand{\ecjp}{E_{{\rm CJ}}'}
\newcommand{\ecsp}{E_{{\rm C}\Sigma}'}

\newcommand{\dej}{\delta E_{\rm J}}
\newcommand{\ddej}{d\mkern-1.0mu E_{\rm J}}

\newcommand{\ddc}{d\mkern-1.0mu C} 

\newcommand{\ddcj}{d\mkern-1.0mu{C_{\rm J}}}

\newcommand{\ddel}{d\mkern-1.0mu E_L}

\newcommand{\px}{\varphi_{\rm ext}}

\newcommand{\be}{\begin{equation}}
\newcommand{\ee}{\end{equation}}

\newcommand{\bigoh}[1]{\mathcal{O}\left( {#1} \right)}

\newcommand{\dg}{^\dagger}

\newcommand{\ave}[1]{\langle #1 \rangle}

\definecolor{burntorange}{rgb}{0.8, 0.33, 0.0}

\newcommand{\zp}{0-$\pi$}
\DeclareMathOperator{\sinc}{sinc}

\begin{document}
\title{Coherence properties of the \zp{} qubit}

\author{Peter Groszkowski$^1$, A.\ Di Paolo$^2$, A.\ L.\ Grimsmo$^2$, A.\ Blais$^{2,5}$,  D.\ I.\ Schuster$^3$, A.~A.~Houck$^4$, and Jens Koch$^1$}

\address{$^1$ Department of Physics and Astronomy, Northwestern University, Evanston, Illinois 60208, USA}
\address{$^2$ Institut quantique and Départment de Physique, Université de Sherbrooke, Sherbrooke J1K 2R1 QC, Canada}
\address{$^3$ Department of Physics and James Franck Institute, University of Chicago, Chicago, Illinois 60637, USA}
\address{$^4$ Department of Electrical Engineering, Princeton University, Princeton, NJ 08540, USA}
\address{$^5$ Canadian Institute for Advanced Research, Toronto, ON, Canada}

\eads{piotrekg@northwestern.edu}
\begin{abstract}
    Superconducting circuits rank among the most interesting architectures for the implementation of quantum information processing devices. The recently proposed \zp{} qubit [Brooks et al., Phys. Rev. A {\bf 87}, 52306 (2013)] promises increased protection from spontaneous relaxation and dephasing. In practice, this ideal behavior is only realized if the parameter dispersion among nominally identical circuit elements vanishes. In this paper we present a theoretical study of the more realistic scenario of slight variations in circuit elements. We discuss how the coupling to a spurious, low-energy mode affects the coherence properties of the \zp{} device, investigate the relevant decoherence channels, and present estimates for achievable coherence times in multiple parameter regimes.   
\end{abstract}

\section{Introduction}
\label{sec:introduction}
Research towards realizing a quantum computer poses a formidable challenge due to the need for a subtle compromise between two conflicting requirements: maximizing coherence by isolating qubits from environmental noise on one hand, and coupling qubits strongly for fast qubit control and readout, on the other hand. Over the last two decades substantial progress has been made in the field of superconducting circuits, where  coherence times have increased by nearly 6 orders of magnitude to milliseconds \cite{devoret2013superconducting}, while gate times stayed in the range of tens of nanoseconds. This impressive improvement is largely due to more advanced qubit designs which minimize the qubit's coupling to unwanted environmental noise sources, such as flux noise \cite{manucharyan2009fluxonium} or charge noise \cite{Koch2007a}, all while keeping the qubit susceptible to control pulses essential for performing gate operations as well as readout.

\begin{figure}
    \centering
   \includegraphics[width=0.6\textwidth]{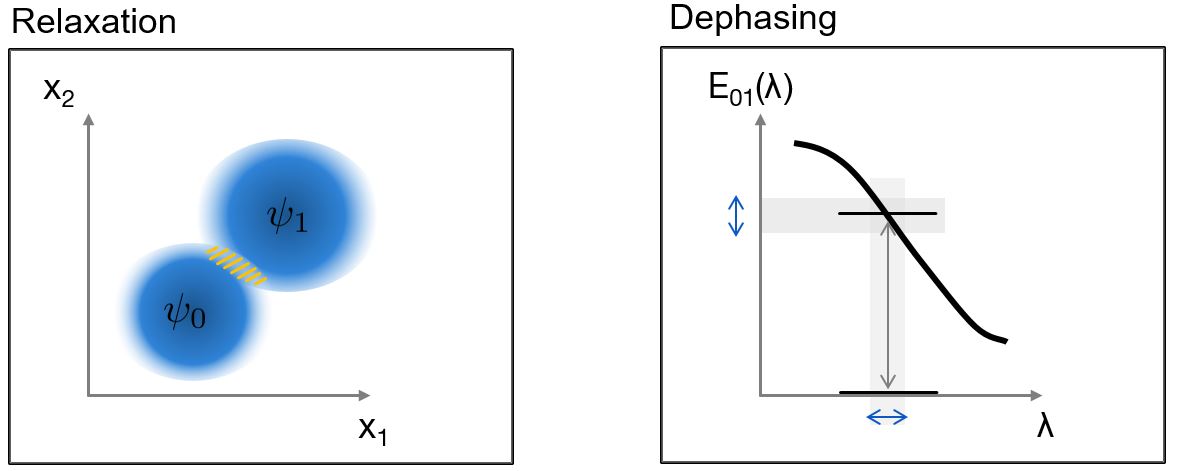}
   \caption{Protection offered by the \zp{} qubit. $T_1$ processes are exponentially suppressed due to nearly disjoint support of the qubit wave functions, i.e., wave functions ``live'' in nearly separate regions of the generalized-variable space; pure-dephasing rates, proportional to the qubit-energy susceptibility $\partial E_{10}/\partial \lambda$ with respect to the noise variable $\lambda$, are exponentially suppressed when nearly degenerate eigenstates of the \zp{} device are used as qubit states.
    \label{fig:T1-Tphi}}
\end{figure}

One particularly interesting design for the next generation of superconducting qubits is the \zp{} qubit, first proposed by Brooks, Kitaev and Preskill (BKP) \cite{Brooks2013}. Conceptually, this circuit exhibits a rudimentary form of topological protection that combines exponential suppression of noise-induced transitions (dissipation) with exponential suppression of dephasing, see figure\ \ref{fig:T1-Tphi}.  The former is achieved by engineering qubit states with disjoint support, the latter by rendering qubit states (nearly) degenerate and exponentially suppressing the sensitivity of the corresponding energies to low-frequency environmental noise.
The circuit underlying the \zp{} qubit consists of four nodes connected by a pair of linear inductors, a pair of capacitors, and a pair of Josephson junctions as shown in figure\ \ref{fig:zeropicircuit}.
Two issues pose challenges to the implementation of the \zp{} design:  first, to achieve the desired regime it is necessary to simultaneously realize large superinductances, large shunting capacitors, and high junction charging energies (very low stray capacitances); second, circuit elements are required to be pairwise identical (no disorder in circuit element parameters) in order to prevent coupling of the qubit to a spurious circuit mode \cite{Dempster2014a}, which we will refer to as the $\zeta$-mode\footnote{This low-energy mode was originally called $\chi$-mode \cite{Dempster2014a}, but is here renamed to avoid confusion with dispersive shifts commonly denoted by ``$\chi$''.}.

While notable increases in accessible inductance values by means of junction-array based superinductances may partially address the first issue \cite{Manucharyan2009,manucharyan2012superinductance,Masluk2012,Bell2012,Pop2014}, some amount of circuit parameter disorder and hence residual coupling to the $\zeta$-mode is unavoidable. In the present work, we theoretically assess the coherence properties expected for realistic \zp{} devices, possible to realize with today's state-of-the art fabrication techniques or as well as in the future.
Specifically, we present calculations of relevant decoherence rates resulting from the qubit's coupling to known noise sources, including both intrinsic sources, such as flux, charge and critical current noise, which couple directly to the qubit's degree of freedom, as well as noise mediated by the coupling to the spurious $\zeta$-mode. We will concentrate our study on three representative parameter sets, which primarily differ in the magnitude of the inductance and are motivated by current experimental capabilities. 

This paper is organized as follows. In the subsequent section, we briefly review the quantization of the \zp{} circuit, properly accounting for parameter disorder, and the coupling of the qubit degree of freedom to the $\zeta$-mode \cite{Dempster2014a}. In section~\ref{sec:circuitParams} we then present \zp{} eigenspectra for the parameter sets considered, and discuss the dependence of spectral properties on the different energy scales. In section~\ref{sec:noiseProcesses} we describe relevant noise processes affecting the \zp{} qubit, and discuss the calculation of decoherence rates. In section~\ref{sec:decoherenceResults}, we present the resulting decoherence rates and identify the processes likely to limit coherence. Finally, we summarize and conclude in section~\ref{sec:conclusions}. 

\section{Hamiltonian of the \zp{} qubit}
\label{sec:System}
We begin by briefly reviewing the circuit of the \zp{} qubit and the corresponding circuit Hamiltonian. As shown in figure \ref{fig:zeropicircuit}(a), the \zp{} circuit consists of two Josephson junctions (Josephson energies $E_{J1,2}$, junction capacitances $C_{J1,2}$) and two large (super-)inductors (inductances $L_{1,2}$), linked to form a loop. The opposing nodes $j=1,\,3$ and $j=2,\,4$ are connected by two large capacitors $C_{1,2}$. As usual \cite{Devoret95,burkard2004multilevel}, we initially employ generalized flux variables $\varphi_j$ for each circuit node $j=1,\ldots,4$. We then switch to physically more meaningful variables $\phi$, $\theta$, $\zeta$, and $\Sigma$ \cite{Dempster2014a} associated with the normal modes of the linearized, non-disordered circuit (see figure \ref{fig:zeropicircuit}(b)):  
\begin{align}
2\phi=(\varphi_2-\varphi_3)+(\varphi_4-\varphi_1),\quad
2\zeta=(\varphi_2-\varphi_3)-(\varphi_4-\varphi_1), \quad
2\theta=(\varphi_2-\varphi_1)-(\varphi_4-\varphi_3),\quad
2\Sigma= \sum_j\varphi_j.
    \label{eq:transf1}
\end{align}
Both the $\phi$-mode and $\theta$-mode involve phase differences across the Josephson junctions and are coupled by the junction nonlinearity, as we will see momentarily. The $\zeta$-mode does not bias the junctions and is therefore, a fully harmonic mode. Finally, the variable $\Sigma$ is cyclic, remains decoupled from the other variables, and can thus be omitted. (Alternatively, one can reach this conclusion by invoking gauge freedom and setting one of the nodes to ground.)

In the absence of disorder among circuit elements, we have $E_{J1}=E_{J2}\equiv E_{J}$ etc., and we can write the symmetric \zp{} Hamiltonian as
\begin{equation}
    H_{\rm sym} = -2\ecj\,\partial_\phi^2-2\ecs\,\partial_\theta^2 -2\ej\cos\theta\,\cos \left(\phi-\frac{\px}{2} \right)+\el\, \phi^2 + H_\zeta,
    \label{eq:symH}
\end{equation}
where
\begin{align}
    H_{\zeta} &=  -2\ec\,\partial_\zeta^2+E_L\, \zeta^2,
    \label{eq:Hzeta1}
\end{align}
is the Hamiltonian for the harmonic $\zeta$-mode. The various parameters are defined as follows: $E_L=\Phi_0^2/2L$ is the inductive energy scale, and $\ec=e^2/2C$, $\ecj=e^2/2\cj$, $\ecs=(1/\ec+1/\ecj)^{-1}$ are the relevant charging energies. Finally, $\varphi_{\rm ext}=2\pi \Phi_{\rm ext} / \Phi_{0} $ is the external magnetic flux through the loop in terms of the magnetic flux quantum $\Phi_0=h/2e$.  
Evidently, the mode decouples in the disorderless case, leaving only the $\phi$ and $\theta$ degrees of freedom to form the effective qubit Hilbert space. (For a detailed discussion of the resulting qubit wave functions and the origin of protection from noise, we refer the reader to reference \cite{Dempster2014a}.) Figure~\ref{fig:zeropicircuit}(c) shows the potential energy of the Hamiltonian from equation~\eqref{eq:symH}.

Once imperfections in fabrication are taken into account, nominally identical circuit elements will acquire slight parameter deviations. It is convenient to introduce the parameter averages $X=\frac{1}{2}(X_{1} + X_{2})$ and relative deviations $d\mkern-1.0mu X = (X_{1} - X_{2})/X$, where $X \in \{\el,\, \ej,\, C,\, \cj \}$. 
Using this notation and employing a leading-order expansion in the capacitive disorder, one can cast the Hamiltonian into the form
$H =  H_{0-\pi} + H_{\zeta} +  H_{\rm int}$, where  
\begin{align}
    H_{0-\pi} &=  H_{\rm sym} + 2 \ecs\, \ddcj\, \partial_\phi\,\partial_\theta + \ej\, \ddej\, \sin \theta\, \sin \left(\phi-\frac{\px}{2} \right) + \bigoh{\ddc^2,\ddcj^2},
    \label{eq:Hzeropi1}
\end{align}
captures the primary \zp{} qubit degrees of freedom, including effects of disorder in junction parameters, 
and 
\begin{align}
    H_{\rm int} &=  2\ecs\, \ddc\, \partial_\theta\,\partial_\zeta+ \el\, \ddel\, \phi\, \zeta  + \bigoh{\ddc^2,\ddcj^2},
\label{eq:Hzetaint1}
\end{align}
describes the coupling between \zp{} and the $\zeta$-mode. As discussed in detail in reference \cite{Dempster2014a}, disorder in junction parameters gives rise to minor corrections to the \zp{} qubit spectrum, but leave the $\zeta$-mode decoupled. Such coupling does arise from disorder in $C$ and $E_L$, and can have important consequences on the coherence of the \zp{} qubit, as we will see in the following sections. For this analysis, it will be helpful to write the Hamiltonian in the product basis comprised of eigenstates of $H_{0-\pi}$ and $\zeta$-mode eigenstates:
\begin{figure}
  \captionsetup[subfigure]{position=top,singlelinecheck=off,topadjust=-5pt,justification=raggedright}
    \centering
    \subfloat[]{   \includegraphics[width=1.4in]{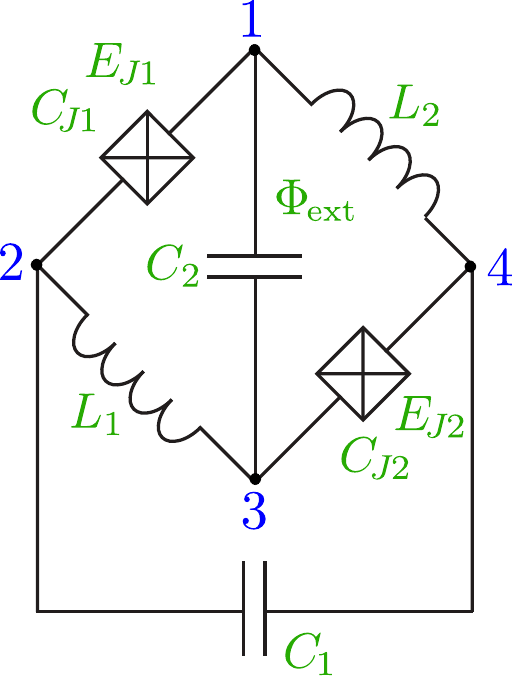}}
    \hspace{1.4cm}
    \subfloat[]{   \includegraphics[width=1.8in]{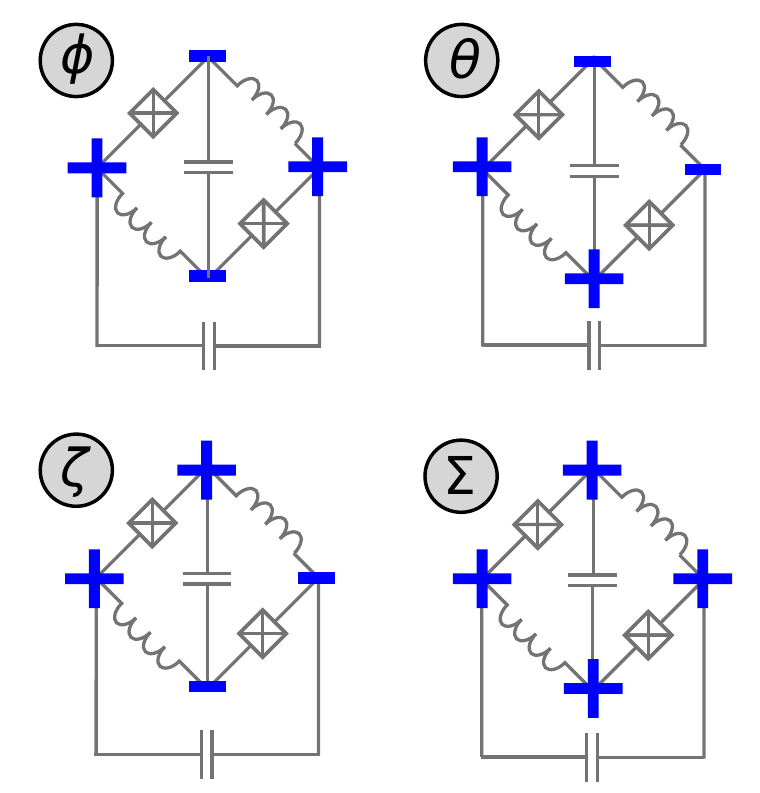}}
    \hspace{1.5cm}
    \subfloat[]{   \includegraphics[width=1.9in]{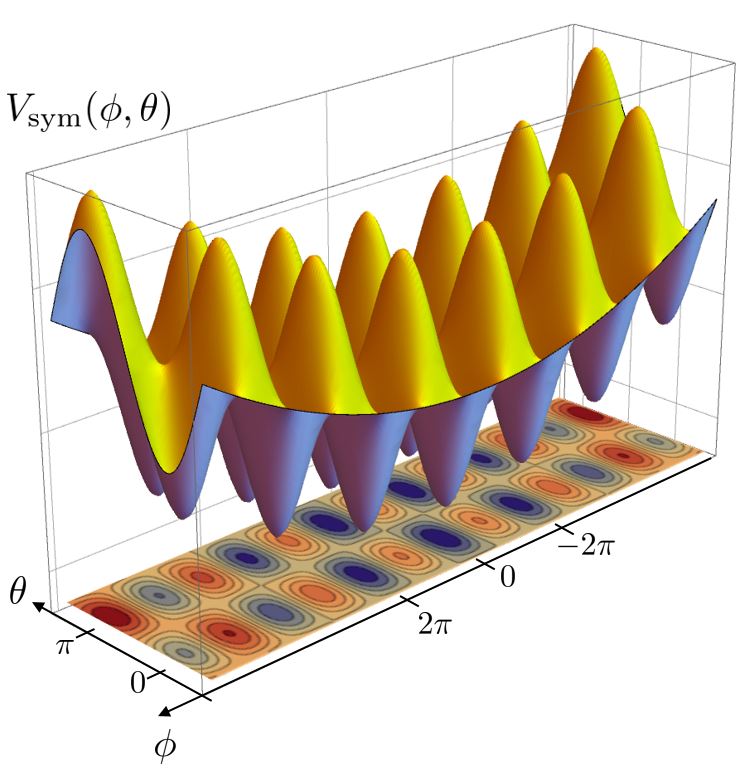}}
    \caption{(a) \zp{} circuit diagram, consisting of two Josephson junctions, two large inductors and two large capacitors. (b) Schematic representation of the normal modes of the linearized circuit (no parameter disorder), which define the new circuit variables $\phi$, $\theta$, $\zeta$ and $\Sigma$. (c) Potential energy of the disorderless \zp{} qubit.}
    \label{fig:zeropicircuit}
\end{figure}
\begin{equation}
    H=\sum_l  \eq_{l}\, \ket{l}\!\bra{l} + \hbar\Omega_\zeta\, a^\dag a +\sum_{l, l'}\left( g_{ll'}\ket{l}\!\bra{l'}\,a +{\rm h.c.}\right),
\label{eq:Hqubitzeta}
\end{equation}
where $a\dg$ ($a$) are the creation (annihilation) operators for the $\zeta$-mode,
$\Omega_{\zeta}=\sqrt{8E_L \ec}/\hbar$ is its angular frequency, $\eq_{l}$ the energy of the $l$th primary \zp{} eigenstate, and  
\begin{equation}
    g_{ll'}=g^\phi_{ll'}+i g^\theta_{ll'} =  \frac{1}{2} \el\, \ddel \left(\frac{ 8 \ec}{\el} \right)^{1/4} \boket{l}{\phi}{l'}
    +i\frac{1}{2} \ddc\, \ecs \left( \frac{32 \el}{\ec} \right)^{1/4}\boket{l}{i\,\partial_\theta}{l'} 
    \label{eq:gs}
\end{equation}
the strength of the coupling that mediates transitions among $H_{0-\pi}$ eigenstates $\ket{l}$, $\ket{l'}$ via emission/absorption of $\zeta$-mode excitations.

In the dispersive limit, where the detunings $\Delta_{ll'}= \eq_{l}- \eq_{l'}- \hbar\Omega_{\zeta}$ are large compared to the coupling, $|g_{ll'}/\Delta_{ll'}|\ll1$,
a Schrieffer-Wolff transformation yields the effective Hamiltonian \cite{Dempster2014a,Zhu2013}
\begin{equation}
    H'=  \sum_{l=0}^\infty (\eq_{l} + \Lambda_{l})\ket{l}\!\bra{l} + \hbar\Omega_{\zeta}\, a^\dag a +\sum_{l} \chi_l\, \ket{l}\!\bra{l}\, a^\dag a
\label{eq:dispH},
\end{equation}
where the ac Stark shifts and Lamb shifts are given by 
\begin{equation}
    \chi_l=\sum_{l'}\abs{g_{ll'}}^2\left(\frac{1}{\Delta_{ll'}}-\frac{1}{\Delta_{l'l}}\right), \qquad \Lambda_{l}=\sum_{l'}\frac{\abs{g_{ll'}}^2}{\Delta_{ll'}},
    \label{eq:chiAndLambda}
\end{equation}
respectively.

So far, we have neglected capacitances between each node and ground. The effect of such ground capacitances depends on their uniformity. If all ground capacitances are identical and the circuit is symmetric, then the effect is minimal: $\Sigma$ remains decoupled and the charging energies $\ec$, $\ecs$ and $\ecj$ are merely renormalized. Node-to-node variations in ground capacitances complicate the situation slightly by inducing coupling between the primary degrees of freedom to the charge operator of $\Sigma$. In the present work, we will focus on the case of small ground capacitances where corrections of this type are negligible.

\section{Eigenspectrum of the \zp{} qubit}
\label{sec:circuitParams}
Our goal is to understand the key coherence properties of the \zp{} qubit. Since these properties have significant dependence on various circuit parameters, we choose three specific parameter sets that explore the balance between coherence times and current fabrication capabilities. Table~\ref{table:parameters} details our choices for inductive, Josephson, and charging energies for parameter sets 1, 2 and 3 (PS1, PS2, PS3). In all cases, we take the relative parameter deviations of nominally identical circuit elements to be at the $5\%$ level (which can be considered pessimistic). While we introduce some variations in all energies, the central scheme behind the parameter choices, is the increasing size of the inductive energy $\el$ as we go from PS1, through PS2, to PS3.  

\begin{figure}[b]
\begin{center}
 \includegraphics[width=2.5in]{{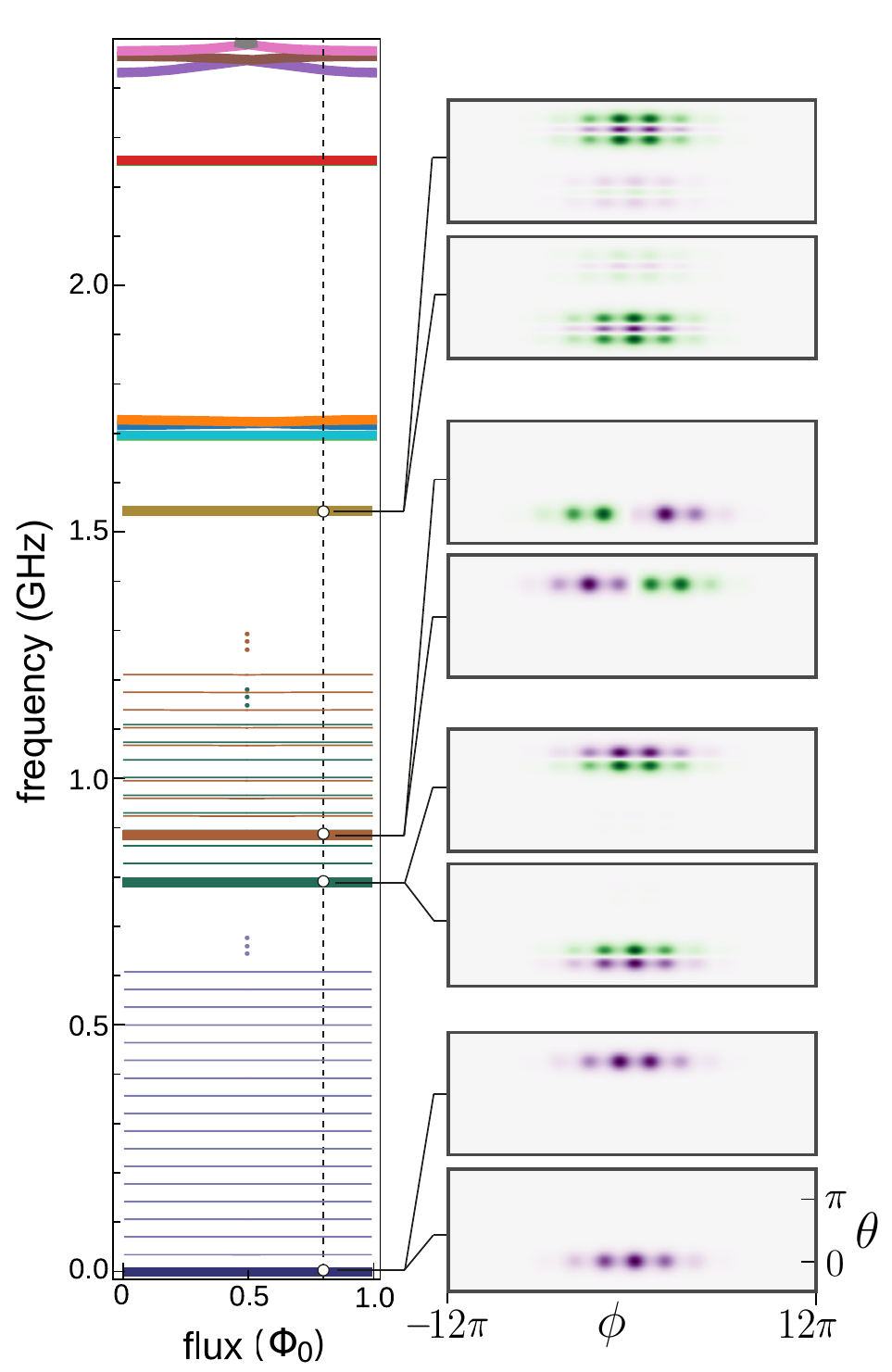}}
\end{center}
\caption{Energy spectrum of parameter set 1 (see table~\ref{table:parameters}), plotted as a function of flux, along with a few selected \zp{} wave functions calculated at $\Phi_{\rm ext} = 0.8\,\Phi_{0}$.
    (Only a limited number of $\zeta$-mode excitations is shown.) 
\label{fig:param1Spec}}
\end{figure}

\begin{figure}
  \captionsetup[subfigure]{position=top,singlelinecheck=off,topadjust=-5pt,justification=raggedright}
    \centering
    \subfloat[]{ \includegraphics[width=2.5in]{{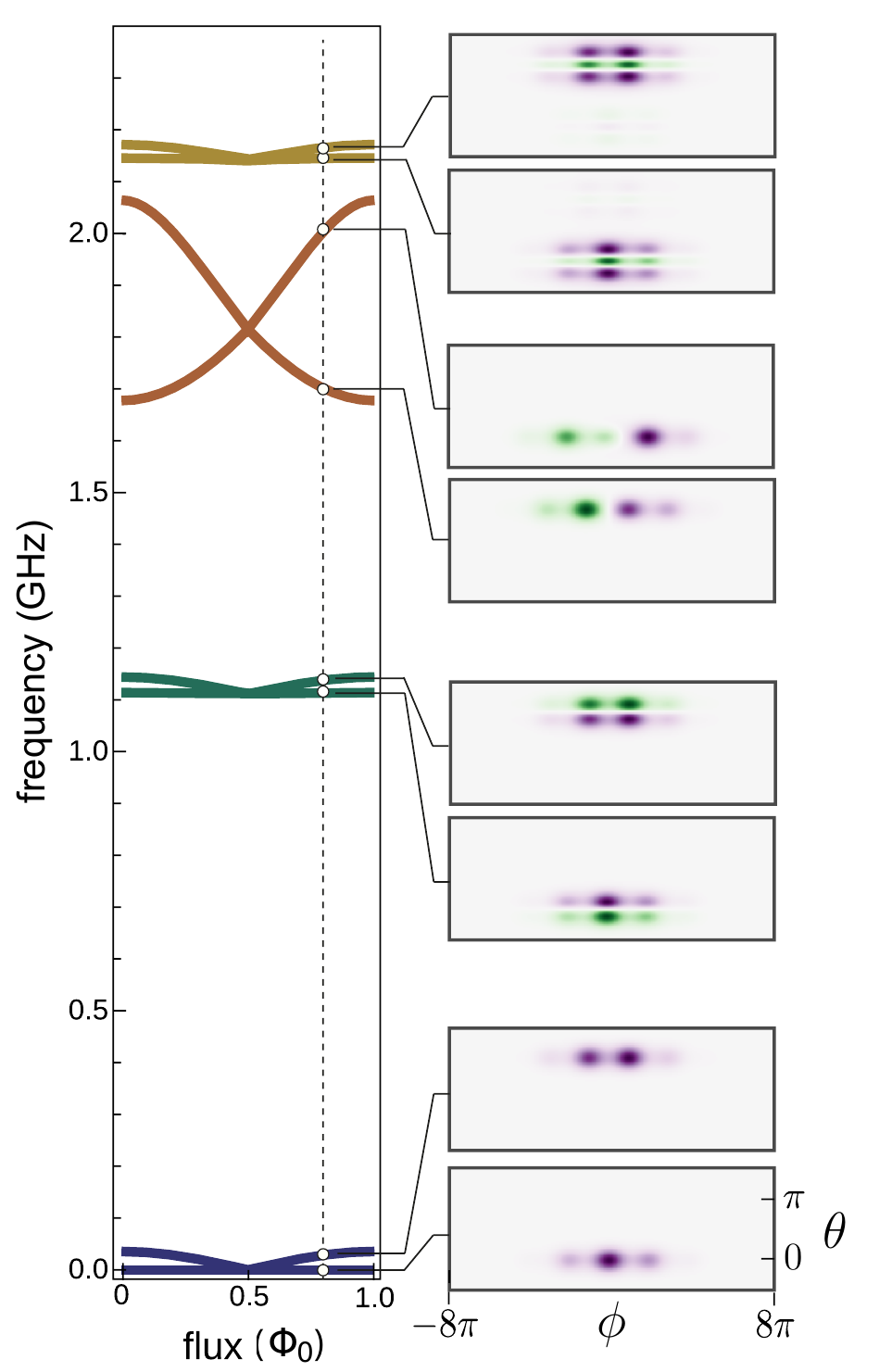}}}
    \hspace{0.2cm}
    \subfloat[]{  \includegraphics[trim=0 0 0 0.15cm, width=3.65in]{{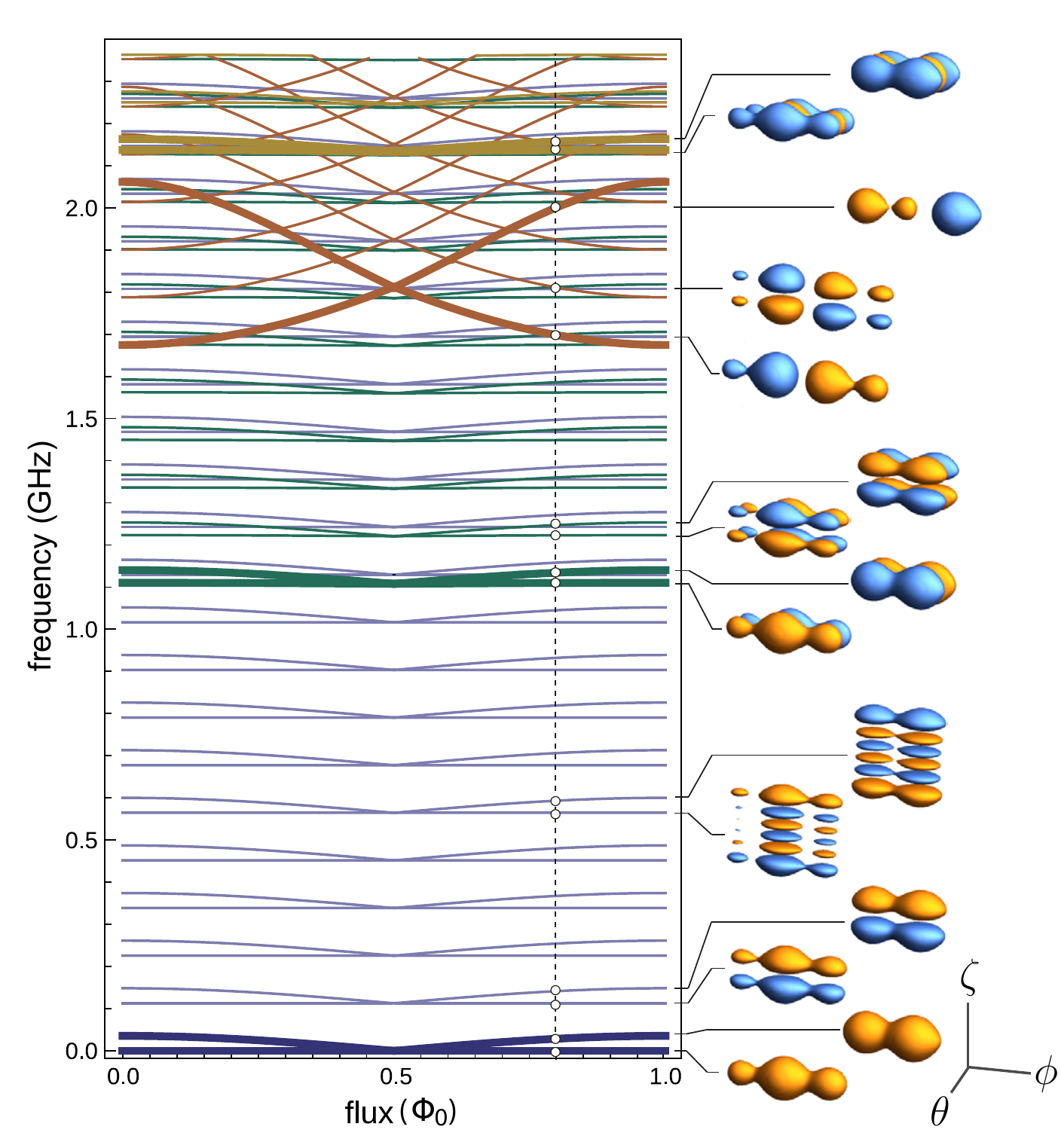}}}\\\hrule
        \subfloat[]{ \includegraphics[width=2.4in]{{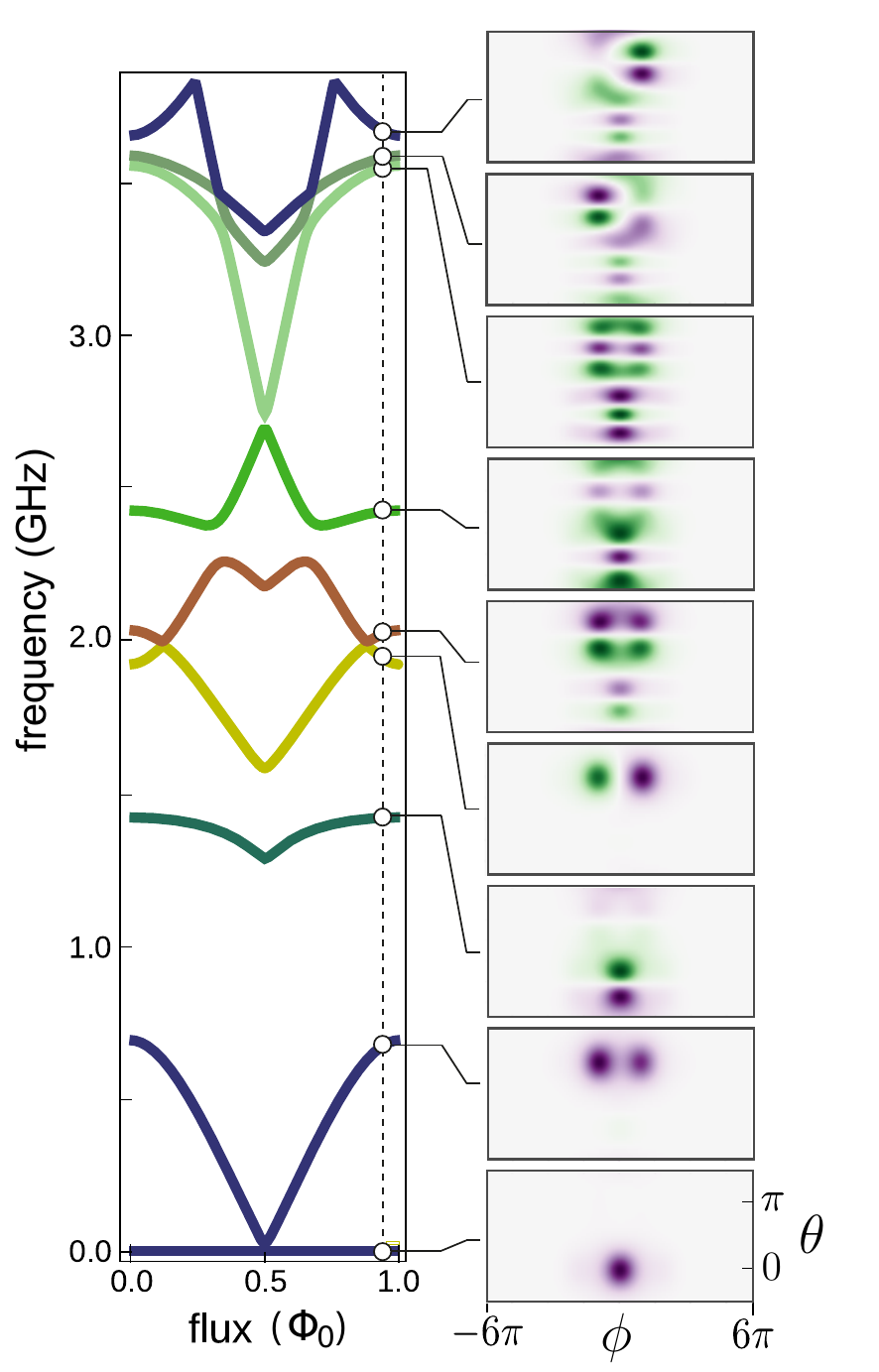}}}
    \hspace{0.2cm}
    \subfloat[]{\includegraphics[trim=0 0 0 -0.53cm, width=3.52in]{{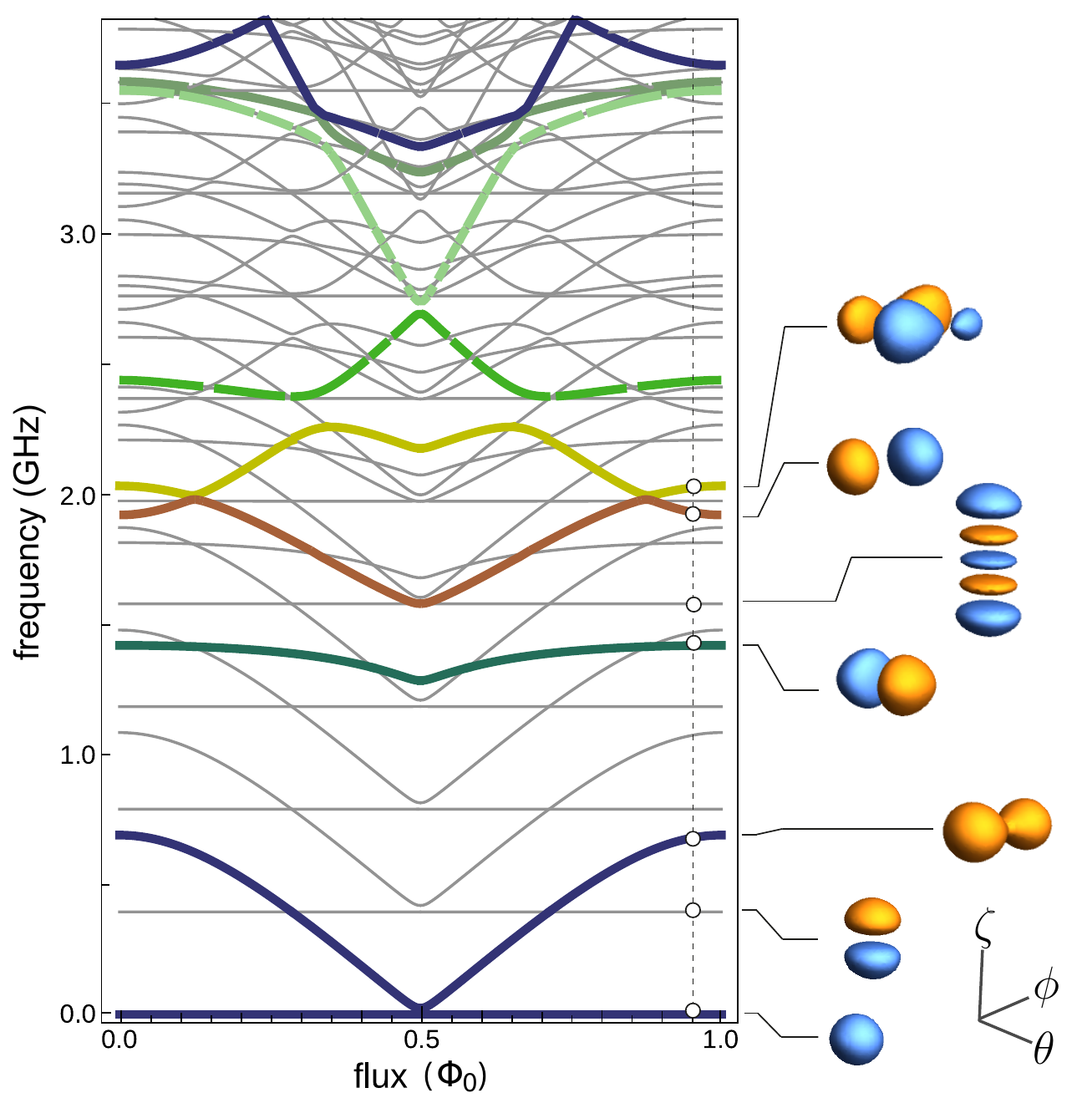}}}
    \caption{Energy spectrum calculated for parameter sets 2 and 3 (see table~\ref{table:parameters}), and  selected wave functions calculated at $\Phi_{\rm ext} = 0.8\,\Phi_{0}$ (PS2) and $0.9\,\Phi_0$ (PS3). (a) PS2  spectrum of $H_{0-\pi}$ for vanishing coupling to the $\zeta$-mode, calculated by setting $\ddel=\ddc=0$ and retaining $\theta$ and $\phi$ degrees of freedom only. (b) Spectrum for PS2 in the presence of disorder in the parameters $\el,\,C$ where the $\phi$, $\theta$ and $\zeta$-mode couple. On the far right: select wave function orbitals, obtained as contour surfaces obeying $\Psi(\theta,\phi,\zeta)=\pm\text{const}$ with orange/blue color indicating positive/negative wave function amplitude.  (c) and (d) show the analogous results for parameter set 3.}
\label{fig:param23Spec}
\end{figure}

Figure~\ref{fig:param1Spec} shows the PS1 energy spectra plotted as a function of flux, as well as a few selected wave functions. Since in this parameter set both $\el$, and $\ec$ are smallest of all we consider, the spectrum is densely populated with levels that mainly correspond to $\zeta$-mode excitations, and hence, only a few of those are explicitly drawn. The plots of wave functions assume a special case where only nonzero disorder in $\ej$ and $\ecj$ is included, (with $\zeta$-mode decoupled), and hence their dependence only in terms of $\theta$ and $\phi$ is presented.
Similarity, figure~\ref{fig:param23Spec} shows the energy spectra and eigenfunctions for both PS2 and PS3, again with a subset of eigenfunctions. The panels (a) and (c) in figure \ref{fig:param23Spec} present the pure \zp{} spectra obtained when the $\zeta$-mode remains decoupled from the $\theta$ and $\phi$ degrees of freedom, as realized when setting $\ddel=\ddc=0$, such that $g_{ll'}$ in equation~\eqref{eq:Hqubitzeta} vanishes. In the (b) and (d) panels, disorder in $E_L$ and $C$ is taken into account, and the spectra show dressed-state excitations of both the \zp{} and the $\zeta$-mode.

\begin{table}
\centering
\begin{tabular}{ccccccc}\hline\hline
   & \multicolumn{2}{c}{Parameter Set 1}  & \multicolumn{2}{c}{Parameter Set 2} & \multicolumn{2}{ c}{Parameter Set 3} \\ \cmidrule(lr){2-3}\cmidrule(lr){4-5}\cmidrule(lr){6-7}
 & [$h\cdot$GHz] & $[\hbar\omega_{p}]$  & [$h\cdot$GHz] & $[\hbar\omega_{p}]$ &  [$h\cdot$GHz] & $[\hbar\omega_{p}]$ \\ \hline
  $\ec$  & 0.02    &0.0005   & 0.04   &0.001  & 0.15     &0.008 \\
  $\ecj$ & 20.0    &0.5      & 20.0   &0.5    & 10.0     &0.5 \\
  $\ej$  & 10.0    &0.25     & 10.0   &0.25   & 5.0      &0.25 \\
  $\el$  & 0.008   &0.0002 & 0.04   &0.001  & 0.13     &0.007 \\\hline\hline
\end{tabular}
\caption{Table of circuit parameters for parameter sets 1, 2 and 3 (PS1, PS2, PS3). Josephson, inductive and capacitive energies are given in units of h $\times$ GHz as well as in units of the plasma energy of the Josephson junctions, $\hbar \omega_{p}=\sqrt{8 \ecj \ej} $, with $\omega_{p}/2\pi = 40\,$GHz for PS1 and PS2, and $\omega_{p}/2\pi = 20\,$GHz for PS3. Disorder in energies and capacitances is assumed to be at the 5\% level, i.e., $d\mkern-1.0mu X = 2 (X_{1} - X_{2})/ (X_{1} + X_{2})=5\%$ for $X \in \{\el, \ej, C, \cj \} $.}
\label{table:parameters}
\end{table}

To discuss the generic aspects of  parameter choices and \zp{} qubit spectra, we first consider the specific case of PS2 [figure \ref{fig:param23Spec}(a) and (b)], and comment on the impact of parameter changes leading to PS1 and PS3 subsequently. Figure \ref{fig:param23Spec}(a) shows that the low-lying eigenstates of $H_{0-\pi}$ are localized close to $\theta=0$ or $\theta=\pi$ (hence the name ``\zp{} qubit"), while spreading over multiple wells in $\phi$ direction. As intended, qubit ground and first excited states are close to being degenerate and are well-separated from higher excitation levels over the entire flux range. We note that the distinct insensitivity of the qubit energy with respect to flux is a crucial feature that distinguishes the \zp{}-qubit physics from that of the effectively 1D double-well physics prevalent for flux qubits \cite{mooij1999josephson}.

Choosing favorable parameters for \zp{} devices is, in part, driven by three central criteria. First, we wish to maximize the state-localization on the $\theta$ axis to realize disjoint-support wave functions, in order to exponentially suppresses all transition matrix elements that enter qubit relaxation/depolarization rates. Such localization can be achieved by rendering the effective mass in $\theta$ direction heavy, $C_J \ll C$, and making the local potential wells deep enough to hold localized states, $\ej\gg\el$ and $\ecs\ll\ej$.
Second, we aim for maximal delocalization along the $\phi$ axis, which suppresses susceptibility to flux variations by the same mechanism responsible for flux insensitivity of metaplasmon energies in fluxonium \cite{koch2009charging}.  As a result, ground and excited states become near-degenerate, and sensitivity to $1/f$ flux noise is suppressed. This regime requires parameters to obey $\el\ll\ecj\ll\ej$.
Third, charge-noise sensitivity of the device is minimized by ensuring $\ecs\ll\ej$, in analogy to the mechanism harnessed for the transmon qubit \cite{Koch2007a}. Decreasing $\el$ even further, as done in PS1, pushes the system closer towards ground-state degeneracy \cite{Brooks2013,Dempster2014a}. However, unless $\el$ decreases pass a certain threshold, one can run into a coherence bottleneck arising from coupling to the $\zeta$-mode (see sections \ref{sec:noiseProcesses} and \ref{sec:decoherenceResults}).

Still concentrating our discussion on PS2, we note that once parameter deviations in $C$ and $\el$ are taken into account, the $\zeta$-mode weakly couples to the primary qubit degrees of freedom. The resulting energy spectrum shown in figure \ref{fig:param23Spec}(b), then includes levels that reflect excitations of the $\zeta$ degree of freedom. However, since this coupling is weak, and \zp{} and $\zeta$-mode excitations are generally off resonance, eigenstates are only weakly dressed and can usually safely be labeled by $\ket{l,n}$ with $l,\,n$ denoting the excitation numbers of the \zp{} and the $\zeta$-mode, respectively. One consequence easily spotted in the spectrum -- especially for frequencies below $\sim 1.1\,$GHz in parameter set 2 -- is that each \zp{} energy level $E_{l,n=0}$ appears ``copied'' at regular intervals set by the $\zeta$-mode frequency, $E_{l,n}\approx E_{l,0}+n\,\hbar \Omega_\zeta$. This physics is also evident in the shapes of the corresponding wave function orbitals presented in figure \ref{fig:param23Spec}(b), which clearly shows the proliferation of nodes along the $\zeta$ axis as the excitation number of the $\zeta$-mode is increased one by one.


Relative to parameter set 2, parameter set 1 contains a lower inductive and charging energies $\el$ and $\ec$. This leads to almost flat spectrum as a function of flux, and hence, this choice of energies represents a ``deep'' \zp{} regime, envisioned in \cite{Brooks2013}
When $\ddel=\ddc=0$, at $\Phi_{\rm ext}=0$, the energy splitting is $(E^{q}_{1} - E^{q}_{0})/h \approx 24\,$kHz, while $(E^{q}_{2} - E^{q}_{0})/h \approx 792\,$MHz. Using such parameters helps in limiting both dephasing due to 1/f noise, but also various relaxation mechanisms. Another feature of this low-$\el$ regime is that the dispersive coupling to the stray $\zeta$-mode is highly suppressed. As we discuss in section~\ref{sec:decoherenceResults}, this limits the dephasing due to thermal shot noise which is relevant, or even dominant, in the other parameter sets we study. The central difficulty of experimentally realizing a circuit using PS1, however, is the ability to build large enough linear superinductance, hence, its near-future prospects may be limited.  

Finally, the parameter choices in PS3 closely match current fabrication capabilities. In particular, the inductive energy scale is increased from PS2 by a factor of more than 3, mitigating the challenge of superinductor fabrication. At the same time, $\ec$ is increased relative to PS2, thus giving rise to an overall upwards shift of the $\zeta$-mode frequency to $\Omega_\zeta/2\pi=\sqrt{8\ec\el}/h\approx 395\,$MHz. The resulting decrease in thermal $\zeta$-mode excitations will play role in our discussion of coherence properties below. As seen in figure \ref{fig:param23Spec}(c) and (d), the change in parameters comes at the cost of increased flux susceptibility and loss of near-degeneracy. Moreover, eigenstates beyond the lowest four are seen to break localization at $\theta=0$ and $\theta=\pi$. In figure~\ref{fig:param23Spec}(d), where coupling to the $\zeta$-mode is included, the spectrum again shows ``copies" of energy levels which arise from the addition of $\zeta$-mode excitations.


\section{Noise channels affecting the \zp{} qubit}
\label{sec:noiseProcesses}
%
To characterize the coherence properties of the \zp{} qubit, we need to identify its most damaging noise channels. 
We thus calculate and compare various depolarization and dephasing rates that originate from coupling to different known noise sources. In our analysis we make the common assumption \cite{Ithier05,Koch2007a,shnirman2002noise,manucharyan2009fluxonium,yan2016flux,you2007low,Martinis03} that noise in different channels is uncorrelated and individual rates can, hence, be calculated separately and then added up to give cumulative rates for depolarization and pure dephasing, similar to the treatment in references \cite{Ithier05,Clerk2007}. We further assume that the interaction with the environment to be sufficiently weak, so that the coupling $V_\lambda$ to the full \zp{} circuit Hamiltonian can be treated perturbatively. In that case, decoherence rates can be calculated either using Fermi's Golden Rule (for relaxation/depolarization), or by studying the effects of $V_\lambda$ on eigenenergies and, in turn, on the time evolution of the off-diagonal elements of the density matrix (for pure dephasing). 

\begin{figure}[b]
  \captionsetup[subfigure]{position=top,singlelinecheck=off,topadjust=-5pt,justification=raggedright}
    \centering
    \subfloat[]{      \includegraphics[width=1.0in]{{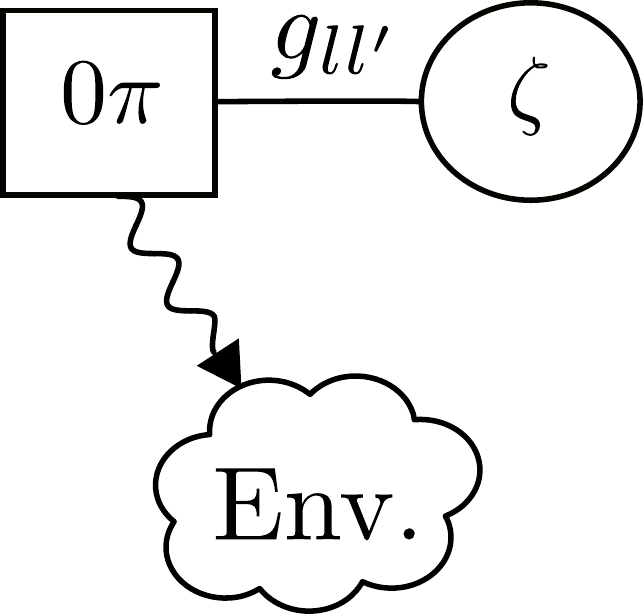}}}
 \hspace{2cm}
 \subfloat[]{      \includegraphics[width=1.5in]{{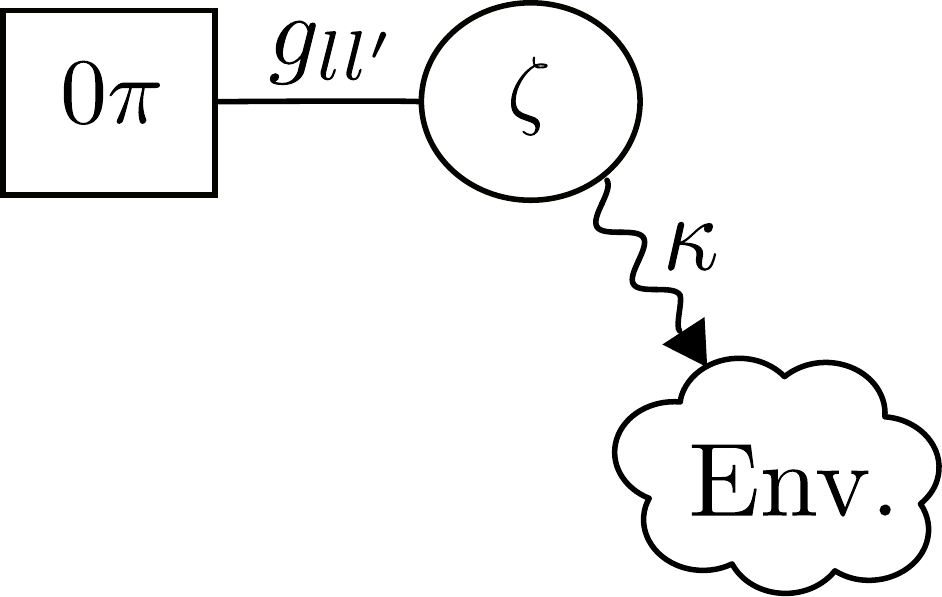}}}
    \caption{Two noise mechanisms contributing to decoherence of the \zp{} qubit. (a) The primary \zp{} qubit degrees of freedom, $\theta$ and $\phi$, can couple directly to a noisy environment. (b) The \zp{} qubit may couple to environmental noise indirectly via the $\zeta$-mode which has an intrinsic lifetime $\kappa$.}
    \label{fig:noiseMechanisms}
\end{figure}

As schematically illustrated in figure~\ref{fig:noiseMechanisms}, we distinguish between two different pathways for decoherence of the \zp{} qubit. First, the primary \zp{} qubit degrees of freedom, $\theta$ and $\phi$, can directly interact with a noisy environment. In this case, we generally find that disorder-induced coupling to the $\zeta$-mode only leads to subdominant corrections to decoherence rates. Second, baths coupled to the $\zeta$-mode can also influence the primary \zp{} qubit degrees of  freedom and lead to indirect decoherence processes. Despite the fact that the interaction $H_\text{int}$ [equation \eqref{eq:Hzetaint1}] is expected to be weak, we will discover that such indirect decoherence processes can play a crucial role in the performance of the \zp{} qubit in some parameter regimes. 

\subsection{Pure dephasing ($T_{\varphi}$)}
\label{ssec:dephasing}
We first consider pure dephasing, the dissipationless loss of phase information. Pure dephasing is quantified by the time scale $T_\varphi$ needed for a quantum superposition to turn into a classically mixed state, observed as the decay time for  off-diagonal elements of the qubit's density matrix, when expressed in the eigenenergy basis. 
Within Bloch-Redfield theory, the total dephasing rate due to a noise channel $\la$, is given by $\Gamma^{\la}_{2} = \frac{1}{2}\Gamma^{\la}_{1} + \Gamma_{\varphi}^{\la}$ \cite{wangsness1953dynamical,geva1995relaxation}, where $\Gamma^{\la}_{1}$ is the depolarization rate (combining relaxation and  thermal excitation of the qubit, see section \ref{ssec:depolarization}) and $\Gamma_{\varphi}^{\la}$ the pure dephasing rate. As usual, we  define corresponding decoherence time scales via $T^{\la}_{2} = 1/\Gamma^{\la}_{2}$,  $T^{\la}_{1} = 1/\Gamma^{\la}_{1}$ and $T_{\varphi}^{\la}=1/\Gamma_{\varphi}^{\la}$ respectively. 

Following \cite{Ithier05,Koch2007a,Martinis03}, we may consider dephasing due to classical noise entering the circuit Hamiltonian in the form of an external parameter $\lambda(t) = \lambda_0 + \dl(t)$. Here, $\dl(t)$ is a noise signal assumed to  arise from a stationary, Gaussian process with zero mean, $\langle \dl(t) \rangle = 0$, and spectral density 
\begin{align}
    S_{\la }(\omega) &=  \int_{-\infty}^{\infty}\rmd t\,  e^{-i \omega t} \ave{ 
    \dl (0)\, \dl (t)} .
    \label{eq:spectralDensitLambda}
\end{align}
The effects of weak noise, can be captured through an operator $V_\lambda$ obtained by Taylor-expanding the Hamiltonian,
\begin{equation}
    H \approx H_0 + \frac{\partial H}{\partial \lambda}\dl(t) + \frac{1}{2}\frac{\partial^2 H}{\partial \lambda^2}\dl^2(t) = H_0 + V_\lambda,
\end{equation}
where $H_0=H(\lambda_0)$, and all derivatives are evaluated at 
$\lambda=\lambda_0$. 
Empirical evidence 
 shows that superconducting qubits are typically exposed to multiple noise 
channels with approximate $1/f$ spectrum
\begin{align}
    S_{\!\la}^{\!1/f}(\omega) = \frac{2\pi A_\la^{2} 
    }{\abs{\omega}^{\gamma}}\qquad (\gamma\approx1),
    \label{eq:spectralDensity1overf}
\end{align}
where $A_\la$ is the noise amplitude for channel $\lambda$ (flux, charge, or 
critical current) \cite{Wellstood1987,Zorin1996,Harlingen04,Yoshihara2006,Pourkabirian2014,Kumar2016,hutchings2017tunable}.
Such noise is most detrimental at low frequencies and thus 
important for qubit dephasing.  The calculation of the corresponding 
pure-dephasing rates $\Gamma^{\la}_{\varphi}$ has been developed in references 
\cite{shnirman2002noise,Makhlin04,Ithier05}, and a brief outline is also presented in 
\ref{app:pureDephasing}. The resulting pure-dephasing time, as measured in a 
Ramsey experiment, is given by
\begin{align}
    T_\varphi^\lambda &= \left\{ 2 A_\la^{2} (\partial_\lambda
        \omega_{\rm ge} )^2\abs{\ln \omega_{\rm ir}t} 
    +2 A_\la^{4} (\partial_\lambda^2 \omega_{\rm ge} 
    )^2\left[\ln^2 (\omega_{\rm uv}/\omega_{\rm ir})+ 
2\ln^2(\omega_{\rm ir}t)\right]\right\}^{-1/2},
\label{eq:dephasingTimeFinal}
\end{align}
where $\omega_{\rm ge}$ is the angular frequency difference between the excited 
and ground states of the qubit, $\omega_{\rm ir}$ and $\omega_{\rm uv}$ 
correspond to the low and high-frequency cutoffs of the noise, and $t$ defines 
the measurement time scale under consideration. The above expression is valid 
in the generic case where noise affects the qubit energy to linear order, as 
well as in the vicinity of ``sweet spots'' \cite{Vion2002} where the linear 
noise susceptibility vanishes, $\partial_\lambda\omega_{\rm ge}=0$. In line 
with references \cite{hutchings2017tunable}, 
\cite{quintana2017observation} and \cite{yan2016flux},  we 
assume that 
$\omega_{\rm ir}/2\pi =1\,$Hz, $\omega_{\rm uv}/2\pi=3.0\,$GHz, and use a 
conservative value of $t=10\,\mu$s in our calculations. Next, we consider 
specific 
$1/f$ noise channels 
known to be important, namely $1/f$ charge \cite{Zorin1996,Pourkabirian2014}, 
flux \cite{Wellstood1987,Yoshihara2006,Kumar2016}, and critical-current noise 
\cite{Harlingen04}.

\begin{figure}[b]
    \centering
   \includegraphics[width=2.4in]{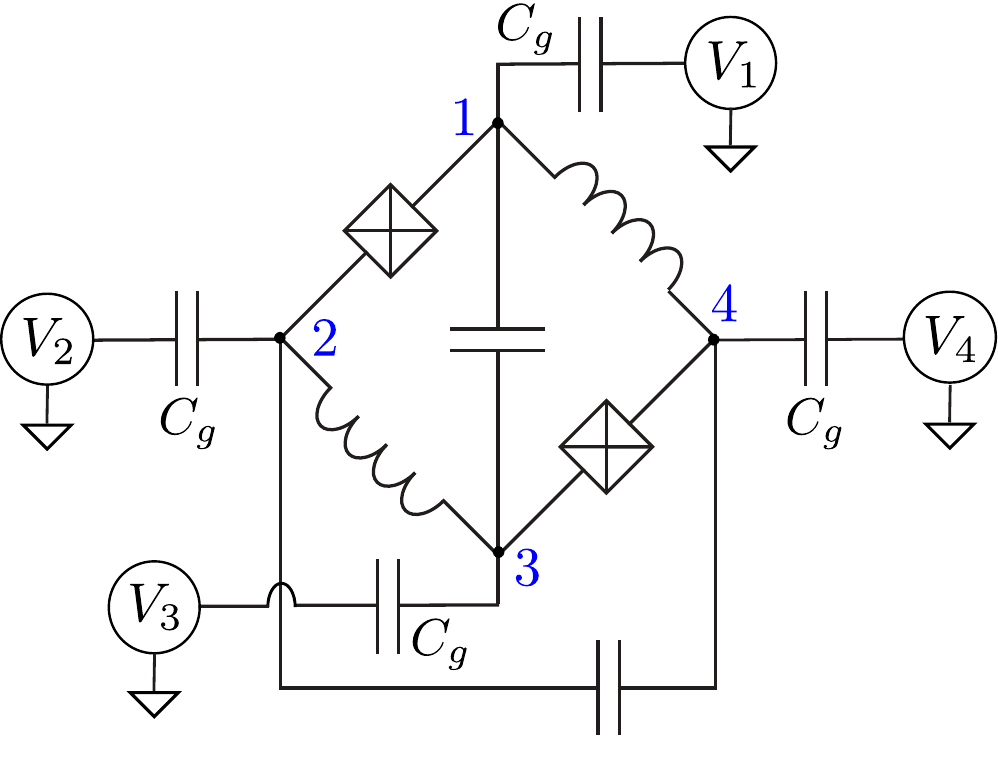}
   \caption{ \zp{} circuit coupled capacitively to voltage sources $V_j(t)$, as 
   used in the modeling of charge noise.}
    \label{fig:zeropicircuitDrive}
\end{figure}

\emph{Charge noise.}--- 
Charge noise can be modeled as a set of noisy voltage sources capacitively 
coupled to the nodes of the circuit, see figure~\ref{fig:zeropicircuitDrive}. 
We assume that the noise signals $V_{j}$ on different circuit nodes are 
independent.  
Repeating the steps of circuit quantization in the presence of these additional capacitive couplings yields a Hamiltonian $H'=T+U$ with 
the same potential energy as previously in $H$, 
\begin{equation}
U= -2\ej\cos\theta\,\cos \left(\phi-\frac{\px}{2} \right)+\el\,\phi^2 + \el\, 
\zeta^{2}  + \ej\, \ddej \sin \theta\, \sin \left(\phi-\frac{\px}{2} \right)  + 
\el\, \ddel\, \phi\, \zeta.
\label{eq:disorderU}
\end{equation}
and a modified expression for the kinetic energy,
\begin{align}
T &= \left(- 2\ecsp\, \partial_{\theta}^{2} - 2 \ecjp\, \partial_{\phi}^{2} - 2 
\ecp\, \partial_{\zeta}^{2}\right)  + \left( 2 \ecsp\, \frac{\ecjp}{\ecj}\, \ddcj\,  
\partial_{\theta}\, 
\partial_{\phi} + 2 \ecsp\, \frac{\ecp}{\ec}\, \ddc\, \partial_{\theta}\, 
\partial_{\zeta}\right)\nonumber\\
&\qquad +\left(  - 4i 
\ecsp\,  n_g^\theta\,  \partial_{\theta}   - 4i 
\ecjp\, n_g^\phi\, \partial_{\phi}  - 4i  \ecp\, 
n_g^\zeta\,  \partial_{\zeta}\right),
\label{eq:drivenH}
\end{align}
Here, primes denote small corrections to charging energies due to the presence 
of the coupling capacitances $C_g$. The effective offset charges $n_g^\phi$, 
$n_g^\theta$, and $n_g^\zeta$ are $1/f$ noise signals, obtained from linear 
superpositions of the fluctuating voltage signals $V_j(t)$. (For details, see 
\ref{app:chargenoise}.). The first group of terms in $T$ comprises 
the kinetic energies of the symmetric \zp{} qubit and of the $\zeta$-mode, the second 
group collects coupling terms originating from capacitive disorder, and the 
third group shows new terms describing the coupling to charge noise. Each 
offset charge may, in principle, consist of an intentional dc bias and $1/f$ 
fluctuations from the environment: $n_g^x(t) = n^x_0 + \delta\mkern-1.0mu 
n_g^x(t)$. Employing the Hamiltonian $H'$, the dephasing due to $n_g^\theta$ 
charge noise can now be calculated 
directly by extracting the $n_g^\theta$ dependence of $\omega_{ge}$, and 
employing equation 
\eqref{eq:dephasingTimeFinal}. We assume a charge-noise amplitude of $A_{n_g^\theta} = 10^{-4} e$ \cite{Zorin1996}. 
Furthermore, we absorb any renormalization of charging energies due to gate capacitance into a 
redefinition of the parameter values given in table~\ref{table:parameters}.

While wave functions of the \zp{} qubit are $2\pi$-periodic in 
$\theta$, they are extended along the $\phi$ and $\zeta$ axis. As a 
consequence, low-frequency charge fluctuations in $\phi$ and $\zeta$ are not 
expected to give rise to significant dephasing 
\cite{koch2009charging,vool2016introduction}. To see this explicitly, we write 
the 
kinetic energy in the form 
$T=\hat{\mathbf{n}}^\top\mathsf{E}\hat{\mathbf{n}} 
+ \mathbf{n}_g^\top\mathsf{D}\hat{\mathbf{n}}$, where 
$\hat{\mathbf{n}}^\top=-i(\partial_\phi,\partial_\theta,\partial_\zeta)$, 
$\mathbf{n}_g^\top=(n_g^\phi,n_g^\theta,n_g^\zeta)$, and $\mathsf{E}$ 
($\mathsf{D}$) is a symmetric (diagonal) 3$\times$3 matrix of energy 
coefficients to be read off of equation \eqref{eq:drivenH}. We may complete the 
square,
\begin{equation}
T=(\hat{\mathbf{n}} + \tfrac{1}{2} 
\mathsf{E}^{-1}\mathsf{D}\mathbf{n}_g)^\top\mathsf{E}(\hat{\mathbf{n}}+\tfrac{1}{2}\mathsf{E}^{-1}\mathsf{D}\mathbf{n}_g)
 -c\,\mathbb{I}
\end{equation}
and drop the irrelevant c-number term, and finally perform a unitary 
transformation using $R = \exp\big[ 
-i\tfrac{1}{2}(\phi,\theta,\zeta)\mathsf{E}^{-1}\mathsf{D}\mathbf{n}_g\big]$,
which produces a momentum shift $ \hat{\mathbf{n}} \to \hat{\mathbf{n}} - 
\tfrac{1}{2} 
\mathsf{E}^{-1}\mathsf{D}\mathbf{n}_g$. We find the resulting Hamiltonian
\begin{align}
    H'' = R^{\dagger} H' R + i \dot{R^\dagger} R 
    &= \hat{\mathbf{n}}^\top\mathsf{E}\hat{\mathbf{n}} 
    -\tfrac{1}{2}(\phi,\theta,\zeta)\mathsf{E}^{-1}\mathsf{D}\dot{\mathbf{n}}_g
     + U 
\label{eq:HeffTransformed1}
\end{align}
and note that the transformation does not affect the boundary conditions for 
the extended variables $\phi$ and $\zeta$ ($L^2$-integrability). Hence, the 
transformation reveals that fluctuations in $n_g^\phi$ and $n_g^\zeta$  
only enter in terms of the time derivative of the noise.
As discussed previously in reference \cite{koch2009charging}, the $1/f$ 
charge-noise spectrum thereby transforms into an Ohmic spectral density,  
$S_{\dot{n}_g}(\omega) = \omega^{2} S_{n_g}^{1/f}(\omega) \sim 
\omega$. The effect of such fluctuations is insignificant for dephasing, since 
$T_\varphi\sim S(0)$ for non-singular noise spectral densities.

\emph{Critical-current noise.}--- Next, we consider $1/f$ noise in the critical 
current $I_c=2\pi\ej/\Phi_0$ characterizing the two Josephson junctions in the 
\zp{} circuit. Microscopically, fluctuations in the critical current are 
suspected to be due to trapping/de-trapping of charges at defect sites in the 
tunneling barrier 
of junctions \cite{Harlingen04,nugroho2013low,Martinis03}. While trapped,
charges block tunneling through a given region of the junction, thus reducing 
the effective junction area. Under suitable condition, the ensemble dynamics of 
many trapping centers can give rise to $1/f$ noise 
\cite{dutta1981low,weissman19881}. In this case, the Josephson energy is the 
Hamiltonian parameter that acquires a fluctuating component, 
$E_\text{J,tot}=\ej+\dej(t)$. Critical-current noise is thus amenable to the 
same treatment as 
charge noise. In our calculations, we use a typical noise amplitude for 
the critical current of $A_{I_{c}}=10^{-7}\,I_{c}$ \cite{Harlingen04,Koch2007a}.

We note that critical-current noise may, in principle, also affect the large 
inductors, if realized as a Josephson junction array. However, for uncorrelated 
noise affecting each of the array's $N_\text{J}\gg1$ junctions independently, 
one finds an overall suppression of the  noise amplitude by a factor of 
$1/\sqrt{N_\text{J}}$  \cite{manucharyan2012superinductance}. We will see that 
critical-current noise is a subdominant noise channel even without this 
suppression, and hence, neglect the effect of such fluctuations on 
superinductances.

\emph{Flux noise.}--- The  third canonical $1/f$ noise source known to affect 
superconducting qubits is $1/f$ flux noise. We model the fluctuations of the 
magnetic flux through the loop enclosed by the two junctions and inductors by 
treating  
$\Phi_{\rm ext}$ as the noisy parameter $\lambda$. Flux noise is ubiquitous in 
current superconducting circuit devices. There is growing evidence that 
fluctuating spins on thin-film surfaces 
\cite{sendelbach2008magnetism,Kumar2016,hutchings2017tunable} may be the 
microscopic origin of this noise. In our calculations of pure dephasing times 
due to flux noise, we make again use of equation \eqref{eq:dephasingTimeFinal} 
with a typical noise amplitude of  $A_{\Phi_{\rm ext}} = 1\,\mu\Phi_{0}$ 
\cite{hutchings2017tunable}. 

\emph{Shot-noise dephasing due to thermal excitations of the $\zeta$-mode.}--- 
The dephasing channels discussed so far are of the direct kind, shown in figure 
\ref{fig:noiseMechanisms}(a). We next analyze an indirect source associated 
with the disorder-induced coupling to the $\zeta$-mode. Since the \zp{} qubit 
is operated in the regime of small $\el$ and $\ec$, the $\zeta$-mode with 
frequency $\Omega_\zeta=\sqrt{8\ec\el}/\hbar$ is generally a low-frequency 
mode, and can be subject to significant thermal excitations. Specifically, for 
the three parameter sets, the $\zeta$-mode frequencies are given by 
$\Omega_{\zeta}/2\pi=36,\,113,\,395\,$MHz, 
leading to average thermal occupation numbers of $n_\text{th}=8.25,\,2.29,\,0.39$,
respectively (with assumed temperature of $T=15\,$mK). Dephasing of the primary \zp{} 
degrees of freedom from thermal fluctuations can be significant when 
operating in the strong dispersive limit, where the qubit-state dependent shift 
of the $\zeta$-mode frequency is large compared to the width of the $\zeta$-mode 
resonance. In that limit, the addition/loss of a single 
$\zeta$-mode excitation number essentially measures the qubit state, leading to 
complete dephasing.
This noise mechanism, referred to as shot-noise dephasing, can be modeled 
within the master equation formalism 
 and produces pure dephasing at the rate \cite{Rigetti2012,Clerk2007}
\begin{equation}
\Gamma^{\rm SN}_{\varphi} = \frac{\kappa_{\zeta} }{2}\, {\rm Re}\, \left[ \sqrt{ 
\left( 1 + \frac{2 i \chi_{01}}{\kappa_{\zeta}}  \right)^{2} +  \frac{8i 
\chi_{01} n_{\rm th}(\Omega_{\zeta})}{\kappa_{\zeta}}  }  - 1 \right].
\label{eq:T2starRateshotnoise}
\end{equation}
Here, $\kappa_{\zeta}$ is the intrinsic lifetime of the harmonic $\zeta$-mode, 
$n_{\rm th}(\omega)=1/[\exp\left( \hbar \omega / k_{B} T \right) - 1] $  the average 
number of thermal photons with (angular) frequency $\omega$ in thermal 
equilibrium at temperature $T$, and $\chi_{01} = \left(\chi_{1} - \chi_{0}\right)/2$ the 
qubit's ac Stark shift due to a single excitation. (See 
equation~\eqref{eq:chiAndLambda} for the definition of $\chi_{l}$.)  
Equation \eqref{eq:T2starRateshotnoise} applies whenever the
\zp{} qubit and $\zeta$-mode are coupled dispersively. It can be further 
simplified in the strong dispersive limit where $\kappa_{\zeta} \gg \chi_{01}  
$, 
and written as 
\begin{equation}
    \Gamma^{\rm SN}_{\varphi} \approx \frac{1}{\kappa_{\zeta}} 4 \chi_{01}^{2} n_{\rm th}(\Omega_{\zeta}) \left(n_{\rm th}(\Omega_{\zeta}) + 1\right) ,
    \label{eq:T2starRateshotnoiseSimp1}
\end{equation}
while in the opposite limit $\chi_{01} \gg \kappa_{\zeta}$, as
\begin{equation}
    \Gamma^{\rm SN}_{\varphi} \approx \kappa_{\zeta} n_{\rm th}(\Omega_{\zeta}).
    \label{eq:T2starRateshotnoiseSimp2}
\end{equation}
From the above equations, we see that both $\chi_{01}$, as well as the thermal occupation $n_{\rm th}(\Omega_{\zeta})$ can play a crucial role in determining the strength of the resulting dephasing rate. 

\subsection{Depolarization ($T_{1}$) }
\label{ssec:depolarization}
Decoherence due to depolarization comprises of processes associated with 
spontaneous transitions between energy eigenstates. Such transitions may occur within the two-level subspace of the \zp{} qubit, or lead to leakage to states outside of this subspace. 
The characteristic time scale for depolarization is 
the $T_{1}$ time \cite{Ithier05}. We define the operator coupling the \zp{} circuit degrees of freedom to noise channel labeled $\la$ as $V_\la = G_\la\, \dl$, where 
$G_\la$ is an operator on the
Hilbert space spanned by $\theta$, $\phi$ and $\zeta$. $\dl$ refers to the bath degrees of freedom and may be an operator acting on the Hilbert space of the bath, or a classical, stochastic variable with appropriately chosen statistics. Using
Fermi's Golden Rule, one obtains the rate for transitions from the initial state $\ket{\psi_i}$ to a 
 final  state $\ket{\psi_f}$ \cite{Ithier05,Clerk2007,schoelkopf2003noise} as
\begin{align}
    \gamma^{\la,\pm}_{i \rightarrow f} = \frac{1}{\hbar^{2}} \abs{\bra{\psi_{f}} G_\la \ket{\psi_{i}} }^{2} S_{\la } \left(\mp \abs{\omega_{fi}} \right).
    \label{eq:transitionRate}
\end{align}
Here, initial and final states are eigenstates of the full \zp{} Hamiltonian, equation \eqref{eq:Hqubitzeta}, with eigenenergy difference $\hbar \omega_{fi} = E_{f} - E_{i}$, and $S_{\la}(\omega)$ is the noise spectral density, see equation \eqref{eq:spectralDensitLambda}. The coupling operator $G_\la$ and spectral density $S_{\la}(\omega)$ depend on the specific noise channel and its statistical properties.
Furthermore, the $\pm$ notation describes whether the rate is upwards ($\gamma^{\la, +}_{i\rightarrow f}$), where $E_{f}>E_{i}$, or downwards ($\gamma^{\la, -}_{i\rightarrow f}$), where $E_{f}<E_{i}$.

We expect to operate the \zp{} qubit in the dispersive regime with respect to the $\zeta$-mode (see equation \eqref{eq:dispH}). In such case, dressed states can be suitably labeled by excitation numbers $n$ and $l$ referring to  $\zeta$-mode and primary \zp{} subspace, respectively: $\ket{\psi_j}=\ket{\psi_{l,n}}$. In practice, we base the assignment of labels $l,\,n$ on the maximum overlap between exact eigenstates of the full Hamiltonian \eqref{eq:Hqubitzeta} and bare product states $\ket{l}_{0\pi}\otimes \ket{n}_\zeta$. (Alternatively, perturbation theory can be used, see \ref{app:purcellRates}.) We may thus write the above transition rates in the form $\gamma^{\la, \pm}_{l,n \rightarrow l',n'} = \gamma^{\la, \pm}_{i \rightarrow f}$.

Since we aim to evaulate the depolarization of the primary \zp{} degrees of freedom, i.e., transitions which change the state index $l$, we define the composite transition rate
\begin{align}
    \Gamma^{\la}_{l \rightarrow l'} = \sum_{n, n'} P_{\zeta}(n)\, \gamma^{\la, \pm}_{l,n \rightarrow l',n'},
    \label{eq:transitionRateQubit}
\end{align}
which includes a summation over all the $\zeta$-mode states $n$ and $n'$, where 
each initial $\zeta$-mode state is weighted by the thermal occupation probability $P_{\zeta}(n) = \left[1 
- \exp(-\hbar \Omega_{\zeta}/k_{B} T) \right] \exp(- n \hbar \Omega_{\zeta}/k_{B} 
T)$, with $k_{B}$ denoting Boltzmann's constant. 
Finally, we define an effective depolarization rate $\Gamma^{\la}_{1}$ and corresponding time $T^{\la}_{1}=1/\Gamma^{\la}_{1}$ for noise channel $\lambda$ as\footnote{Often, depolarization rates are exclusively based on transitions within the two-level subspace \cite{Ithier05}. In the \zp{} qubit, transitions to states outside this subspace can be dominant and must be included. 
}
\begin{align}
    \Gamma^{\la}_{1} = \Gamma^{\la}_{1 \rightarrow 0} + \Gamma^{\la}_{0 \rightarrow \rm up} + \Gamma^{\la}_{1 \rightarrow \rm up}.
    \label{eq:Gamma1}
\end{align}
Here, $\Gamma^{\la}_{1 \rightarrow 0} $ is the ordinary qubit relaxation rate, and $\Gamma^{\la}_{0 \rightarrow \rm up}$,  $\Gamma^{\la}_{1 \rightarrow \rm up}$ are the excitation rates from ground and first excited state to all higher levels. For the \zp{} qubit, we find that upward transitions to states outside the two-level subspace typically dominate over the downward rate $\Gamma_{1\rightarrow 0}^{\la}$, even at low temperatures. This is precisely due to the disjoint-support of the eigenstates with $l=0,1$ and the resulting exponential suppression of the corresponding matrix elements in equation~\eqref{eq:transitionRate}. We elaborate on this further in section~\ref{sec:decoherenceResults}. 

\emph{Depolarization from critical-current noise.}--- Based on these considerations, we can assess the effects of critical-current noise on qubit depolarization. Similar to section~\ref{ssec:dephasing}, we expand the critical current into a static and a fluctuating part,  $I_{c,\text{tot}} = I_{c} + \delta I_{c}$. Keeping terms up to leading order, we can write the interaction  $V_{I_c}$ as\footnote{In the regime of weak parameter disorder considered here, we may neglect the fact that the \zp{} circuit has two independent junctions, and instead associate a single random noise process with the mean critical current.}
\begin{align}
    V_{I_c} &=  \frac{\partial H}{\partial {I_{c}}} \delta I_{c}  =  G_{I_{c}} \delta I_{c} 
    =  \left[ - \frac{\Phi_{0}}{\pi} \cos \theta\, \cos\left( \phi - \frac{\varphi_{\rm ext}}{2} \right)  +  \frac{\Phi_{0}}{2\pi} \, \ddej \sin\theta \,\sin\left( \phi - \frac{\varphi_{\rm ext}}{2} \right)  \right] \delta I_{c}.
    \label{eq:vNoiseCC}
\end{align}
Employing equations~\eqref{eq:spectralDensity1overf}, \eqref{eq:transitionRateQubit} and \eqref{eq:Gamma1}, this enables us to calculate the depolarization rate $\Gamma_{1}^{I_{c}}$ due to $1/f$ critical-current noise.

\emph{Depolarization from flux noise.}--- In analogous fashion, we characterize  depolarization due to flux noise. Identifying $\lambda$ as the external flux $\Phi_{\rm ext}$, and assuming $\Phi_{\rm ext,tot} = \Phi_{\rm ext} + \delta \Phi_{\rm ext}$, we obtain the coupling operator\footnote{We stress that there is an ambiguity in expression~\ref{eq:vNoiseFlux} that arises from the choice of flux grouping with different terms of the Hamiltonian. The details related to such flux grouping may be covered in a future publication. }
\begin{align}
    V_{\Phi_\text{ext}} &=  \frac{\partial H}{\partial \Phi_{\rm ext}} \delta \Phi_{\rm ext} =  G_{\Phi_{\rm ext}}\, \delta \Phi_{\rm ext} 
    =  \left[ - \frac{2\pi\ej}{\Phi_{0}} \cos \theta \sin \left( \phi - \frac{\varphi_{\rm ext}}{2} \right)  - \frac{\pi \ej}{ \Phi_{0}} \ddej \sin \theta \cos \left( \phi - \frac{\varphi_{\rm ext}}{2} \right)  \right] \delta \Phi_{\rm ext}.
    \label{eq:vNoiseFlux}
\end{align}
For flux nose, two different noise channels may be considered: flux noise due to current-fluctuations in the flux-bias line, as well as $1/f$ noise. For the former, we consider fluctuations in magnetic flux due to Ohmic current noise in the flux-bias line  which couples to the \zp{} circuit via a mutual inductance $M$ \cite{Koch2007a}. The spectral density for such current noise can be described by
\begin{align}
    S_{I}^{\rm Ohm}(\omega) = \frac{2 \hbar \omega }{R} \left[ 1 + \coth\left(\frac{\hbar \omega} { 2 k_{B} T }\right) \right],
    \label{eq:spectralDensityCurrentOhmic}
\end{align}
where $R$ is taken as $50\,\Omega$. This leads to the flux noise spectral density of $S_{\Phi_{\rm ext}}(\omega) = M^{2} S^{\rm Ohm}_{I}(\omega)$. We will assume a mutual inductance between the qubit loop and the biasing line to be $M=1000\,\Phi_{0}/{\rm A }$. Together, this allows us to calculate the flux-noise depolarization rate $\Gamma_{1}^{\Phi_{\rm ext}, {\rm Ohm}}$. The analysis of $1/f$ intrinsic flux noise proceeds in a straightforward way from equations~\eqref{eq:spectralDensity1overf} and \eqref{eq:vNoiseFlux}, leading to a depolarization rate $\Gamma_{1}^{\Phi_{\rm ext}}$. 

\emph{Purcell depolarization via $\zeta$-mode.}--- Depolarization of the qubit may also occur due to processes analogous to Purcell decay, since the \zp{} qubit's $\theta$ and $\phi$ degrees of freedom are coupled to the harmonic $\zeta$-mode which, itself, is subject to intrinsic decay with rate $\kappa_{\zeta}$. The resulting relaxation and excitation rates are enhanced or suppressed depending how dispersive the coupling to the $\zeta$-mode is. We show in \ref{app:purcellRates} that the resulting rates for upward and 
downward transitions can be written as 
\begin{align}
    \Gamma^{{\rm Purcell}}_{l\rightarrow l'} &= \kappa_{\zeta} n_{\rm th}(\omega^{\rm q}_{l'l}) \sum_{n, n'} P_{\zeta}(n)  \abs{ \bra{\psi_{l',n'}} a\dg \ket{\psi_{l, n}} }^{2}   \approx \kappa_{\zeta}\,n_{\rm th}(\omega^{\rm q}_{l'l}) \frac{\abs{g_{ll'}}^{2}}{\abs{E_{l}^{q} - E_{l'}^{q} + \hbar \Omega_{\zeta}}^{2}} 
    \label{eq:purcellRatePlus}
\end{align}
in the case of $E^{q}_{l'}>E^{q}_{l}$, and 
\begin{align}
    \Gamma^{{\rm Purcell}}_{l\rightarrow l'} &= \kappa_{\zeta} \left[ n_{\rm th} (\omega^{\rm q}_{ll'} )  + 1 \right] \sum_{n, n'} P_{\zeta}(n)  \abs{ \bra{\psi_{l',n'}} a \ket{\psi_{l, n}} }^{2}   \approx \kappa_{\zeta} \left[ 1 +  n_{\rm th}(\omega^{\rm q}_{ll'}) \right] \frac{\abs{g_{ll'}}^{2}}{\abs{E_{l}^{q} - E_{l'}^{q} - \hbar \Omega_{\zeta}}^{2}} 
    \label{eq:purcellRateMinus}
\end{align}
in the case of $E^{q}_{l'} < E^{q}_{l}$. In the above expressions, we use  $\omega^{\rm 
q}_{ll'} = (E^{\rm q}_{l} - E^{\rm q}_{l})/\hbar$, and sum over $\zeta$-mode occupation numbers with the appropriate thermal weights, analogous to our previous treatment in equation~\eqref{eq:transitionRateQubit}. The approximations given in equations \eqref{eq:purcellRatePlus} and \eqref{eq:purcellRateMinus} can be obtained with perturbation theory (see \ref{app:purcellRates}). 
Summation as indicated in equation~\eqref{eq:Gamma1} then allows us to obtain the effective depolarization rate due to the $\zeta$-mode mediated Purcell effect, $\Gamma^{\rm Purcell}_{1}$.

\section{Calculated coherence times}
\label{sec:decoherenceResults}

\begin{figure}
    \captionsetup[subfigure]{position=top,singlelinecheck=off,topadjust=-5pt,justification=raggedright}
    \centering
    \subfloat[]{    \hspace*{-0.50cm}    \includegraphics[width=2.40in]{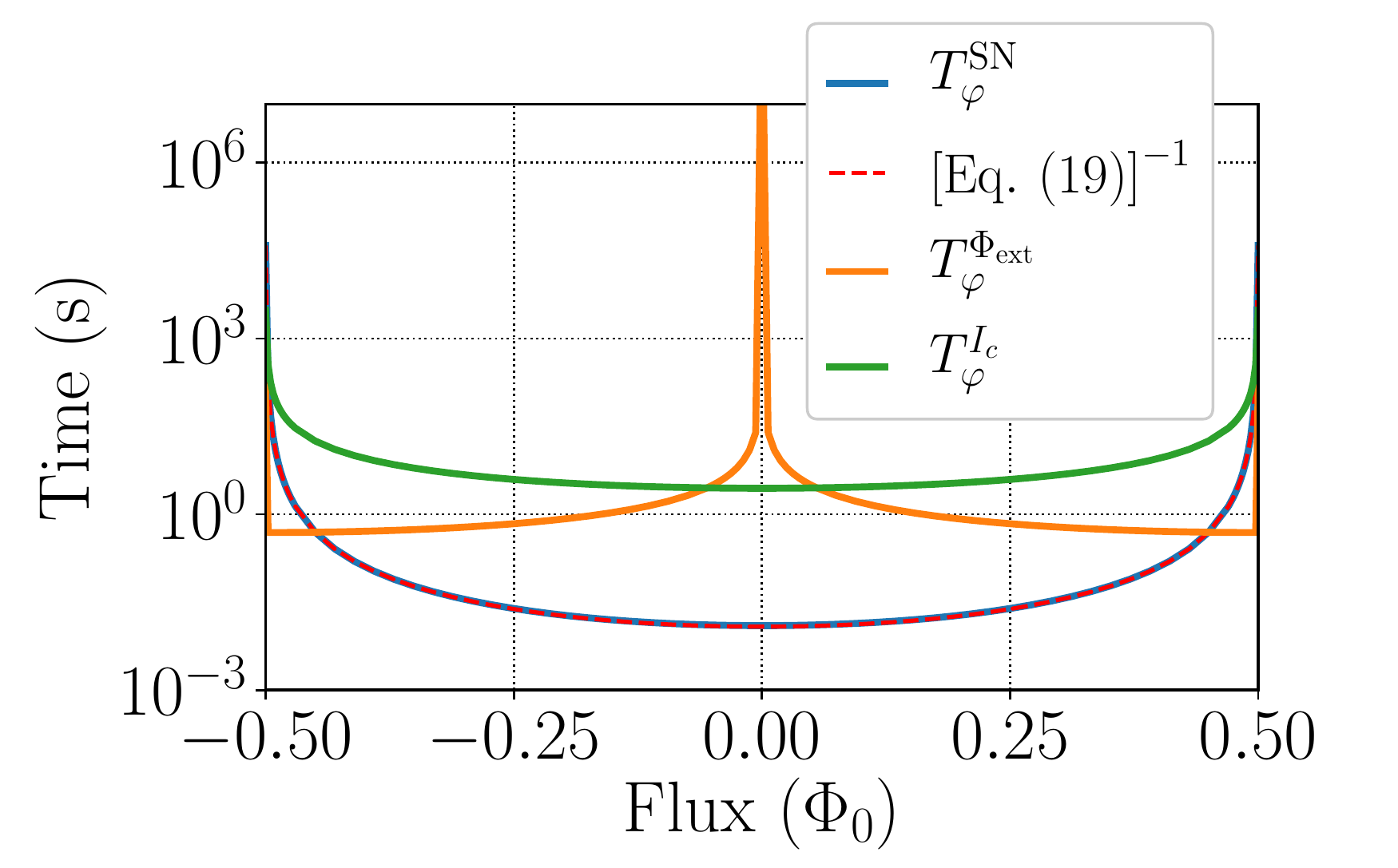}}
    \hspace*{-0.10cm}
    \subfloat[]{    \hspace*{-0.50cm}    \includegraphics[width=2.40in]{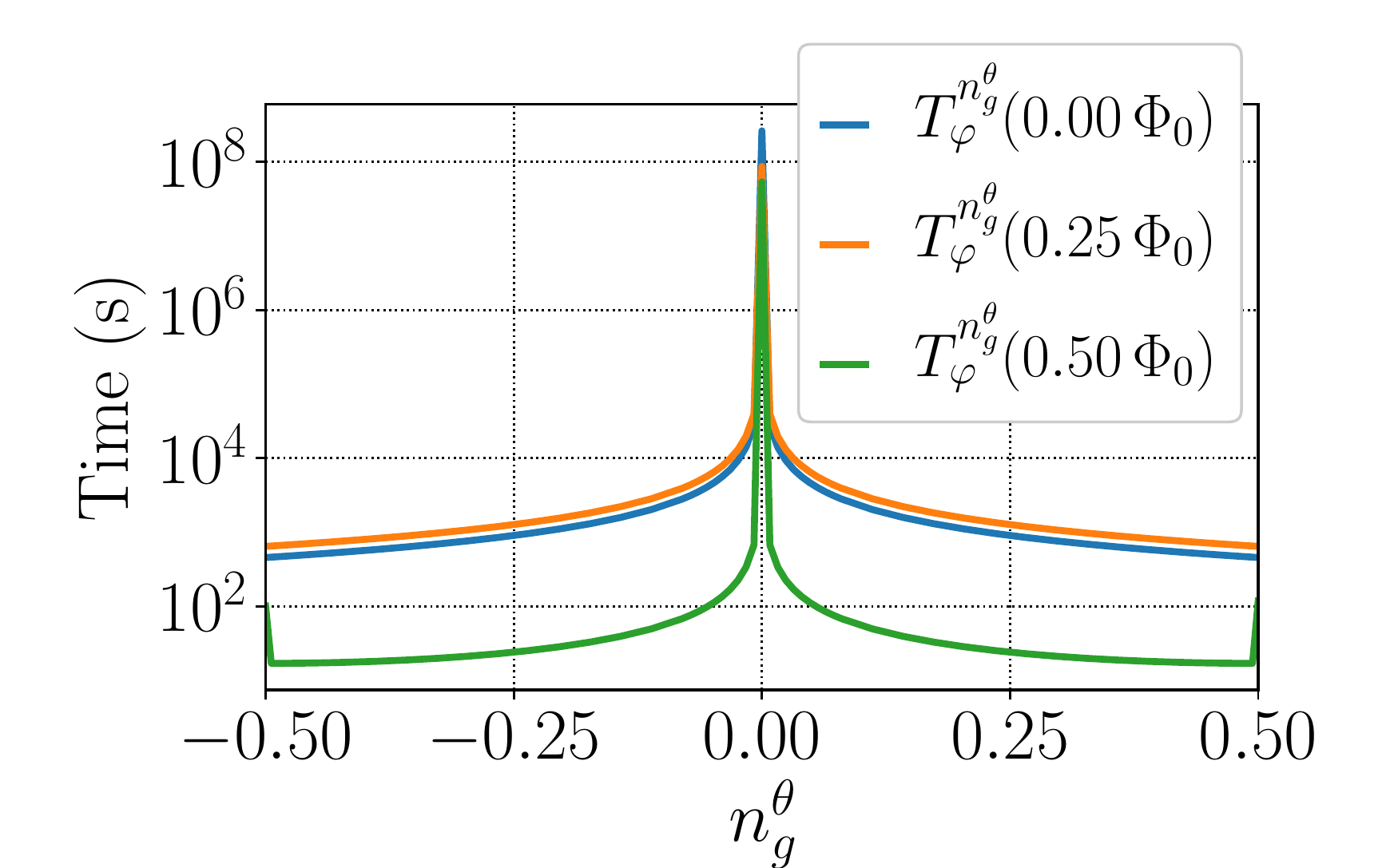}} 
    \hspace*{-0.10cm}
    \subfloat[]{   \hspace*{-0.50cm}     \includegraphics[width=2.40in]{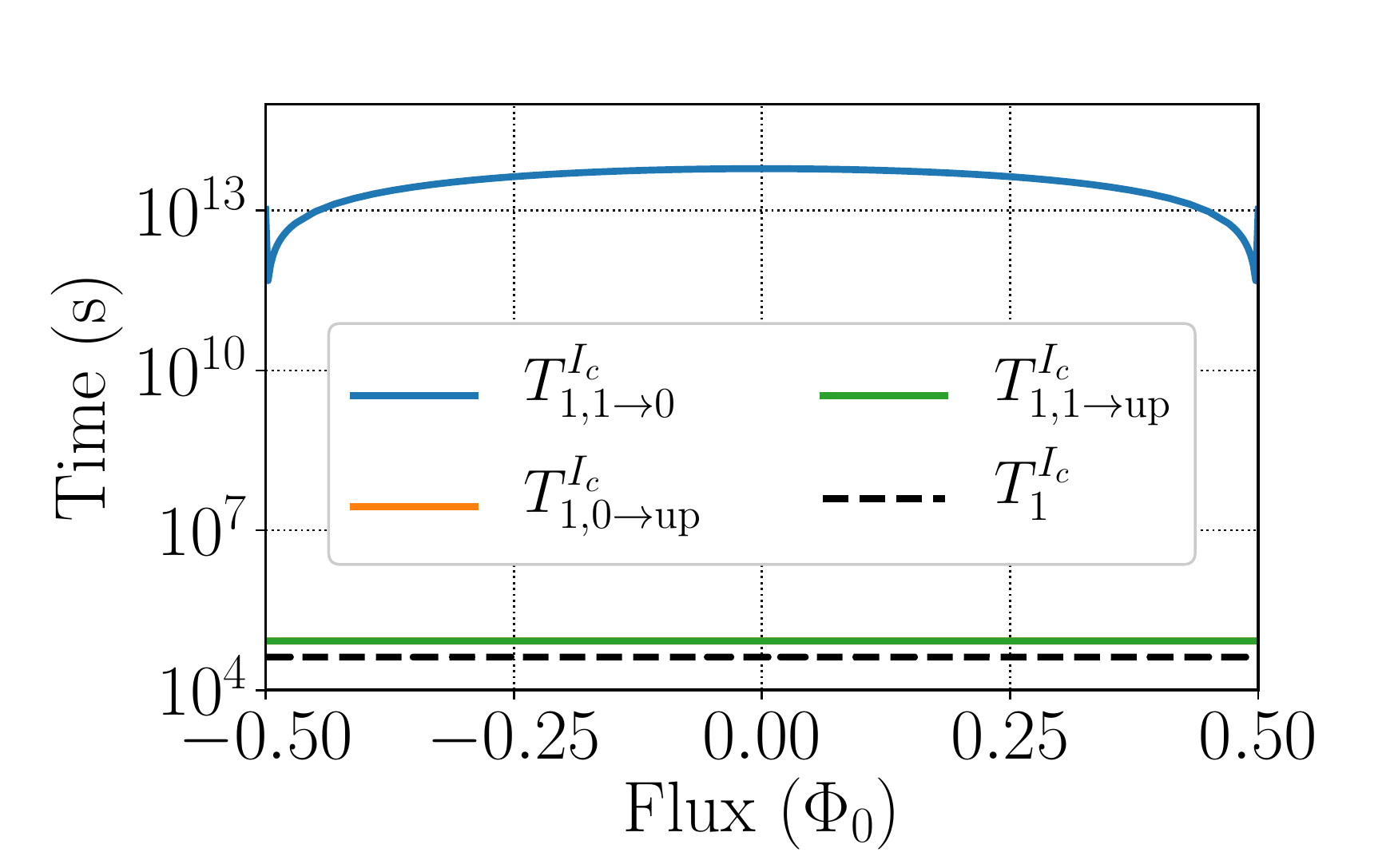}} \\
    \vspace{-0.5cm}
    \subfloat[]{    \hspace*{-0.50cm}    \includegraphics[width=2.40in]{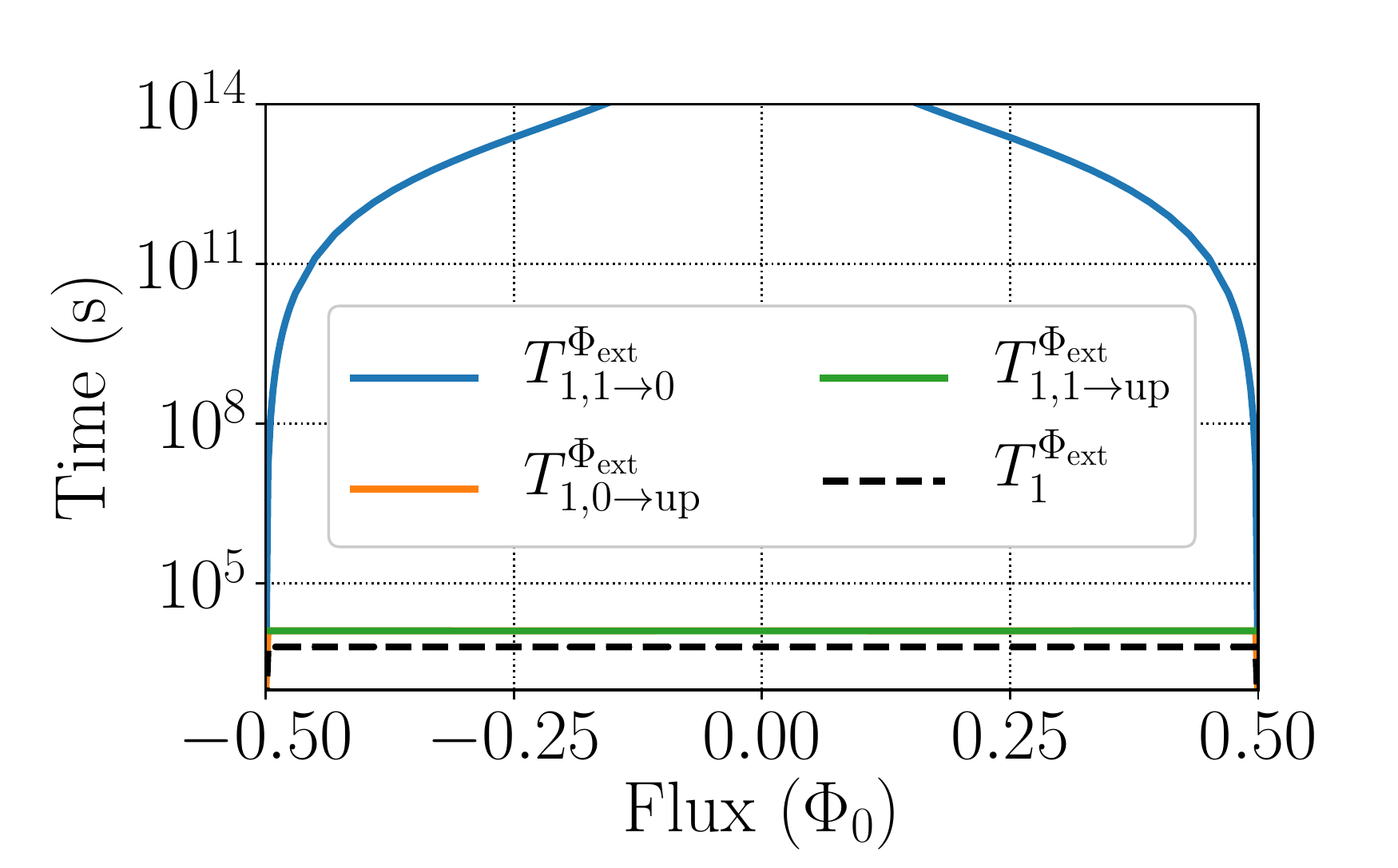}} 
    \hspace*{-0.10cm}
    \subfloat[]{   \hspace*{-0.50cm}     \includegraphics[width=2.40in]{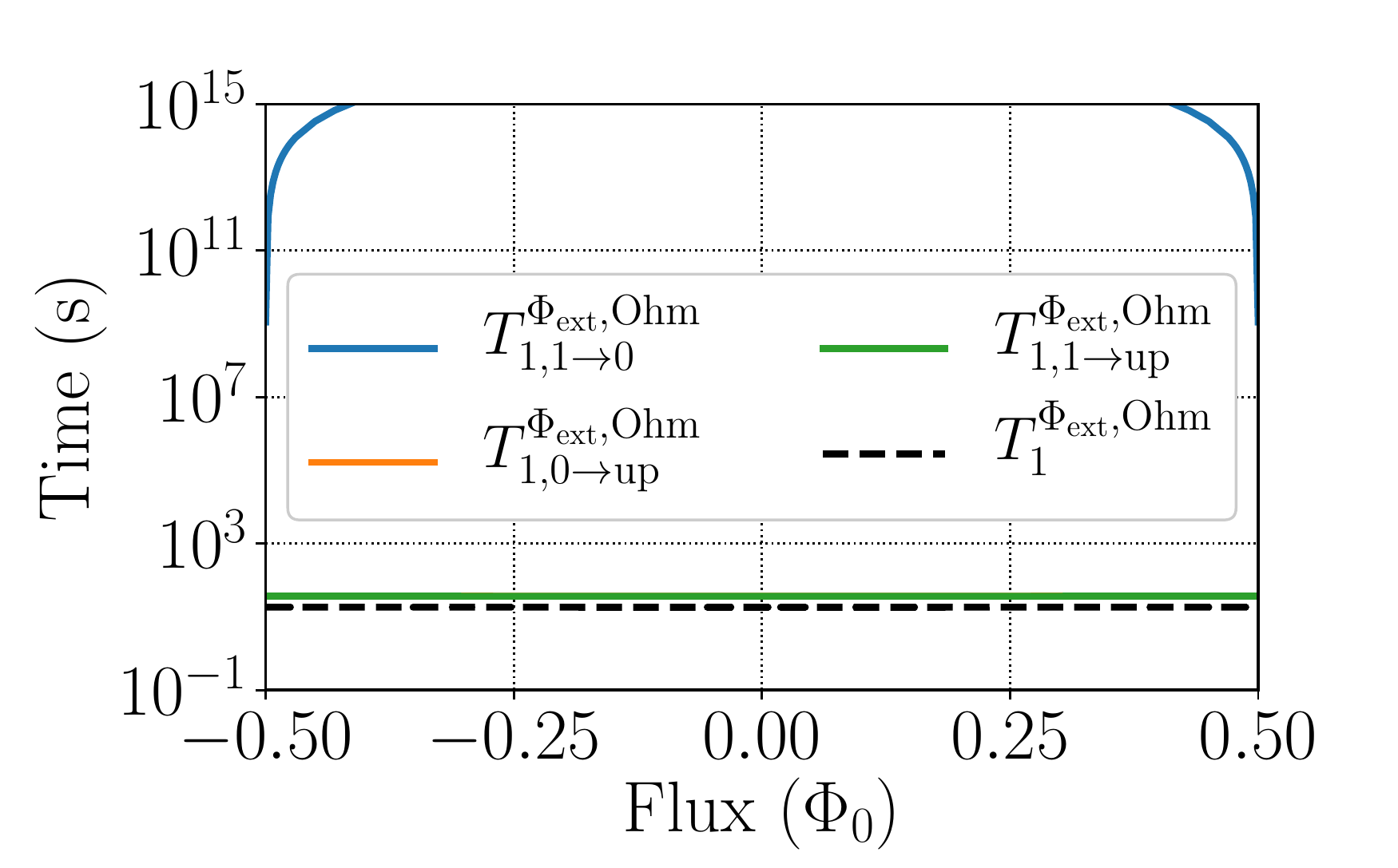}}
    \hspace*{-0.10cm}
    \subfloat[]{   \hspace*{-0.50cm}      \includegraphics[width=2.40in]{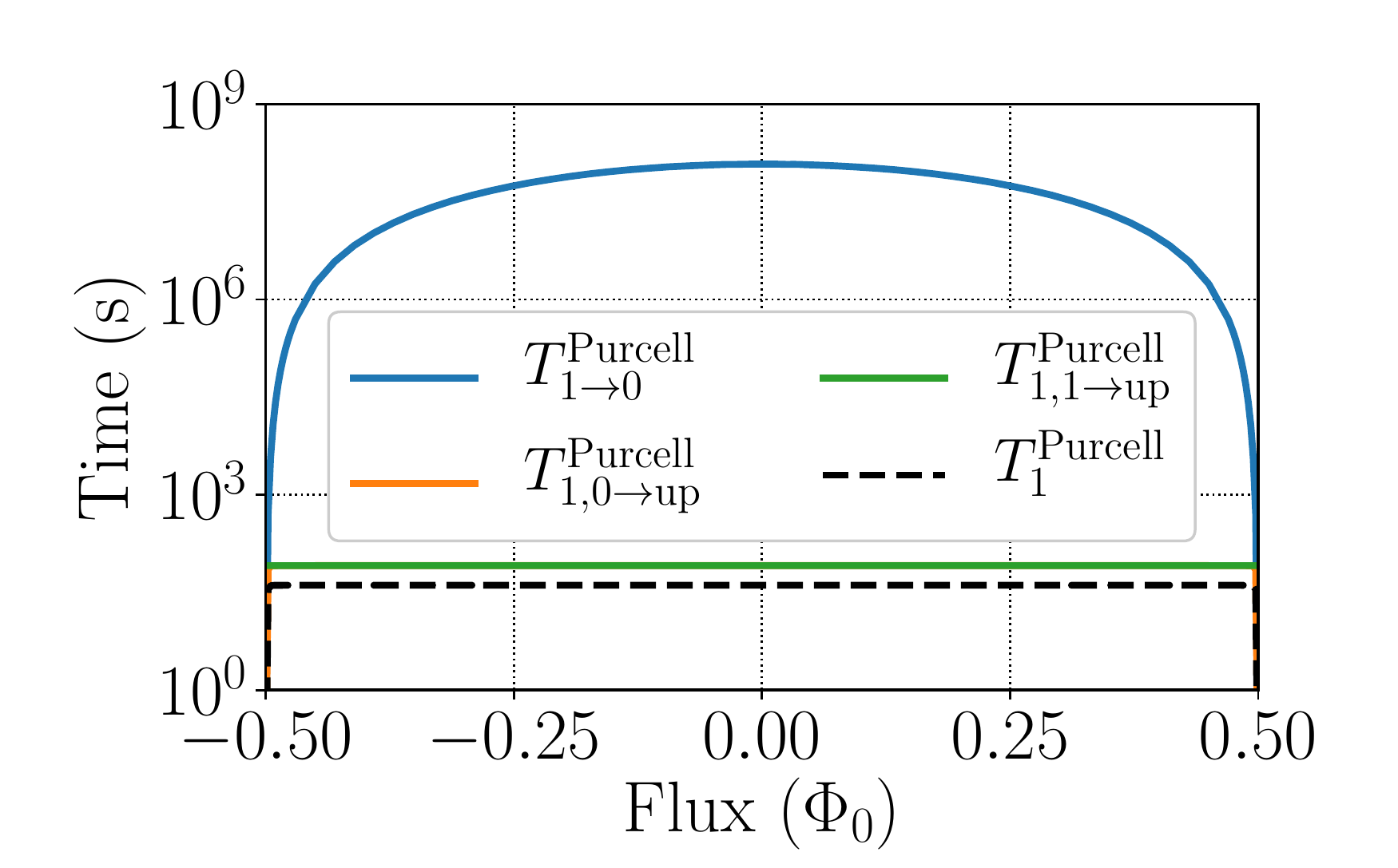}}
    \caption{Calculated coherence times for parameter set 1. (a) Pure dephasing times due to $1/f$ flux noise $\tphiflux$ (orange curve),  $1/f$ critical current noise $\tphicc$ (green curve) as well as shot noise $\tphisn$ (blue curve), with its approximation $\kappa_{\zeta} \left[  4 \chi_{01}^{2} n_{\rm th}(\Omega_{\zeta})  \left(n_{\rm th}(\Omega_{\zeta}) + 1 \right) \right]^{-1} $ from equation~\eqref{eq:T2starRateshotnoiseSimp1} (dashed red line), valid when $\chi_{01} \ll \kappa_{\zeta}$. (b) Pure dephasing time due to $1/f$, charge noise $\tphing$ along the $\theta$ direction, plotted as a function of $n_{g}^{\theta}$ and calculated at $\Phi_{\rm ext} = 0.0\,\Phi_{0}$ (blue curve), $\Phi_{\rm ext} = 0.25\,\Phi_{0}$ (orange curve), and $\Phi_{\rm ext} = 0.50\,\Phi_{0}$ (green curve). (c) $T_{1}$ due to $1/f$ critical current noise. (d) $T_{1}$ due to $1/f$ flux noise. (e) $T_{1}$ due to biasing flux line noise. (f) Purcell depolarization time. Plots in (c)--(f), show transition times of states $1$ to $0$ (blue curves), $0 \rightarrow {\rm upwards}$ (orange curves), $1 \rightarrow {\rm upwards}$ (green curves), and finally effective (combined) times (dashed black curves). See main text for analysis. }
     
    \label{fig:resultsRatesPS1}
\end{figure}

\begin{figure}
    \captionsetup[subfigure]{position=top,singlelinecheck=off,topadjust=-5pt,justification=raggedright}
    \centering
    \subfloat[]{    \hspace*{-0.50cm}    \includegraphics[width=2.40in]{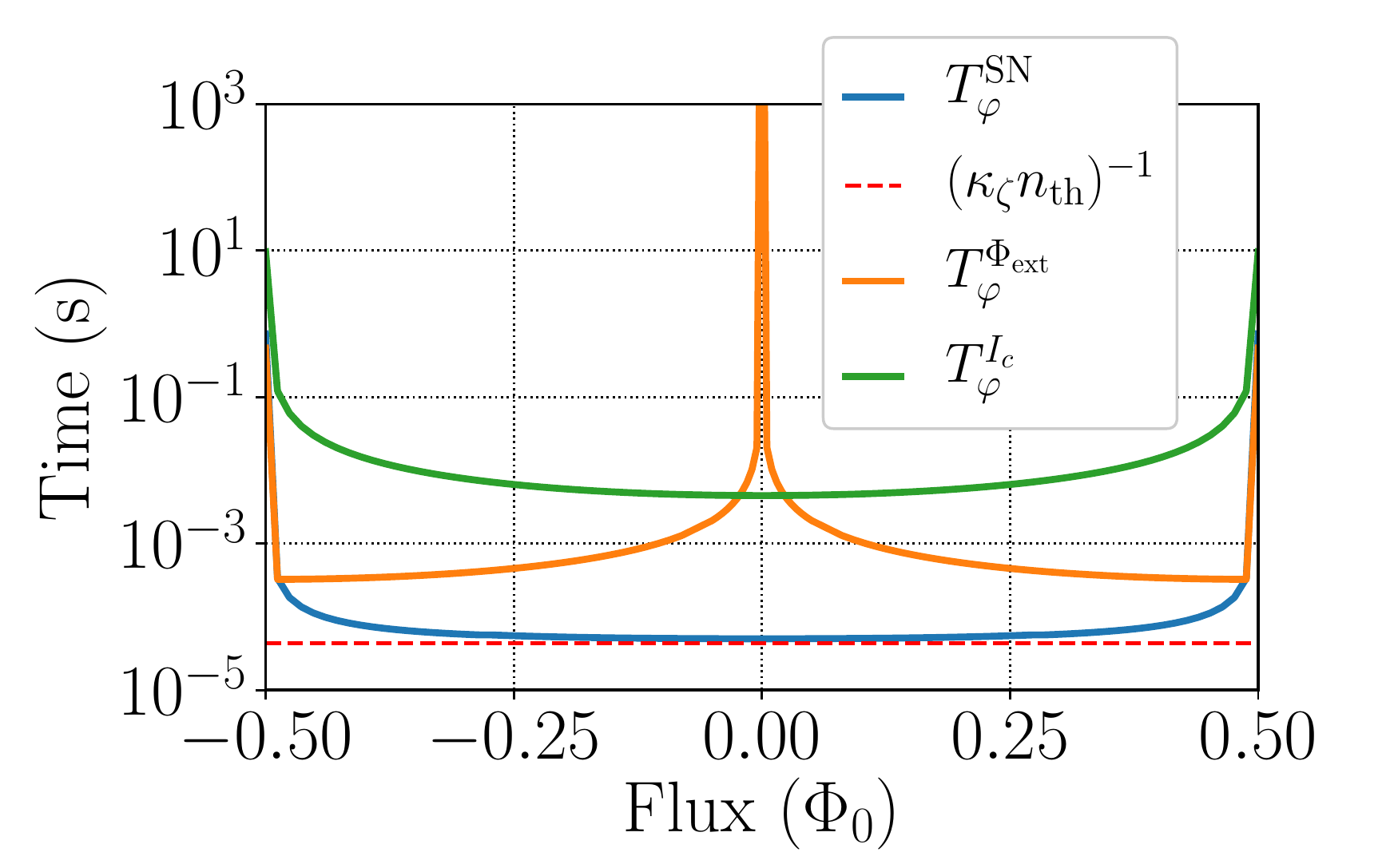}}
    \hspace*{-0.10cm}
    \subfloat[]{    \hspace*{-0.50cm}    \includegraphics[width=2.40in]{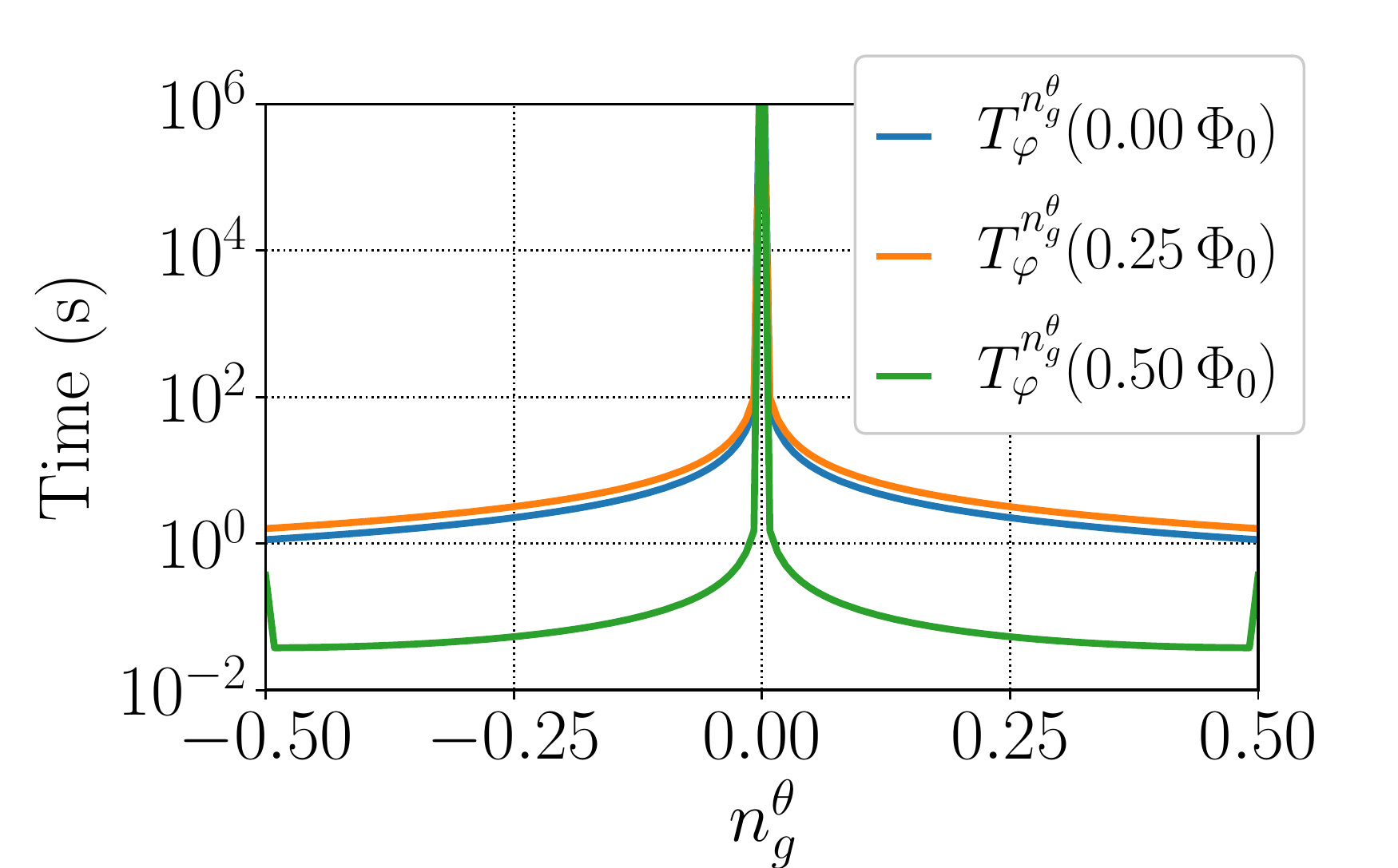}} 
    \hspace*{-0.10cm}
    \subfloat[]{   \hspace*{-0.50cm}     \includegraphics[width=2.40in]{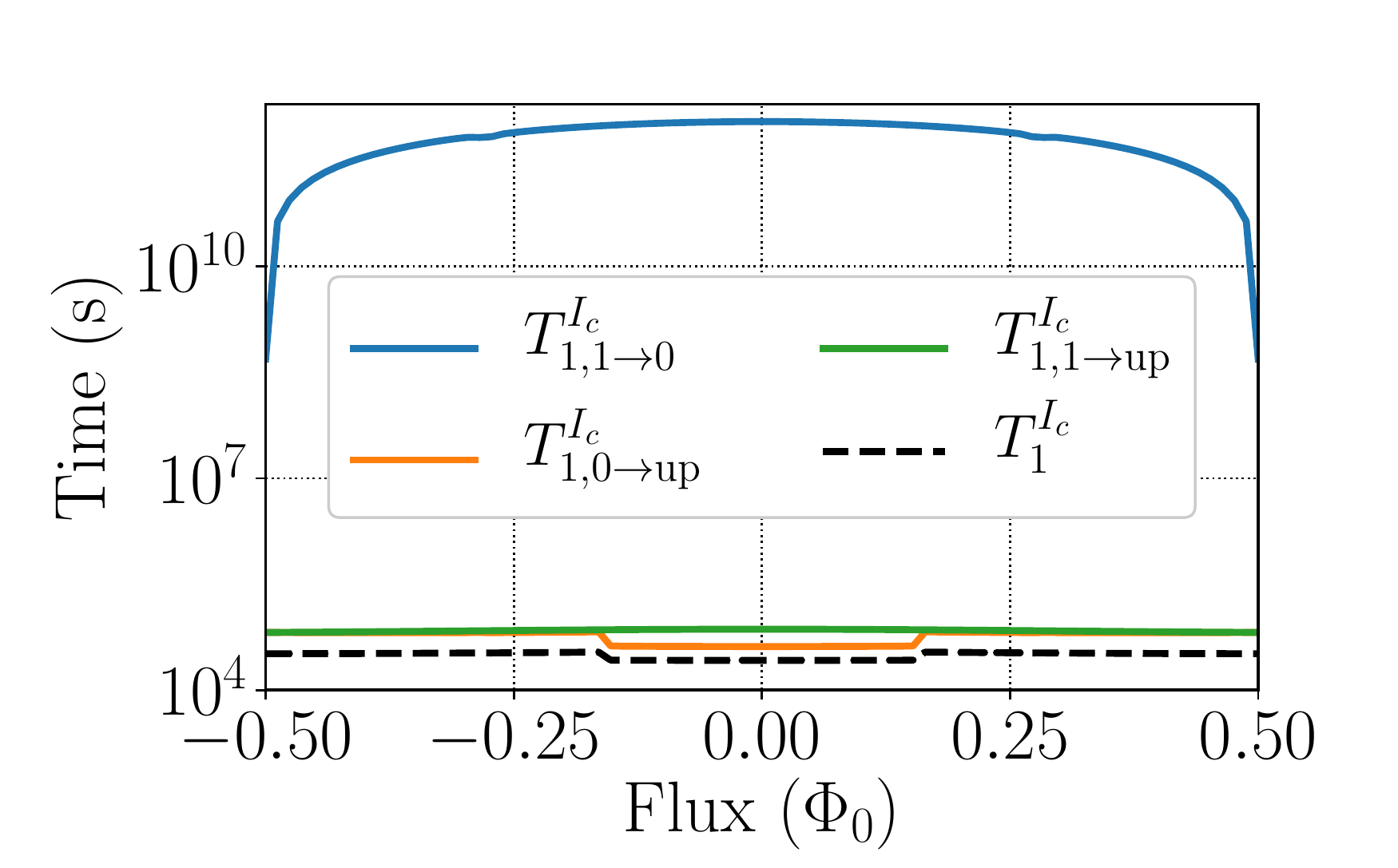}} \\
    \vspace{-0.5cm}
    \subfloat[]{    \hspace*{-0.50cm}    \includegraphics[width=2.40in]{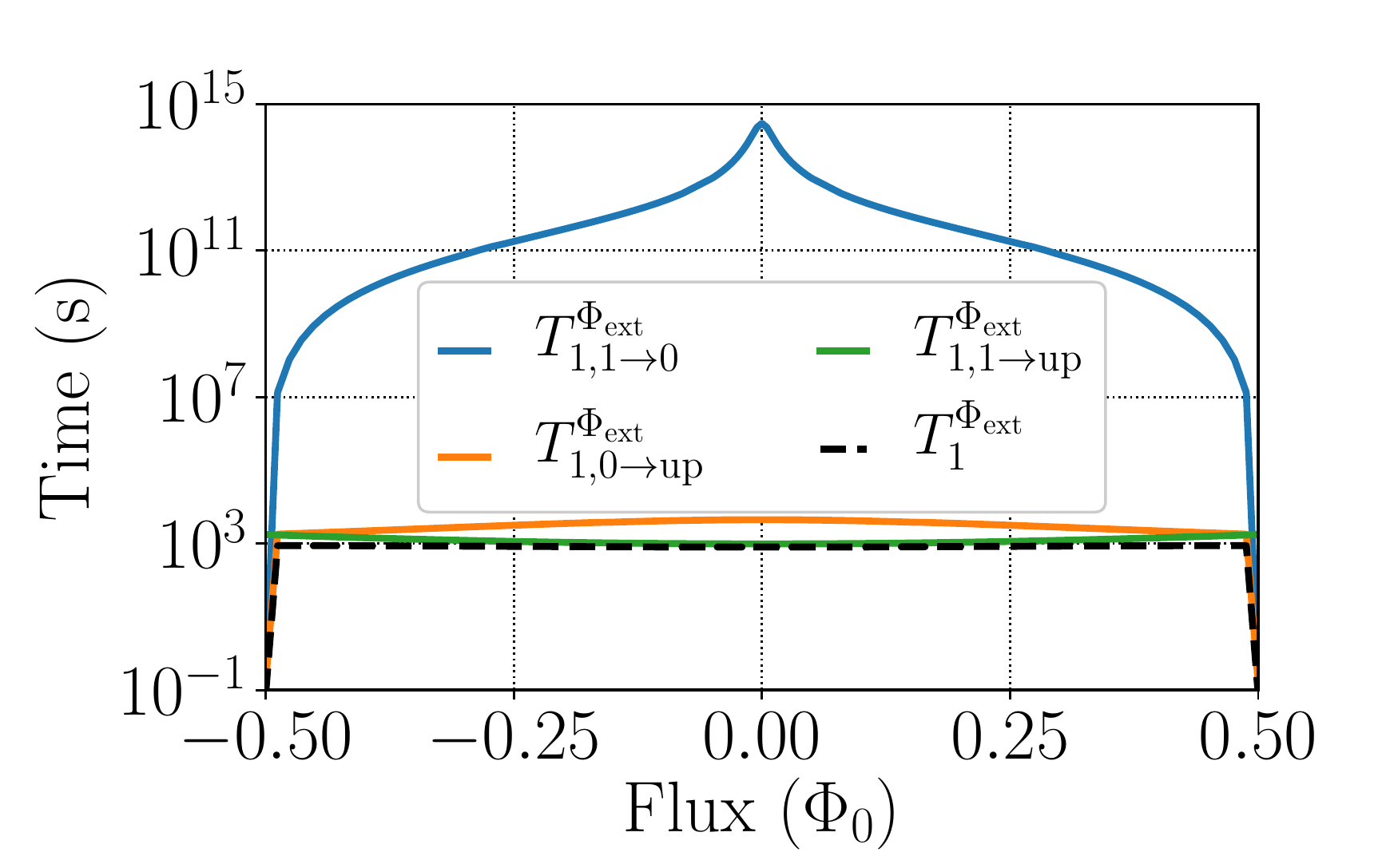}} 
    \hspace*{-0.10cm}
    \subfloat[]{   \hspace*{-0.50cm}     \includegraphics[width=2.40in]{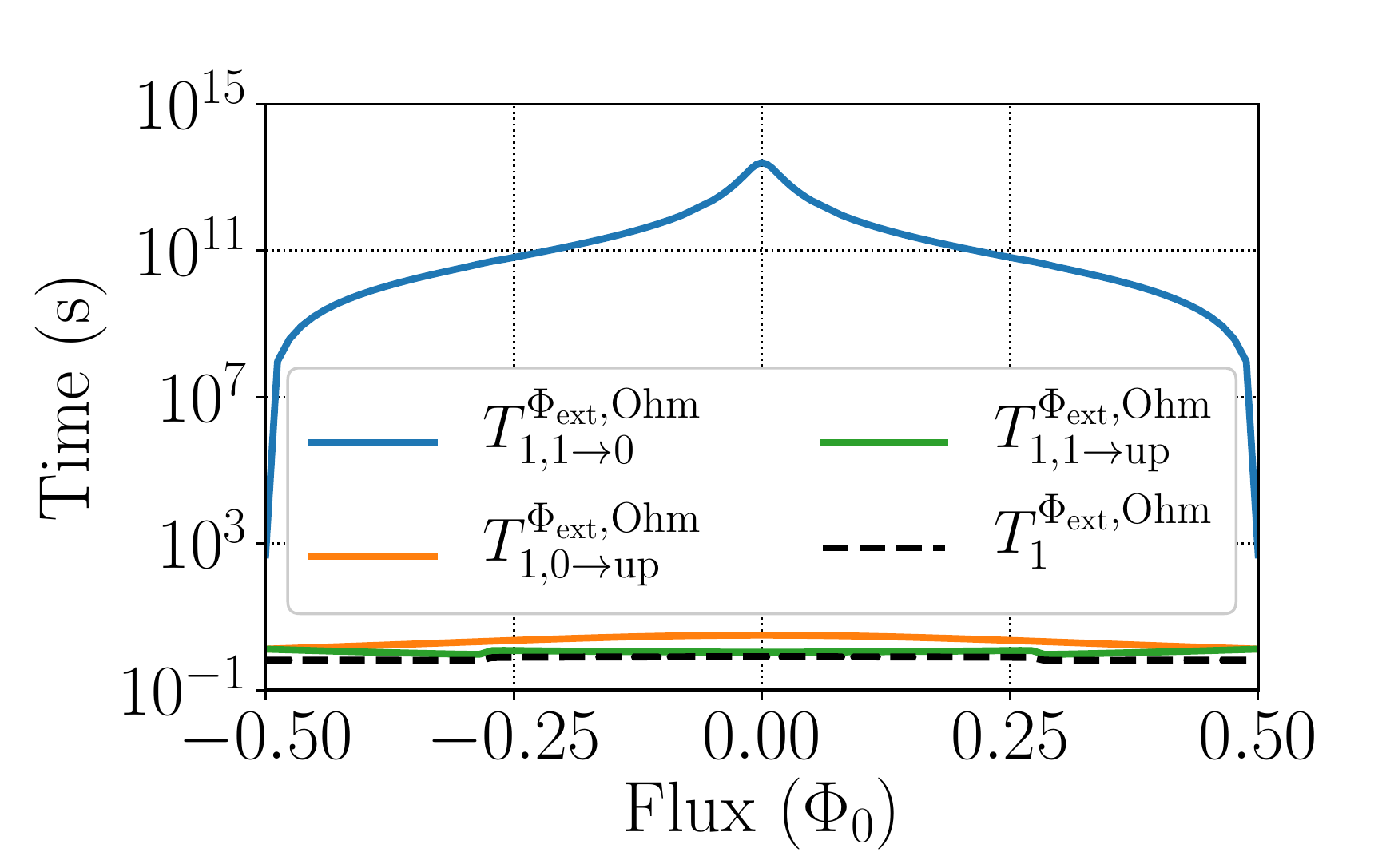}}
    \hspace*{-0.10cm}
    \subfloat[]{   \hspace*{-0.50cm}      \includegraphics[width=2.40in]{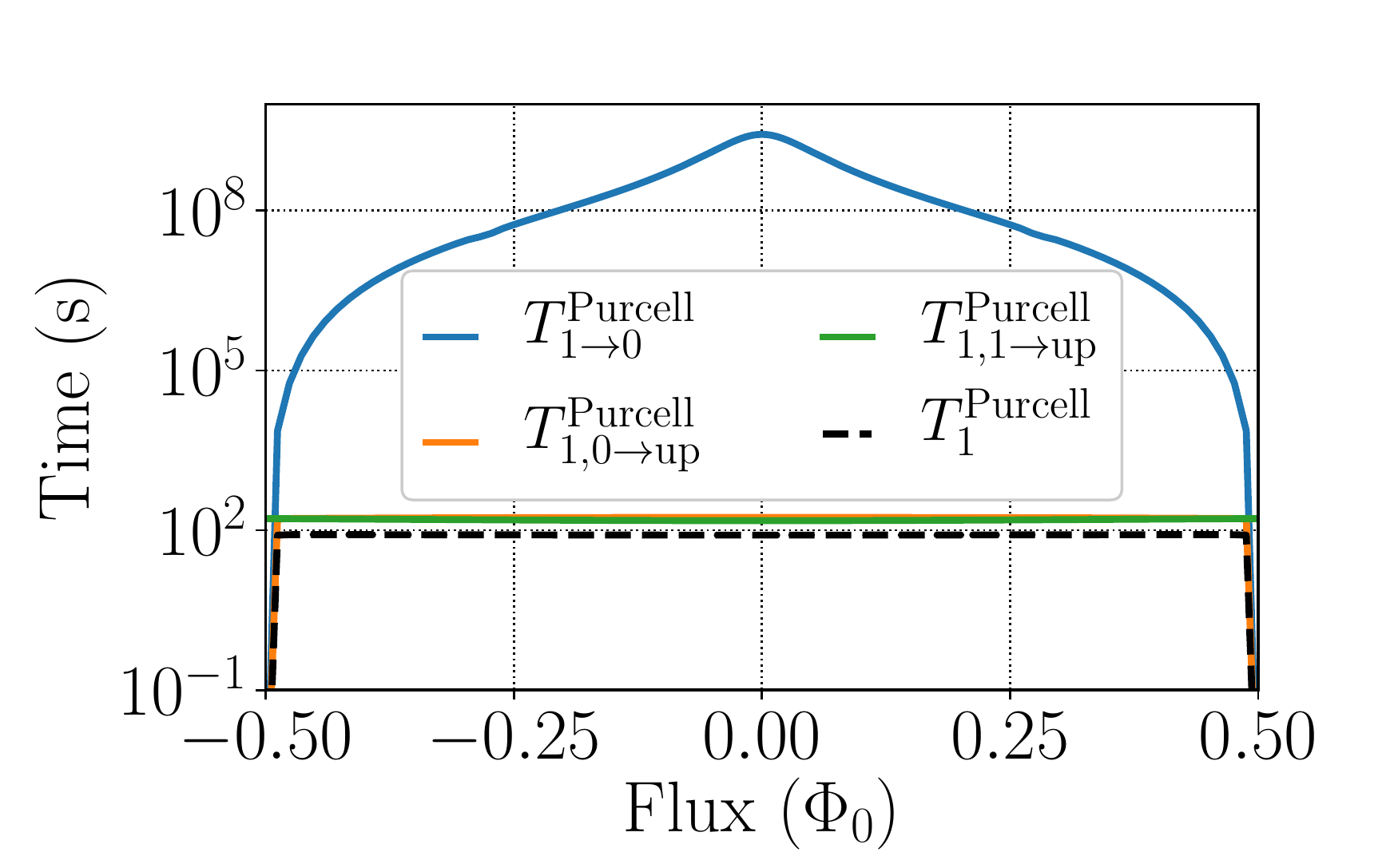}}
    \caption{Calculated coherence times for parameter set 2. (a) Pure dephasing times due to $1/f$ flux noise $\tphiflux$ (orange curve),  $1/f$ critical current noise $\tphicc$ (green curve) as well as shot noise $\tphisn$ (blue curve), with its approximation $1/\kappa_{\zeta} n_{\rm th}(\Omega_{\zeta})$ (dashed red line), valid when $\chi_{01} \gg \kappa_{\zeta}$. (b) Pure dephasing time due to $1/f$, charge noise $\tphing$ along the $\theta$ direction, plotted as a function of $n_{g}^{\theta}$ and calculated at $\Phi_{\rm ext} = 0.0\,\Phi_{0}$ (blue curve), $\Phi_{\rm ext} = 0.25\,\Phi_{0}$ (orange curve), and $\Phi_{\rm ext} = 0.50\,\Phi_{0}$ (green curve). (c) $T_{1}$ due to $1/f$ critical current noise. (d) $T_{1}$ due to $1/f$ flux noise. (e) $T_{1}$ due to biasing flux line noise. (f) Purcell depolarization time. Plots in (c)--(f), show transition times of states $1$ to $0$ (blue curves), $0 \rightarrow {\rm upwards}$ (orange curves), $1 \rightarrow {\rm upwards}$ (green curves), and finally effective (combined) times (dashed black curves). See main text for analysis. }
    \label{fig:resultsRatesPS2}
\end{figure}

\begin{figure}
    \captionsetup[subfigure]{position=top,singlelinecheck=off,topadjust=-5pt,justification=raggedright}
    \centering
    \subfloat[]{    \hspace*{-0.50cm}    \includegraphics[width=2.40in]{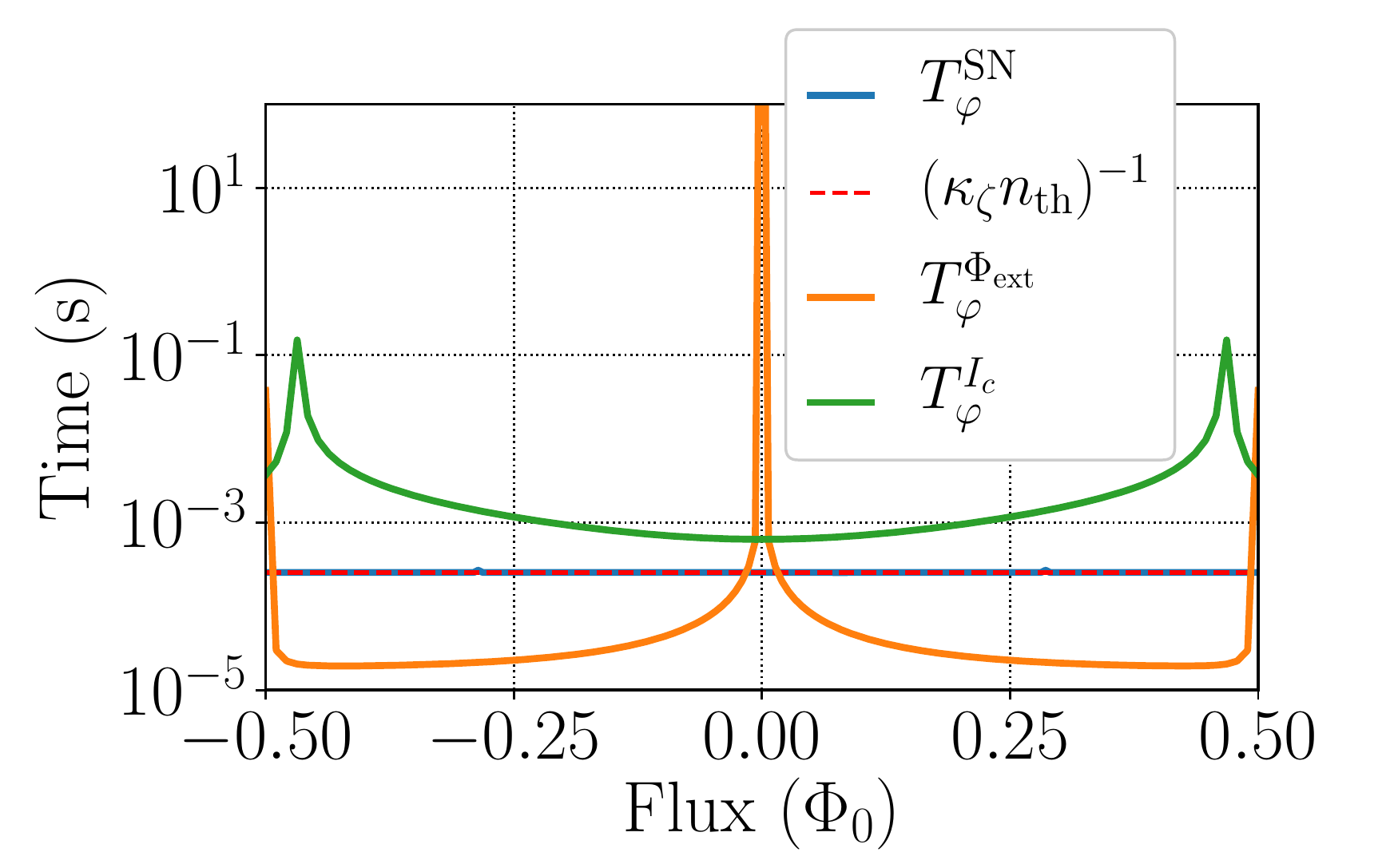}}
    \hspace*{-0.10cm}
    \subfloat[]{    \hspace*{-0.50cm}    \includegraphics[width=2.40in]{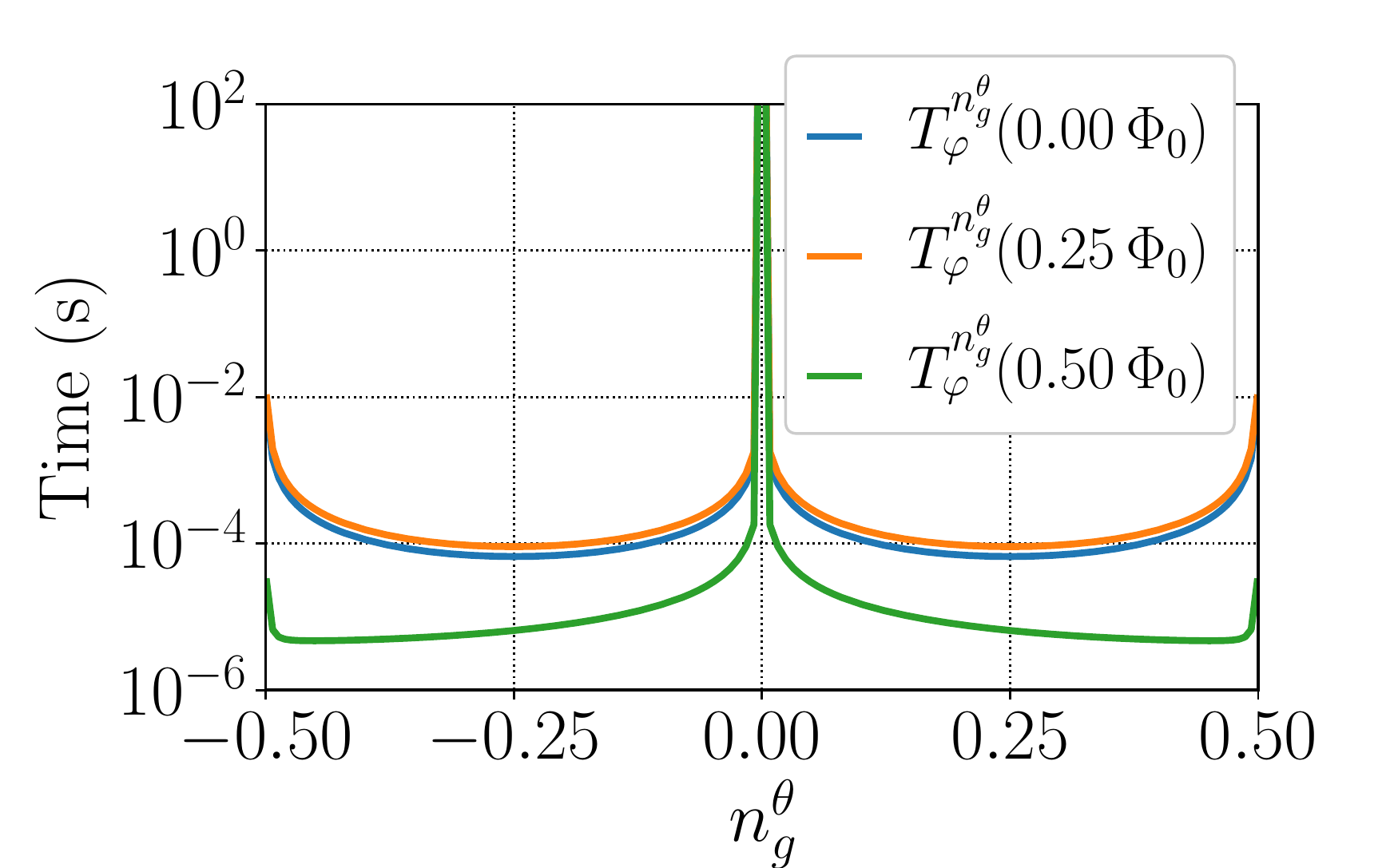}} 
    \hspace*{-0.10cm}
    \subfloat[]{   \hspace*{-0.50cm}     \includegraphics[width=2.40in]{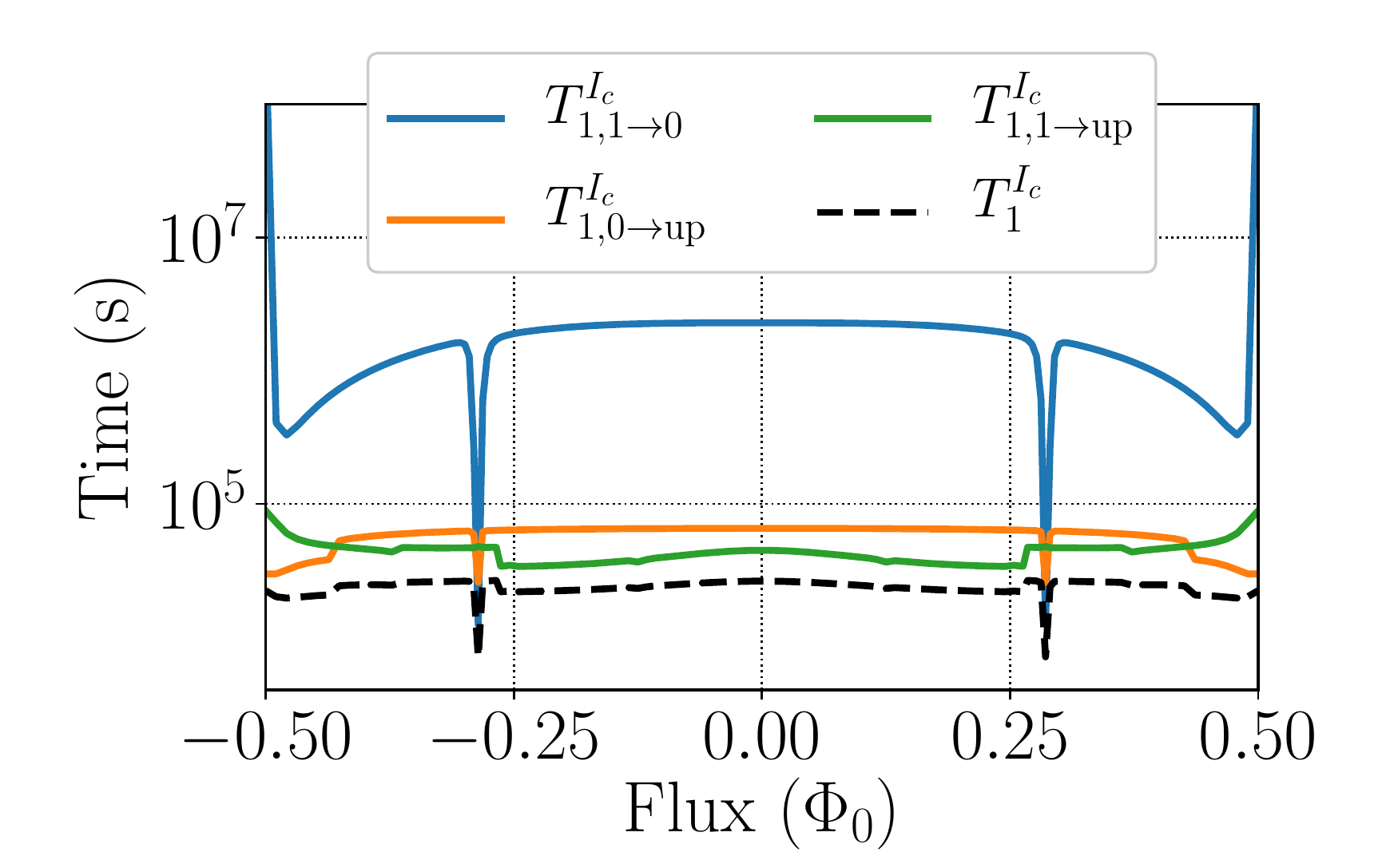}} \\
    \vspace{-0.5cm}
    \subfloat[]{    \hspace*{-0.50cm}    \includegraphics[width=2.40in]{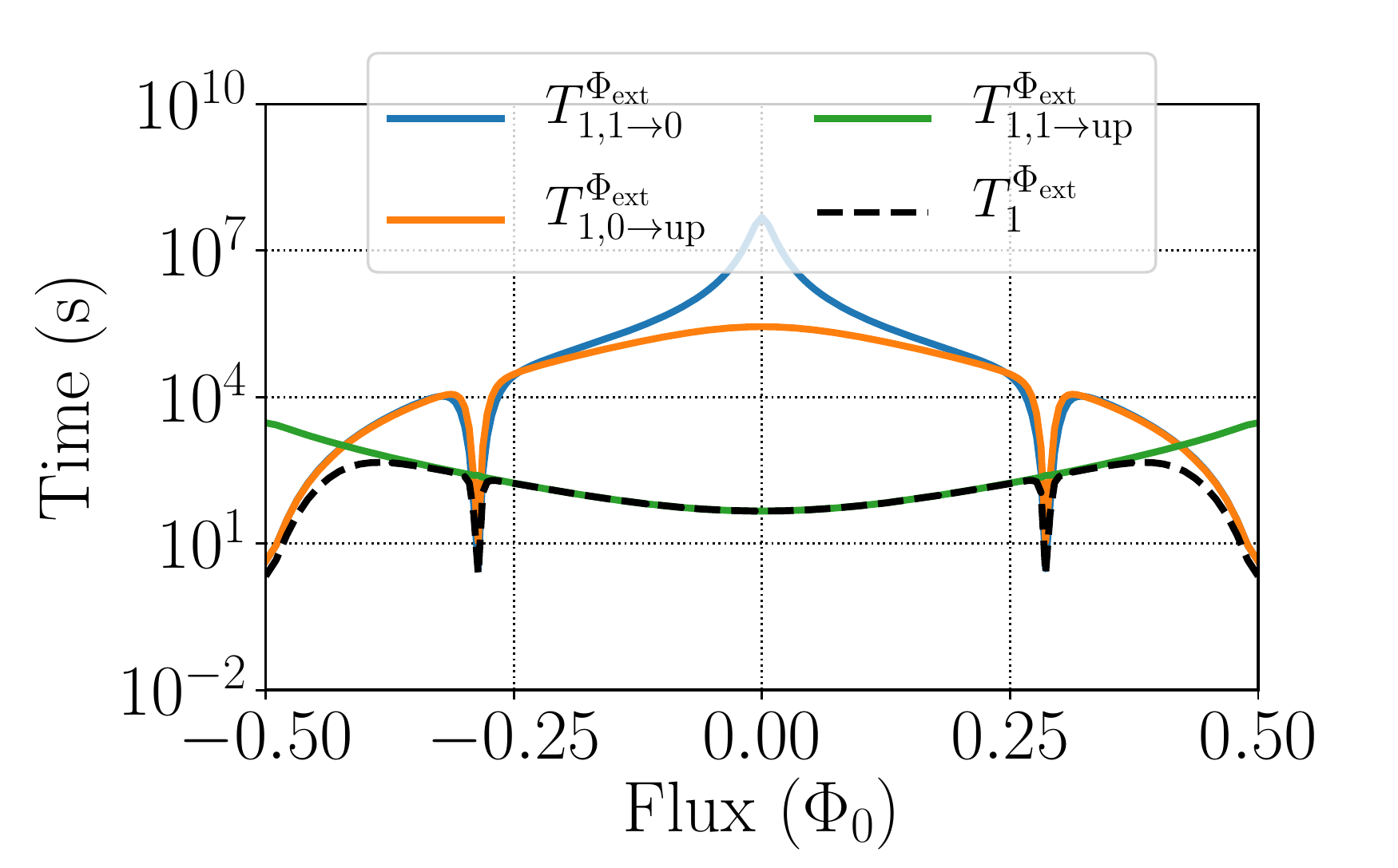}} 
    \hspace*{-0.10cm}
    \subfloat[]{   \hspace*{-0.50cm}     \includegraphics[width=2.40in]{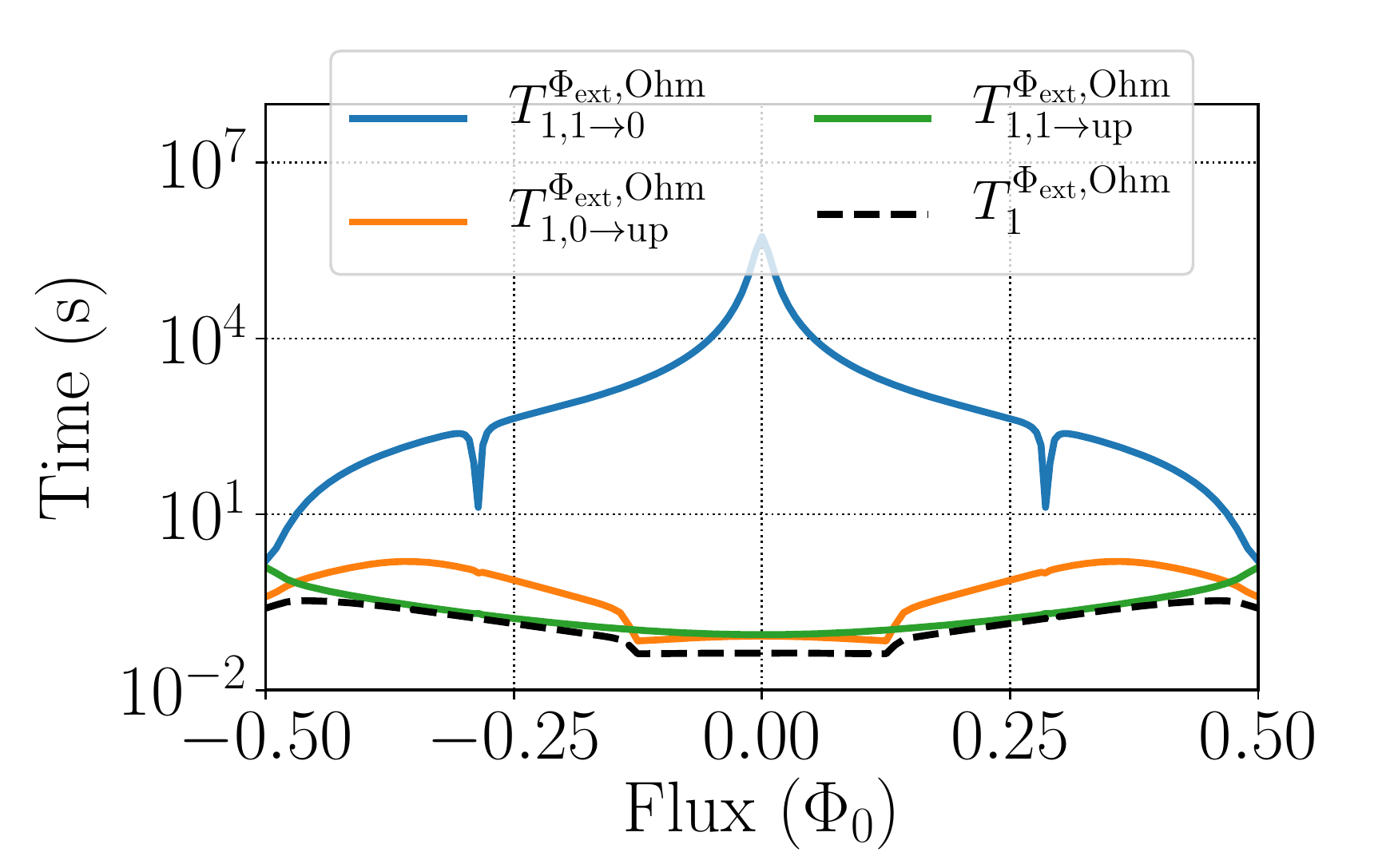}}
    \hspace*{-0.10cm}
    \subfloat[]{   \hspace*{-0.50cm}      \includegraphics[width=2.40in]{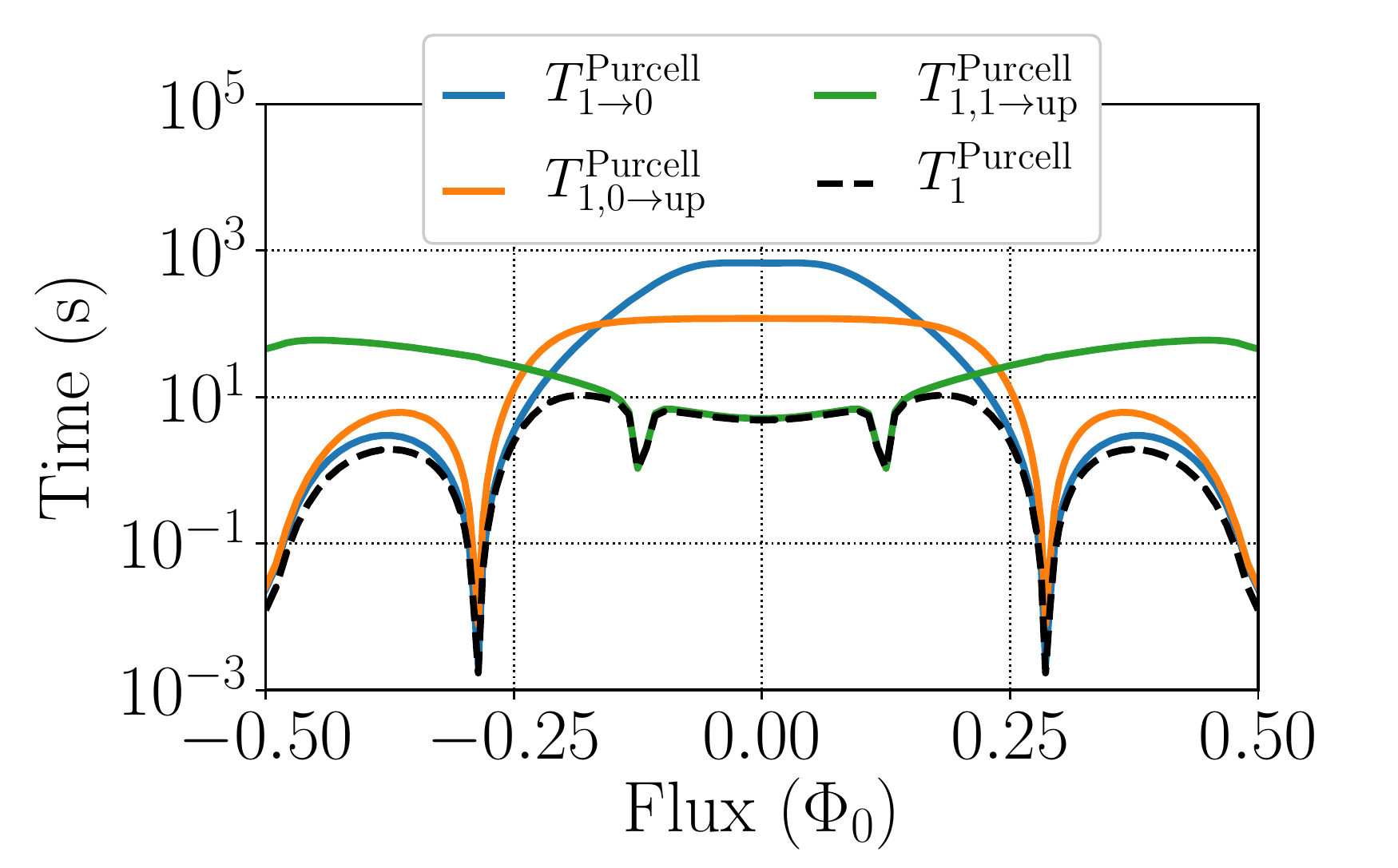}}
    \caption{Calculated coherence times for parameter set 3. (a) Pure dephasing times due to $1/f$ flux noise $\tphiflux$ (orange curve),  $1/f$ critical current noise $\tphicc$ (green curve) as well as shot noise $\tphisn$ (blue curve), with its approximation $1/\kappa_{\zeta} n_{\rm th}(\Omega_{\zeta})$ (dashed red line), valid when $\chi_{01} \gg \kappa_{\zeta}$. (b) Pure dephasing time due to $1/f$, charge noise $\tphing$ along the $\theta$ direction, plotted as a function of $n_{g}^{\theta}$ and calculated at $\Phi_{\rm ext} = 0.0\,\Phi_{0}$ (blue curve), $\Phi_{\rm ext} = 0.25\,\Phi_{0}$ (orange curve), and $\Phi_{\rm ext} = 0.50\,\Phi_{0}$ (green curve). (c) $T_{1}$ due to $1/f$ critical current noise. (d) $T_{1}$ due to $1/f$ flux noise. (e) $T_{1}$ due to biasing flux line noise. (f) Purcell depolarization time. Plots in (c)--(f), show transition times of states $1$ to $0$ (blue curves), $0 \rightarrow {\rm upwards}$ (orange curves), $1 \rightarrow {\rm upwards}$ (green curves), and finally effective (combined) times (dashed black curves). See main text for analysis. }
    \label{fig:resultsRatesPS3}
\end{figure}

The coherence times calculated using expressions from section~\ref{sec:noiseProcesses}, for parameter sets 1, 2 and 3 (see section~\ref{sec:circuitParams}) are shown in figures \ref{fig:resultsRatesPS1}, \ref{fig:resultsRatesPS2} and \ref{fig:resultsRatesPS3} respectively. 
Panels (a) present pure dephasing times vs.\ flux, namely: $\tphiflux$ due to $1/f$ flux noise (orange curve), $\tphicc$ due to $1/f$ critical-current noise (green curve), as well as $\tphisn$ due to shot noise from the $\zeta$-mode coupling (blue curve). The approximate expressions for $\tphisn$, from equation~\eqref{eq:T2starRateshotnoiseSimp1} in the case of PS1, and from equation~\eqref{eq:T2starRateshotnoiseSimp2} in the case of PS2 and PS3 (dashed red line) are also presented for comparison. 
Panels (b) show the pure dephasing time $\tphing$ due to $1/f$ charge 
noise as a function of offset charge $n_g^\theta$. The three curves correspond to three different values of external flux: $\Phi_{\rm ext} = 0.0$ (blue curve), $\Phi_{\rm ext} = 
0.25\,\Phi_{0}$ (orange curve), and $\Phi_{\rm ext} = 0.50\,\Phi_{0}$ (green 
curve). Panels in (c)--(f) outline the relevant $T_{1}$ depolarization times: (c)  depolarization from $1/f$ critical current noise; (d) depolarization due to $1/f$ flux noise; (e) depolarization due to Ohmic noise in the flux-bias line; (f) depolarization due to $\zeta$-mode mediated Purcell processes. 
All plots in (c)--(f) show inverse rates for transitions between 
states $1$ to $0$ (blue curves), $0\rightarrow {\rm upwards} $ (orange curves), 
$1\rightarrow  {\rm upwards}$ (green curves), and finally effective (combined) times (dashed black curves). 

Figure~\ref{fig:resultsRatesPS1} shows that the PS1, which corresponds to the ``deep \zp{} limit'' is the best performing of the parameter sets that we study. However, as was discussed in section~\ref{sec:circuitParams}, it may not be easily experimentally realizable, mainly due to difficulties in building large linear inductors. Hence, below, besides discussing the numerical results in detail for all three parameter sets, we outline how under some circumstances actually decreasing the circuit inductance (increasing $\el$), and therefore going away from the ``deep \zp{} limit'', may be also beneficial to the overall coherence properties of the \zp{} qubit.  

In both PS1 and PS2, the pure dephasing is dominated by $\zeta$-mode shot noise. 
Even the relatively small amount of disorder ($5\%$ in $\el$ and/or $\ec$) causes the primary qubit degrees of 
freedom to couple to the $\zeta$-mode, which for PS1 and PS2 has a low frequency of $\Omega_{\zeta}/2\pi \approx 36\,$MHz and $\Omega_{\zeta}/2\pi \approx 113\,$MHz respectively. At a temperature of $T=15\,{\rm mK}$, this corresponds to a thermal occupations of $n_{\rm th}(\Omega_{\zeta}) \approx 8.25$ and $n_{\rm th} (\Omega_{\zeta}) \approx 2.29$ photons. In PS1, the dispersive shift $\chi_{01}$ is, however, much smaller than $\kappa_{\zeta}$, and a small $\chi_{01}$ dominates $\tphisn$ (see equation~\eqref{eq:T2starRateshotnoiseSimp1}), which is not particularily damaging, as even in worst case, at $\Phi_{\rm ext}=0$, $\tphisn \approx 20\,$ms. In PS2, on the other hand, over most of the flux range, the $\chi_{01}$ is larger than the $\zeta$-mode decay rate -- taken here as $\kappa_{\zeta}=1/100\,\mu{\rm s}$. There, we observe that $\tphisn$ is well approximated by the asymptotic expression $1/ \kappa_{\zeta} n_{\rm th}(\Omega_{\zeta}) \approx 43\,\mu$s over most of the flux range, in which case the shot noise rate is doinated by the thermal photon count occupying the $\zeta$-mode (we discuss the interplay between $\chi_{01}$ and $n_{\rm th} (\Omega_{\zeta})$ in more detail below).   
 

For PS3, as shown in figure~\ref{fig:resultsRatesPS3}(g), $\tphisn$ is no longer the bottleneck across the full flux range. Here $\tphisn$, as in PS2 away from $\Phi_{\rm ext}=0.50\,\Phi_{0}$, is still dominated by $n_{\rm th} (\Omega_{\zeta})$, but both $\el$ and $\ec$ are over $3$ times larger than in PS2, leading to an increased $\zeta$-mode frequency $\Omega_{\zeta}/2\pi \approx395\,$MHz and therefore decreased corresponding thermal occupation of $n_{\rm th} (\Omega_{\zeta})\approx 0.39$ photons. This results in an approximate $\tphisn \approx 254\,\mu$s, but comes at the cost of increased flux-noise sensitivity (as well as charge-noise sensitivity, see below). Away from the flux sweet spot, this can produce a $\tphiflux$ as low as $20\,\mu$s. This unfavorable behavior is a consequence of the large energy-flux dispersion, easily observed in figure~\ref{fig:param23Spec}(c) and (d). Near the sweet spot, however, the flux noise is subdominant and shot-noise dephasing quantified by $\tphisn$ remains the limiting factor. Therefore, perhaps somewhat surprisingly, as long as qubit operation is performed in the vicinity of zero flux, actually increasing $\el$ and $\ec$ (from that of PS2 to PS3) can be beneficial to the qubit's effective pure dephasing time. While decreasing disorder ultimately also mitigates shot-noise sensitivity, we find that disorder levels as small as 
$\sim 2\%$ and $1/\kappa_{\zeta}=100\,\mu$s still lead to significant dispersive shifts $\chi_{01} \gg \kappa_{\zeta}$ (for $\el$ and $\ec$ of PS2 and PS3) away from $\Phi_{\rm ext}=0.5\,\Phi_{0}$. If $\el$ cannot be decreased as done in PS1, the resolution to this challenge is to either further decrease the thermal population of the $\zeta$-mode, or to decrease $\kappa_{\zeta}$ itself. 

Indeed, one key result of our calculations is the non-trivial dependence of shot-noise sensitivity on $\el$. 
In PS1, both $\el$ and $\ec$ are decreased relative to their values in PS2, by factors of $5$ and $2$ respectively, which leads to a substantial mitigation of shot noise. The origin of the observed improvement is subtle, as decreasing $\ec$ and $\el$ also decreases $\Omega_{\zeta}$, which actually increases the thermal population of the $\zeta$-mode and could make shot noise even more damaging. However, while $n_{\rm th}(\Omega_{\zeta})$ gets larger, beyond a certain threshold, the dispersive shift $\chi_{01}$ decreases very rapidly. Specifically, the dispersive shifts $\chi_0$ and $\chi_1$ for the two qubit states become essentially identical, thus rendering $\zeta$-mode shot noise ineffective for dephasing.

\begin{figure}
    \centering
    \includegraphics[width=3.4in]{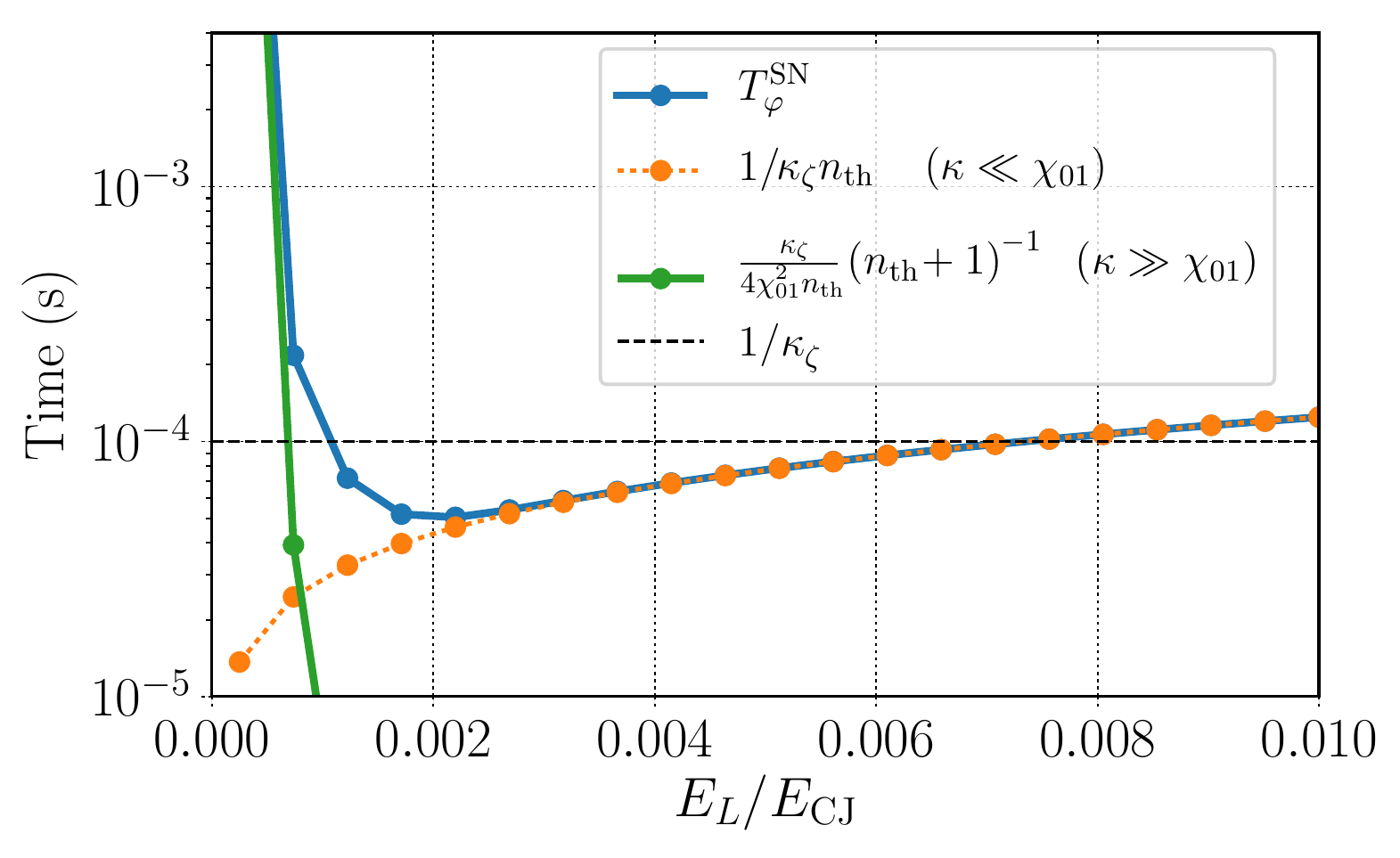}
    \caption{Plot of dephasing time $\tphisn$ due to $\zeta$-mode shot noise as a function of $\el/\ecj$ (blue curve) for PS2. $\el$ is varied while keeping all other energies fixed. At $\el/\ecj \approx 0.0021 $ (i.e. $\el = 0.042\,$GHz), the dephasing time $\tphisn$ reaches a minimum. For $\el/\ecj \gg 0.0021 $, $\tphisn$ can be approximated by $1/\kappa_{\zeta} n_{\rm th} (\Omega_{\zeta})$ (orange curve), and increasing $\el$ is beneficial because it (slowly) decreases the thermal population of the $\zeta$-mode. For $\el/\ecj \ll 0.0021 $, $\tphisn$ can be approximated using equation~\eqref{eq:T2starRateshotnoiseSimp1} (green curve). In that case, the dispersive shift decreases at a faster rate than $n_{\rm th} (\Omega_{\zeta})$, leading to an overall increase in $\tphisn$. The lifetime $1/\kappa_{\zeta}$ is shown for comparison (black dashed line). }
    \label{fig:PS2shotNoiseVsEl}
\end{figure}
To illustrate this effect in more detail, figure~\ref{fig:PS2shotNoiseVsEl} presents $\tphisn$ as a function of $\el$ while keeping all other parameters fixed to the values of PS2. The plot shows that  $\tphisn$ goes through a minimum at $\el/\ecj \approx 0.0021 $ (or equivalently $\el = 0.042\,$GHz). For $\el/\ecj$ well above the minimum at $\el/\ecj = 0.0021 $, $\tphisn$ can be approximated by $1/\kappa_{\zeta} n_{\rm th} (\Omega_{\zeta})$, and increasing $\el$ is beneficial because it decreases the thermal population of the $\zeta$-mode. This is consistent with the benefit we observe when increasing $\el$ (and $\ec$) from the values of PS2 to PS3. In the opposite limit, $\el/\ecj \ll 0.0021 $, $\tphisn$ can be approximated using equation~\eqref{eq:T2starRateshotnoiseSimp1}. Since the dispersive shift decreases at a faster rate than $n_{\rm th}(\Omega_{\zeta})$ increases, the sensitivity to $\zeta$-mode shot noise is actually reduced and $\tphisn$ gets larger, matching the observations made for PS1. Hence, this leads us to believe, that it may be beneficial to keep decreasing $\el$, but only when beyond the threshold corresponding the to the minimum of $\tphisn$. 

As can be expected from the energy-flux dependence (see figures~\ref{fig:param1Spec} and \ref{fig:param23Spec}(a) and (b)), in both PS1 and PS2, the qubit is well-protected from $1/f$ flux noise near the sweet spots at $\Phi_{\rm  ext} = 0$. This is mainly due to the $\el$ being small enough, which contributes to the localization of the \zp{} qubit wave functions in the $\theta=0$ and $\theta=\pi$ potential energy wells, and lead to near-degeneracy as well as suppressed flux dispersion. Pure dephasing due to critical current fluctuations, by contrast, has its flux sweet spot close to half-integer flux, and constitutes the second most dominant pure dephasing mechanism at $\Phi_{\rm ext}=0$, the natural operating point for the \zp{} qubit. Panels (b) of figures~\ref{fig:resultsRatesPS1}, \ref{fig:resultsRatesPS2} and \ref{fig:resultsRatesPS3} show the effects of charge noise. 
For PS1 and PS2, dephasing due to charge noise is weak, and at $\Phi_{\rm ext}=0$, in worst case, away from charge sweet spots, the dephasing times exceed  
$\tphing=500\,$s and $\tphing \approx 1\,$s respectively.
For PS3, charge noise can become a limiting factor away from charge sweet spots, as seen in figure~\ref{fig:resultsRatesPS3}(b) if the qubit is biased near $\Phi_{\rm ext}=0.5\,\Phi_{0}$. Here, the charge-noise sensitivity is increased by the larger charging energy $\ec$ as well as the decreased Josephson energy $\ej$ which in total reduce the ratio $\ej/\ecs \sim \ej/\ec$. Altogether, this increases the energy-charge dispersion (not explicitly shown) and leads to the reduction in dephasing time. Since in practice it may be difficult to limit stray charge offsets, in PS3, one might need to operate the qubit as close to $\Phi_{\rm ext}=0$ as possible, where $\tphing \geq 10^{-4}\,$s. Alternatively, a more detail optimization of PS3 would be possible where $\ec$ could be decreased, while $\el$ further increased. This could potentially limit $\tphing$, while still minimizing the impact of shot noise (over PS2).

Depolarization times $T_{1}$ from critical-current and flux noise are shown in  
panels (c)-(e), while from Purcell effect, in panels (f) of figures~\ref{fig:resultsRatesPS1}, \ref{fig:resultsRatesPS2} and \ref{fig:resultsRatesPS3}. For all three parameter sets, the effective (combined) results (black, dashed lines) are large, with values exceeding $10\,$s (PS1), $500\,$ms (PS2), $40\,$ms (PS3) at $\Phi_{\rm ex}=0$. We note that the relaxation rates from the first excited to the ground state (blue curves) are substantially smaller when compared to excitation rates towards higher states (orange and green curves). This, ``by-design'' behavior is a result of the significant suppression of all matrix elements between ground 
and first excited states of the qubit. Figures~\ref{fig:param1Spec} and \ref{fig:param23Spec} show that for all parameters sets we study, the two lowest eigenfunctions exhibit strong localization along the $\theta$ direction -- even for PS3 away from $\Phi_{\rm ext} = 0.5\,\Phi_{0}$, where the near-degeneracy of the states is lost. As a result, upwards transitions leaking out of the two-level qubit subspace are much more likely than ordinary relaxation/excitation processes within it. 
We also see that the $\tone$ results for PS1 and PS2 are generally flat, while in the case of PS3, we observe not just more variation as a function of flux, but also abrupt dips, especially near $\Phi_{\rm ext} \approx 0.29\,\Phi_{0}$. The increased flux variation has to do with a much larger dependence of the wave functions on changes in flux, which is mainly a result of an increased $\el$. The dips correspond to anticrossings between the states of the qubit and the $\zeta$-mode. In the case of $\tonepurcell$, for example, right at, or very near such dips, we expect the dispersive approximation to break down. There, the qubit is no loner protected by its detuning from the $\zeta$-mode, which results in rates that increase the effective depolarization \cite{sete2014purcell}. 


\begin{figure}
    \centering
    \includegraphics[trim=-2.75cm 0 0 0, width=5.1in]{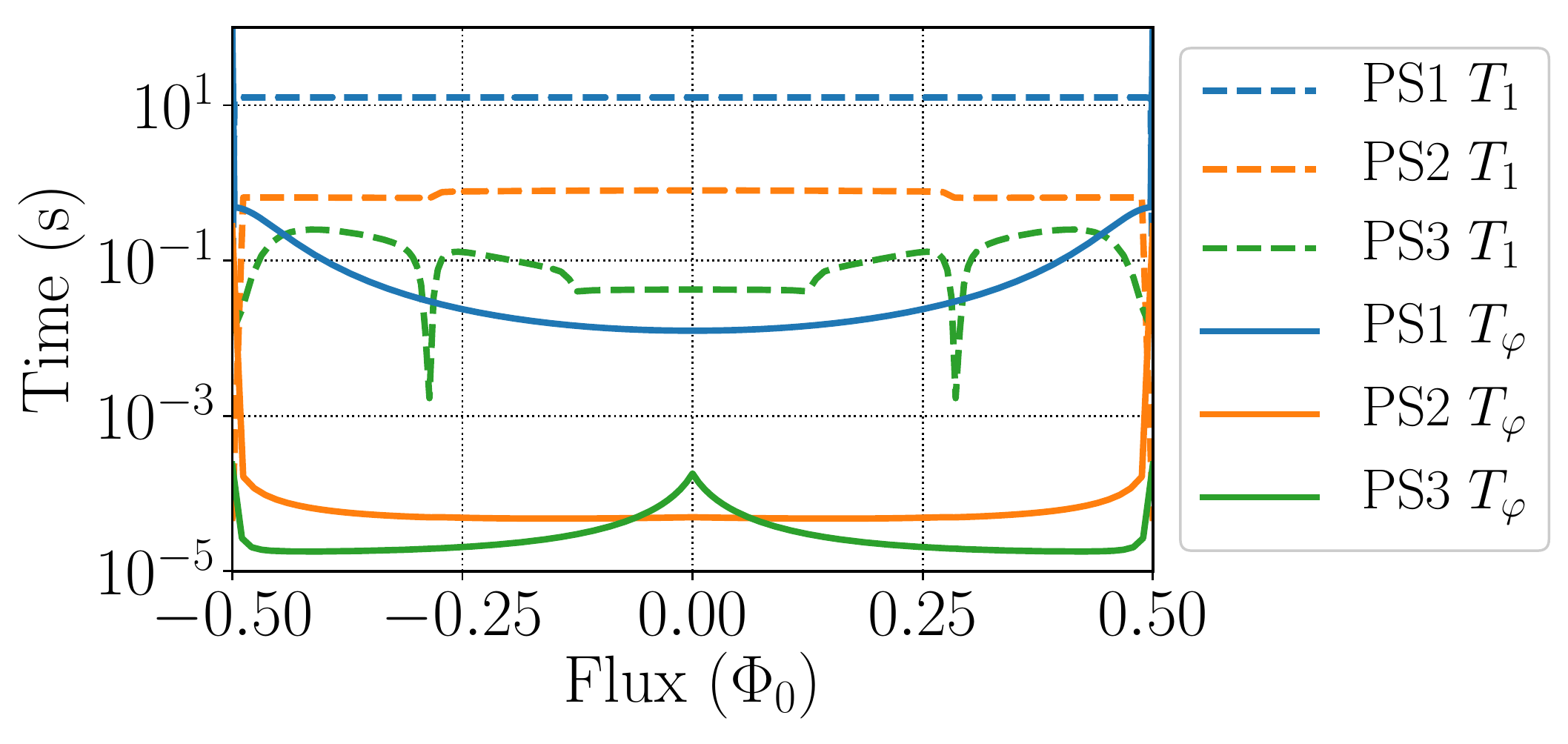}
    \caption{Effective coherence times for PS1 (blue), PS2 (orange), and PS3 (green). The displayed pure-dephasing times $T_{\varphi}$, and depolarization times $T_{1}$ are obtained from the cumulative rates combining all processes described in section~\ref{sec:noiseProcesses}.}    
    \label{fig:effectiveRates}
\end{figure}
Figure~\ref{fig:effectiveRates} summarizes our results with a plot of the total coherence times for PS1 (blue), PS2 (orange), and PS3 (green). The displayed pure-dephasing times\footnote{In the case of the combined $\tphi$, the charge noise rate $1/\tphing$ is not includeded in the calculations. Its inclusion, however, would have minimal (i.e visually indistinghishable) impact on the the result, except in PS3, at $\Phi_{0}=0.5\,\Phi_{0}$ and the near charge bias of $n_{g}^{\theta}=0.5$.} $T_{\varphi}$, and depolarization times $T_{1}$ are obtained from the cumulative rates combinining all noise processes described in section~\ref{sec:noiseProcesses}. At the zero-flux working point, we find: 
$\tphi \approx 20\,$ms and $\tone \approx 10\,$s for PS1, 
$\tphi \approx 50\,\mu$s and $\tone \approx 500\,$ms for PS2,
and $\tphi \approx 200\,\mu$s and $\tone \approx 40\,$ms for PS3. 
These rates confirm that the \zp{} qubit is a promising device benefitting from intrinsic protection. This applies especially to the ``deep \zp{} limit'' exemplified by PS1 and envisioned by BKP  \cite{Brooks2013}. Future work on superinductors based on Josephson junction arrays and high-inductance materials will have to explore ways to reach the needed high inductance values. In the meantime, PS2 and PS3 show that the effect of intrinsic protection can already be reaped with intermediate parameter choices accessible with current capabilities in superinductor fabrication.


\section{Conclusions}
\label{sec:conclusions}

We have studied the coherence properties of the \zp{} qubit and presented calculations of coherence times for three representative sets of 
circuit parameters. We find that  the inductive energy $\el$ has a key impact on the coherence properties: despite spurious coupling to the low-frequency $\zeta$-mode, very large inductances currently beyond experimental capabilities could indeed realize the promise of an intrinsically protected superconducting qubit.  

In the absence of disorder in circuit parameters, the $\zeta$-mode remains decoupled and the \zp{} qubit is expected to be well-protected against noise-induced transitions leading to depolarization, and against fluctuations in qubit  energies leading to pure dephasing. 
Once disorder in the inductive or charging energies ($\el$, $\ec$) is present, the coupling of the primary qubit degree of freedom to the low-energy, harmonic $\zeta$-mode introduces additional decoherence channels that can change the optimal parameter landscape of the qubit. Even with a moderate amount of disorder of a few percent, the thermal population of the $\zeta$-mode can lead to significant shot-noise dephasing of the qubit.
In particular, we found that in the case of parameter set 2, $\tphisn$ has a minimum around $\el^{\rm min} \approx 0.042\,$GHz. For $\el > \el^{\rm min}$, the shot-noise rate is dominated by the thermal occupation of the $\zeta$-mode, and hence can be minimized by making $\Omega_{\zeta}$ larger. This comes at a cost of larger flux dispersion and, hence, enhanced sensitivity to $1/f$ flux noise which can become the limiting factor. In the opposite regime of $\el < \el^{\rm min}$ (large inductance limit), the  rise of the $\zeta$-mode thermal occupation $n_{\rm th}(\Omega_{\zeta})$ is compensated by a dramatic decrease in the qubit's dispersive shift $\chi_{10}$, leading in fact to an overall reduction in the shot-noise dephasing rate -- see equation~\eqref{eq:T2starRateshotnoiseSimp1}. The \zp{} qubit is generally found to behave well with respect to depolarization processes across the parameter sets we considered. 

The effective (combined) pure dephasing and depolarization rates at $\Phi_{\rm ext}=0$ were found to be 
$\tphi \approx 20\,$ms and $\tone \approx 10\,$s for PS1, 
$\tphi \approx 50\,\mu$s and $\tone \approx 500\,$ms for PS2,
and $\tphi \approx 200\,\mu$s and $\tone \approx 40\,$ms for PS3. 
We believe that further optimization might lead to even more favorable results, motivating future research into  experimentally realizing even larger superinductances. In summary, we conclude that the coupling to the spurious $\zeta$-mode does not invalidate the prospects of intrinsic noise protection in \zp{} qubits. We predict that noise protection is at work even in the regime of modest, currently accessible superinductances, rendering the \zp{} qubit an attractive candidate for next-generation superconducting qubits.  

\section{Acknowledgments}
\label{sec:acknowledgments}
We acknowledge valuable discussions with Andy C.~Y.~Li, and the hospitality from the ICTS-TIFR (JK). ADP acknowledges support from the Fundaci\'on Williams en Argentina and the Bourse d'excellence de 3e cycle, Facult\'e des Sciences, Universit\'e de Sherbrooke. This work was supported by the Army Research Office under Grant no.\ W911NF-15-1-0421 and NSERC. This research was undertaken thanks in part to funding from the Canada First Research Excellence Fund.

\appendix

\section{Pure dephasing due to classical noise}
\label{app:pureDephasing}

In this appendix, we review the derivation of pure dephasing rates. We retain terms up to second order in the noise coupling, so that the full crossover from linear noise susceptibility to second-order susceptibility at sweet spots \cite{Vion2002} can be evaluated. Our treatment here is in part based on previous work published in references  
\cite{Ithier05,Makhlin04}.

We consider an external parameter $\lambda(t) = \lambda_0 + \dl 
(t)$ subject to a classical noise signal $\dl(t)$ arising from a stationary, 
Gaussian process with a mean $\langle \dl(t) \rangle = 0$ and given noise power 
spectrum $S(\omega)=\int_{-\infty}^\infty \rmd t\,\rme^{-\rmi\omega t}\langle 
\dl(t)\dl(0) \rangle$. The system Hamiltonian depends parametrically on the 
external parameter, $H=H(\lambda(t))$, and we assume that the effect of noise is 
sufficiently small to allow an expansion in powers of $\dl$, 
\begin{equation}
H = H_{0} + \frac{\partial H}{\partial \lambda}\dl(t) + 
\frac{1}{2}\frac{\partial^2 H}{\partial \lambda^2}\dl^2(t) + \mathcal{O}(\dl^3) 
\approx H(\lambda_0) + V_\lambda(t), 
\end{equation}
where $H_{0} = H(\lambda_0)$, and the derivatives are evaluated at 
$\lambda=\lambda_0$.
To analyze how the noise terms $V_\lambda(t)$ affect the phase coherence of 
the system, it is convenient to switch to the interaction picture, in which 
states and operators take the usual form $\ket{\bar{\psi}(t)}=\rme^{\rmi H_0 
t}\ket{\psi(t)}$ and $\bar{X}(t)=\rme^{\rmi H_0 t}X\rme^{-\rmi H_0 t}$. 
We further employ the eigenbasis $\{\ket{n}\}$ of $H_0$ to express the state 
$\ket{\bar{\psi}(t)}$ in terms of the probability amplitudes 
$c_n(t)=\bket{n}{\bar{\psi}(t)}$.
In the interaction picture, the time-dependent Schr\"odinger equation 
thus takes the form
$ i \frac{\rmd}{\rmd t}c_n(t) = \boket{n}{\bar{V}_\la(t)}{\bar{\psi}(t)}=\sum_{n'} \boket{n}{\bar{V}_\la(t)}{n'}c_{n'}(t) $.
In general, the noise operator $\bar{V}_\la(t)$ incorporates both longitudinal and transverse terms,
\begin{equation}
\bar{V}_\la(t) = \sum_n v_n(t)\ketb{n}{n} + \sum_{n\not=m} 
v_{n,m}(t)\ketb{n}{m},
\end{equation}
where the former is responsible for pure dephasing, while the latter introduces transitions among different states. In the following discussion we concentrate on pure dephasing, and, hence ignore the transverse portion of the Hamiltonian. In such case, the system of differential equations for $c_n(t)$ decouples, and we find 
\begin{equation}
\ket{\bar{\psi}_n(t)} = \exp\left(-\frac{i}{\hbar} \int_0^t \rmd 
t'\,v_n(t)\right)\ket{n},
\end{equation}
for the time evolution of the initial state $\ket{n}$. As expected, 
the longitudinal coupling only affects the phase of the state.
Next, we make use of the decomposition of the noise 
 into contributions of first and second order, 
\begin{equation}
v_n(t) = \boket{n}{\partial_\lambda H}{n} \dl(t) + \frac{1}{2} 
\boket{n}{\partial_\lambda^2 H}{n} \dl^2(t) =
d_n\,\dl(t) + D_n\, \dl^2(t).
\end{equation}
(Again, derivatives are evaluated at $\lambda=\lambda_0$.)
The first-order coefficient $d_n$ can be written as
\begin{equation}
d_n = \bra{n}  \Big(\partial_\lambda{\textstyle \sum_m} E_m(\lambda) 
\ketb{m(\lambda)}{m(\lambda)}\Big) \ket{n}
= \partial_\lambda E_n  + E_n(\lambda_0)\Big( \bra{n} 
\partial_\lambda \ket{n(\lambda)} + {\rm c.c.}\Big)
= \partial_\lambda E_n,
\end{equation}
where all derivatives are evaluated at $\lambda=\lambda_0$, and the term proportional to $E_n(\lambda_0)$ on the right-hand side is zero, 
since $\ket{n(\lambda)}$ is normalized. The second order coefficient is
\begin{equation}
D_n =  \partial_\lambda\boket{n}{\partial_\lambda 
H}{n} = \partial^2_\lambda E_n,
\end{equation} evaluated at $\lambda=\lambda_0$.

To extract the pure dephasing times, we consider a Ramsey-type experiment, starting in an initial superposition $\ket{\bar{\psi}(0)}=(\ket{0}+\ket{1})/\sqrt{2}$. The pure dephasing time is related to the decay of off-diagonal elements of the density matrix in the relevant $2\times2$ subspace,
\begin{equation}
\rho(t) = \frac{1}{2}\left(
\begin{array}{cc} 1 & \rho_{01}(t) \\ \rho_{01}^*(t) & 1\end{array}
\right) \qquad {\rm with} \qquad
\rho_{01}(t) = \exp\left( -i\, \partial_\lambda \omega_{01}\int_0^t \rmd t'\, 
\dl(t') -i\,{\textstyle \frac{1}{2}} \partial_\lambda^2 \omega_{01}\int_0^t 
\rmd t'\, \dl^2(t')\right).
\end{equation}
where $\omega_{01}=(E_0-E_1)/\hbar$, and its derivatives are evaluated at 
$\lambda=\lambda_0$.
Upon averaging over noise realizations $\dl(t)$, the phase factor $\rho_{01}$ 
approaches zero at long times, $\lim_{t\to\infty} \langle 
\rho_{01}\rangle(t)=0$. We will see that the details of this decay depend on 
the noise power spectrum $S_{\la}(\omega)$. However, in common cases the decay 
occurs on some characteristic time scale $T_\varphi$, the pure dephasing time. 
To proceed, we note that the exponent of $\rho_{01}$ is a Gaussian random 
variable, such that $\langle \rme^{i Y}\rangle = \rme^{-\langle Y^2 
\rangle/2}$, which lets us write the noise average as
\begin{equation}
\langle \rho_{01} \rangle (t) = \exp \left(
-{\textstyle \frac{1}{2}}(\partial_\lambda \omega_{01})^2\int_0^t \rmd 
t_1\int_0^t \rmd t_2\, \langle \dl(t_2-t_1)\dl(0)\rangle
-{\textstyle \frac{1}{4}}(\partial_\lambda^2 \omega_{01})^2\int_0^t \rmd 
t_1\int_0^t \rmd t_2\, \langle \dl^2(t_2-t_1)\dl^2(0)\rangle
\right).
\label{eq:rhonm}
\end{equation}
Here, we have used $\langle \dl(t_2-t_1)\dl^2(0)\rangle=0$. Next, we treat the 
the integrals from first and second order contributions:
\[
I_1(t) = \int_0^t \rmd t_1\int_0^t \rmd t_2\, \langle \dl(t_2-t_1)\dl(0)\rangle 
= \int_0^t \rmd t_1\int_0^t \rmd t_2 \int_{-\infty}^\infty 
\frac{\rmd\omega}{2\pi} e^{i\omega (t_2-t_1)}S_{\la}(\omega) = t^2 
\int_{-\infty}^\infty \frac{\rmd\omega}{2\pi} {\rm sinc}^2\left(\frac{\omega 
t}{2}\right)\,S_{\la}(\omega).
\]
For the second-order expression, we apply Wick's theorem to obtain $\langle 
\dl^2(t_2-t_1)\dl^2(0)\rangle = \langle \dl^2(0) \rangle^2 + 2 \langle 
\dl(t_2-t_1)\dl(0)\rangle^2$. Also noting that $\langle \dl^2(t_{2}-t_{1}) 
\rangle^2= \langle \dl^2(0) \rangle^2$, we find
\begin{align}
I_2(t) &=  \int_0^t \rmd t_1\int_0^t \rmd t_2\, \langle 
\dl^2(t_2-t_1)\dl^2(0)\rangle = t^2\sigma_\la^4 + 2t^2\int_{-\infty}^\infty 
\frac{\rmd\Omega}{2\pi}\int_{-\infty}^\infty 
\frac{\rmd\omega}{2\pi}\sinc^2\left(\frac{\left(\Omega + \omega\right) t}{2}\right) 
S_{\la}(\omega)\,S_{\la}(\Omega),
\end{align}
where $\sigma_\la^2 = \langle \dl^2(0) \rangle = 
\int_{-\infty}^\infty\frac{\rmd\omega}{2\pi} S_{\la}(\omega)$.

\subsection{Behavior for long times $t\gg t_c$}
Let us assume that the correlations $\langle \dl(t)\dl(0)\rangle$ are negligible for times exceeding some characteristic time scale $t_c$. 
(This implies that the noise power spectrum falls off significantly at the 
frequency scale $\omega_c=2\pi/t_c$.) If $t\gg t_c$, then the function 
$\sinc^2(\omega t/2)$ in the integrands above filters out the part of the 
noise power spectrum close to $\omega=0$. Specifically, we have
\begin{equation}
\lim_{t\to \infty} \frac{t}{\pi}\sinc^2 \left(\frac{\omega t}{2}\right) = 
\delta(\omega).
\end{equation}
Thus, if the noise power spectrum is well-behaved at low frequencies (see 
discussion of $1/f$ noise below) and $t\gg t_c$, we obtain
\begin{equation}
\langle \rho_{01}\rangle(t) = \exp \left[ -\frac{1}{4}(\partial_\lambda 
\omega_{01})^2 S_{\la}(0)t
-\frac{1}{4}(\partial_\lambda^2 \omega_{01})^2\sigma_\la^4 t^2
-\frac{1}{4}(\partial_\lambda^2 \omega_{01})^2 t \int_{-\infty}^\infty 
\frac{\rmd\Omega}{2\pi}S_{\la}^2(\Omega)\right].
\end{equation}
As a concrete example, let us consider a Gaussian two-time correlator and 
corresponding noise-power spectrum,
\begin{equation}
\langle \dl(t)\dl(0)\rangle = \sigma_\la^2 
\exp\left(-\frac{t^2}{2t_c^2}\right)\qquad \Rightarrow \qquad
S_{\la}(\omega) = \sqrt{2\pi} t_c \sigma_\la^2 
\exp\left(-\frac{t_c^2\omega^2}{2}\right).
\end{equation}
Evaluating the integral over $S_{\la}^2(\Omega)$ yields
\begin{equation}
\langle \rho_{01}\rangle(t) = \exp \left[ -\frac{1}{4}(\partial_\lambda 
\omega_{01})^2S_{\la}(0)t
-\frac{1}{4}(\partial_\lambda^2 \omega_{01})^2\sigma_\la^4 t^2
-\frac{1}{4}(\partial_\lambda^2 \omega_{01})^2 \sqrt{\pi}\sigma_\la^4 t_c t 
\right],
\end{equation}
so that for $t\gg t_c$ we obtain the final result
\begin{equation}
\langle \rho_{01}\rangle(t) \simeq 
\exp \left[ -\frac{1}{4}S_{\la}(0)\,t \left\{ (\partial_\lambda \omega_{01})^2
+\frac{1}{\sqrt{2}}(\partial_\lambda^2 \omega_{01})^2 \sigma_\la^2 \right\} 
\right].
\end{equation}
(Note: it is possible to construct functions $\omega_{01}(\lambda)$ that have 
very large 
curvature at some extrema, but are essentially flat with a very small slope in 
other regions. In this case, it is not clear that ``sweet spot'' operation is 
actually ``sweet''.) 
The decay of the off-diagonal element is a simple exponential, and we define 
the pure-dephasing time as the inverse of the decay rate as usual. This yields
\begin{equation}
T_\varphi = 4\,\bigg/\left[ S_{\la}(0)\, \left\{ (\partial_\lambda \omega_{01})^2
+ (\partial_\lambda^2 \omega_{01})^2 \sigma_\la^2/\sqrt{2} \right\} 
\right].
\end{equation}

\subsection{$1/f$ noise}
For $1/f$ noise, $S_\lambda(\omega)$ is singular for $\omega\to0$, and the noise variance diverges logarithmically. As a result, infrared and ultraviolet regularizations are needed, and are commonly introduced by appropriate cutoffs at $\omega_{\rm ir}$ and 
$\omega_{\rm uv}$. (Note that certain quantities may depend on the type of 
cutoff chosen, i.e., abrupt or ``soft'' \cite{Ithier05}.)
Returning to equation~\eref{eq:rhonm} and evaluating the integral $I_1(t)$ for 
the noise spectrum $S_{\la}(\omega)=2 \pi A_\la^2/|\omega|$, leads to
\begin{equation}\label{eq:I1}
I_1(t) = 8 (2\pi A_\la^{2})\int_{\omega_{\rm ir}}^\infty 
\frac{\rmd\omega}{2\pi}\frac{1}{\omega^3}\sin^2 \left(\frac{\omega t}{2}\right) 
\simeq 2 A_\la^{2}\abs{\ln \omega_{\rm ir}t}\,t^2,
\end{equation}
where we have extracted the leading log-divergent term for $\omega_{\rm ir}\to0$ and assumed 
$t\ll \omega_{\rm ir}^{-1}$. For the second-order contribution, the leading log-divergent contribution is
\begin{equation}\label{eq:I2}
I_2(t) = 4 A_\la^4 \ln^2 \left(\omega_{\rm uv}/\omega_{\rm ir}\right)\,t^2 
+ 
8A^4\ln^2(\omega_{\rm ir}t)\,t^2.
\end{equation}
Equations \eref{eq:I1} and \eref{eq:I2} imply that the decay of the 
off-diagonal elements of the density matrix follows a Gaussian (up to 
logarithmic corrections):
\begin{equation}
\langle \rho_{01}\rangle(t) \sim \exp \left\{ -A_\la^{2} \left(\partial_\lambda 
\omega_{01} \right)^2\abs{\ln \omega_{\rm ir}t}\,t^2
-A_\la^4 \left(\partial_\lambda^2 \omega_{01}\right)^2\left[\ln^2 
\left(\omega_{\rm uv}/\omega_{\rm ir}\right)+ 
2\ln^2(\omega_{\rm ir}t)\right]\,t^2 \right\}.
\end{equation}
Therefore, using the standard variation of the Gaussian as a measure of the dephasing time, we obtain 
\begin{equation}
T_\varphi = \left\{ 2 A_\la^{2} \left(\partial_\lambda
\omega_{01} \right)^2\abs{\ln \omega_{\rm ir}t} 
+2 A_\la^{4} \left(\partial_\lambda^2 \omega_{01} \right)^2\left[\ln^2 
\left(\omega_{\rm uv}/\omega_{\rm ir}\right)+ 
2\ln^2(\omega_{\rm ir}t)\right]\right\}^{-1/2},
\end{equation}
which is equation~\eqref{eq:dephasingTimeFinal} shown in the main text.

\section{Capacitive coupling to circuit nodes}
\label{app:chargenoise}
The analysis of capacitive coupling to the \zp{} nodes, shown in figure 
\ref{fig:zeropicircuitDrive}, proceeds by including the gate capacitances $C_g$ 
and external voltage signals $V_j(t)$ ($j=1,\ldots,4$) in the circuit 
Lagrangian. The transformation of variables $\varphi_j \to 
\phi,\,\theta,\,\zeta,\,\Sigma$, is accompanied by defining analogous 
superpositions of external voltage signals $V_j \to 
V_\phi,\,V_\theta,\,V_\zeta,\,V_\Sigma$, namely $2 
V_{\phi}=(V_2-V_3)+(V_4-V_1)$ etc., see equation \eqref{eq:transf1}. After
Legendre transform of the circuit Lagrangian, one finds that the charging energies 
are renormalized due to the presence of gate capacitances. Denoting the 
renormalized  
capacitances $\cjp=(\cj + \cg/2)$, $\cp=(C + \cg/2)$ and $\csp = (\cs + 
\cg/2)$, we can write the renormalized charging energies [see equation 
\eqref{eq:drivenH}] as $\ecjp=e^{2}/2\cjp$, 
$\ecp=e^{2}/2\cp$, and $\ecsp=e^{2}/2\csp$ respectively. In the final 
expression of the kinetic energy, equation \eqref{eq:drivenH}, the fluctuating 
voltages are compactly written in terms of effective offset charges. If we 
define the offset charges associated with each linearized-mode variable by 
$\bar{n}_g^x=\frac{\cg V_x}{2e}$, with $x \in \{\theta, 
\phi, \zeta \}$, then the effective offset charges used in equation 
\eqref{eq:drivenH} are given by
\begin{align}
    n_g^\theta = \bar{n}_g^\theta  -  \frac{1}{2}\frac{\ecjp}{\ecj}\, \ddcj\, \bar{n}_g^\phi - 
    \frac{1}{2}\frac{\ecp}{\ec}\, \ddc\, \bar{n}_g^\zeta,\qquad
    n_g^\phi =  \bar{n}_g^\phi  - \frac{1}{2} \frac{\ecsp}{\ecj} 
    \ddcj\, \bar{n}_g^\theta , \qquad
    n_g^\zeta =  \bar{n}_g^\zeta  - \frac{1}{2} \frac{\ecsp}{\ec} 
    \ddc\, \bar{n}_g^\theta . 
\end{align}
These expressions show that disorder in the capacitances $C$ and $C_{J}$ leads 
to non-trivial 
``mixing'' between the circuit degrees of freedom $\theta$, $\phi$ and $\zeta$ 
and the corresponding voltages -- a fact that may be of importance when 
performing \zp{} qubit gates by driving capacitively coupled resonators~\footnote{To be discussed in a future publication.}.

\section{Purcell depolarization via the $\zeta$-mode}
\label{app:purcellRates}

In this appendix we review the derivation of relaxation and excitation 
rates associated with the Purcell effect. In the context of the 
\zp{} qubit, Purcell depolarization may occur due to the coupling of the primary \zp{} degrees of freedom (variables $\phi$ and $\theta$) to the lossy $\zeta$-mode. The Hamiltonian for \zp{} circuit interacting with a bath can be written as $H  = H_{\rm sys} + H_{\rm int} + H_{\rm bath}$
where the individual contributions are:
\begin{equation}
    H_{\rm sys} = \sum_{k} E_{k}^{\rm sys} \ket{\psi_{k}^{\rm sys}} \bra{\psi_{k}^{\rm sys}}, \,\,\,\,\,\,\,\,
    H_{\rm bath} = \sum_{k} \hbar \omega_{k} b_{k}\dg b_{k}, \,\,\,\,\,\,\,\,
    H_{\rm int} = \sum_{k} \hbar \lambda_{k} \left( a b_{k}\dg  + a\dg b_{k}\right).
    \label{eq:purcHparts}
\end{equation}
Here, $H_{\rm sys}$ is the full \zp{} circuit Hamiltonian, including the $\zeta$-mode. The latter couples linearly to 
a bath via $H_{\rm int}$, where $a$ and $b_k$ correspond to the lowering 
operators of the $\zeta$-mode and bath modes, respectively. Using Fermi's Golden Rule, we find that $H_{\rm int}$ induces transitions among the eigenstates of 
$H_{\rm sys} + H_{\rm bath}$ with a rate
\begin{equation}
    \gamma_{i \rightarrow f} = \frac{2\pi}{ \hbar}  \delta \left(E_{i} - E_{f} \right) \abs{ \bra{\psi_{f}} H_{\rm int} \ket{\psi_{i}} }^{2}.
    \label{eq:purcFermiGoldenRule}
\end{equation}
The states
\begin{equation}
\ket{\psi_{i}} = \ket{\psi_{i}^{\rm sys}}  \bigotimes_{k}  \ket{m_{k}} \,\,\,\,\,\,\,\, {\rm and} \,\,\,\,\,\,\,\, 
\ket{\psi_{f}} = \ket{\psi_{f}^{\rm sys}}  \bigotimes_{k}  \ket{m_{k}'},
\label{eq:purcPsii}
\end{equation}
are the initial and final eigenstates of $H_{\rm sys} + H_{\rm bath}$, and $E_{i}$ and $E_{f}$ are the corresponding eigenenergies.
Substituting these expressions into equation~\eqref{eq:purcFermiGoldenRule} and simplifying leads to
\begin{equation}
    \gamma_{i, \{m_{k}\} \rightarrow f, \{m_{k}'\}} = \frac{2\pi}{ \hbar}  \delta \left(E_{i} - E_{f} \right)  \sum_{k} \hbar^{2}\abs{\lambda_{k}}^{2} \left[ \abs{ \bra{\psi_{f}^{\rm sys}} a\dg \ket{\psi_{i}^{\rm sys}} }^{2}  m_{k}\, \delta_{m_{k}',m_{k}-1}   +  \abs{ \bra{\psi_{f}^{\rm sys}} a \ket{\psi_{i}^{\rm sys}} }^{2}  (m_{k} + 1) \delta_{m_{k}',m_{k}+1}   \right] \prod_{k' \ne k} \delta_{m_{k'}', m_{k'}},
    \label{eq:purcFermiGoldenRule2}
\end{equation}
where $\{m_{k}\}$ and $\{m_{k}'\}$ denote the initial and final configuration of the bath modes. 
\begin{figure}
	\captionsetup[subfigure]{position=top,singlelinecheck=off,topadjust=-5pt}
	\centering
    \subfloat[]{  \hspace*{-0.20cm}   \includegraphics[trim=0 0 0 3cm, clip=true, width=2.2in]{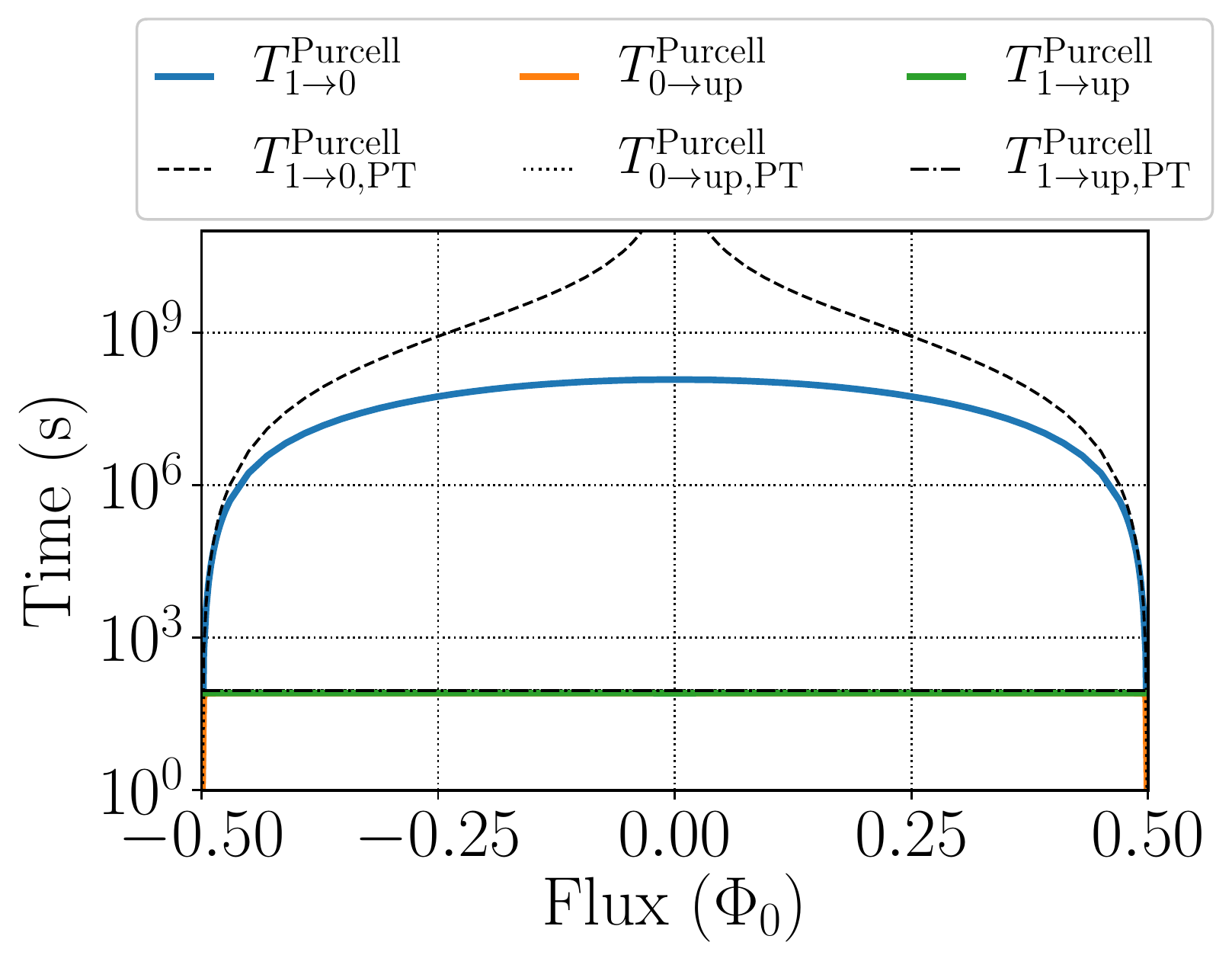}}
	\subfloat[]{  \hspace*{-0.20cm}     \includegraphics[trim=0 0 0 3cm, clip=true, width=2.2in]{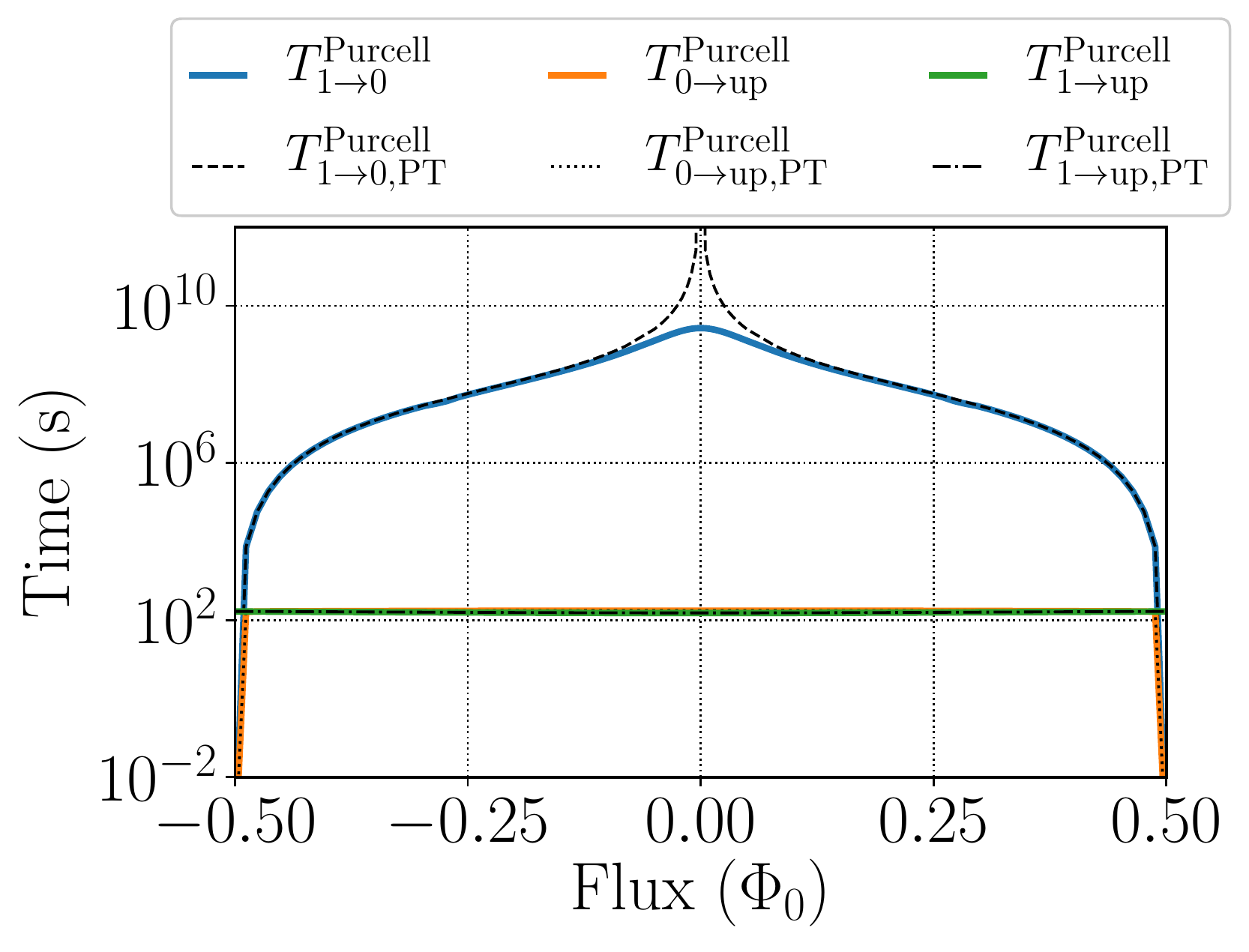}} 
	\subfloat[]{  \hspace*{-0.20cm}     \includegraphics[trim=0 0 0 3cm, clip=true, width=2.2in]{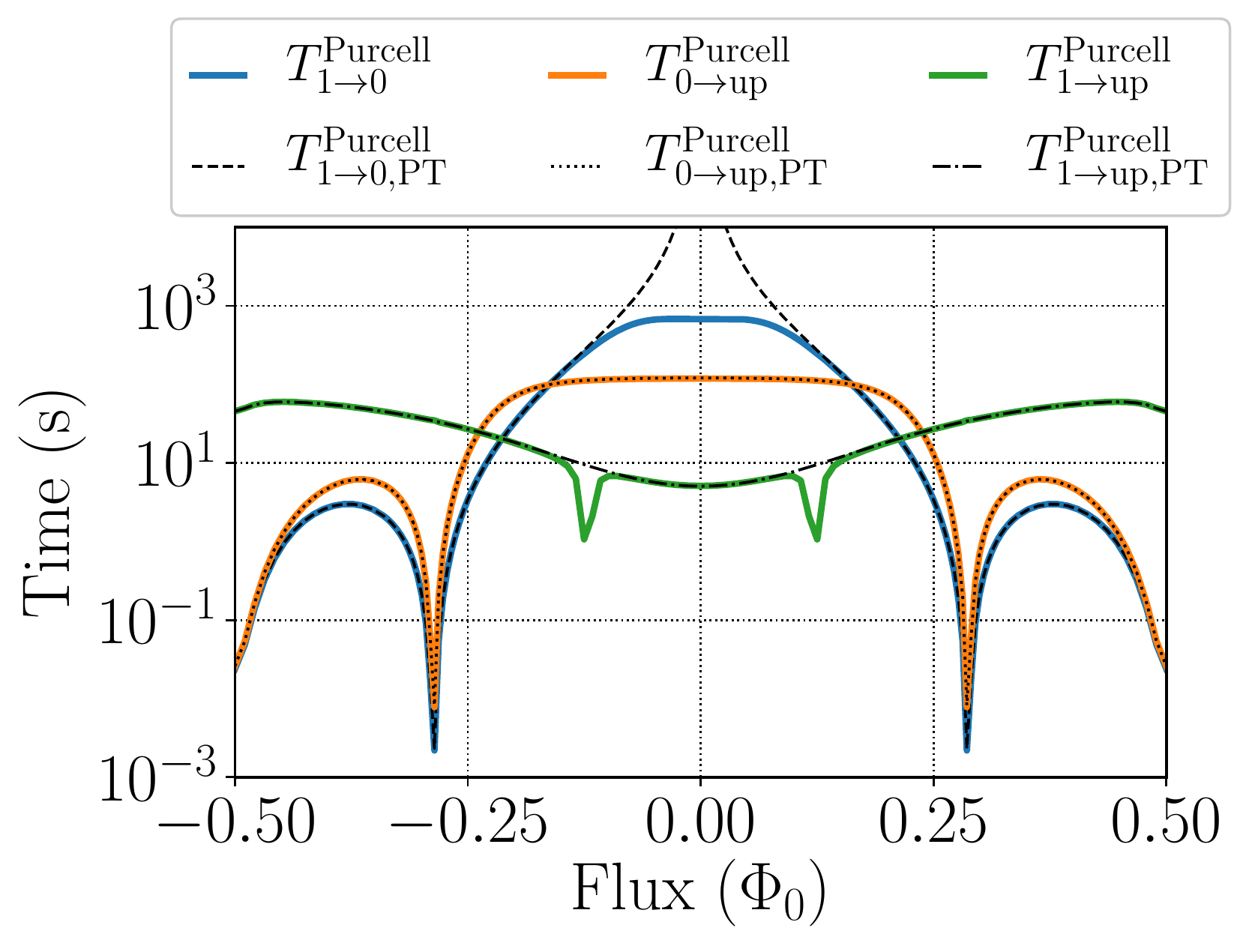}}  \\
    \vspace{-0.35cm}
	\subfloat{   \includegraphics[width=6.4in]{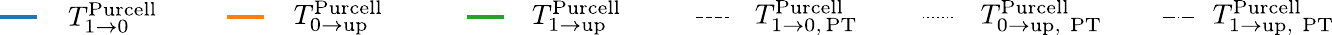}} 
    \caption{Purcell depolarization times obtained for parameters (a) PS1, (b) PS2 and (c) PS3. Results obtained by numerically are shown in solid, colored lines, while using perturbation theory in black, dashed, dotted and dashed-dotted lines.}
	\label{fig:purcellfullPt}
\end{figure}
Next, we note that the bare energies of $\ket{\psi_{i}}$ and $\ket{\psi_{f}}$ can be written as  
\begin{equation}
    E_{i} = E_{i}^{\rm sys} + \hbar \sum_{m_{k}} m_{k} \omega_{k} \,\,\,\,\,\,\,\, {\rm and} \,\,\,\,\,\,\,\, 
    E_{f} = E_{f}^{\rm sys} + \hbar \sum_{m_{k}'} m_{k}' \omega_{k}.
    \label{eq:purcEnergiesT}
\end{equation}
To obtain the effective rate for the transition $i\to f$, we sum over all initial and final states of the bath, weighting initial states by their probability of occurrence $P(\{m_{k}\})$, as appopriate for a bath in a thermal state at temperature $T$. With this, we obtain
\begin{align}
    \Gamma_{i \rightarrow f} =   \sum_{ m_{k},  m_{k}'} P(\{m_{k}\}) \gamma_{i, \{m_{k}\} \rightarrow f, \{m_{k}'\}} 
&=   2\pi \hbar \sum_{k}  \abs{\lambda_{k}}^{2}  \delta \left(E_{i}^{\rm sys} - E_{f}^{\rm sys} + \hbar \omega_{k}  \right)   \abs{ \bra{\psi_{f}^{\rm sys}} a\dg \ket{\psi_{i}^{\rm sys}} }^{2}  n_{\rm th}(\omega_{k})  \nonumber \\
    &\quad + 2\pi \hbar \sum_{k}  \abs{\lambda_{k}}^{2}  \delta \left(E_{i}^{\rm sys} - E_{f}^{\rm sys} - \hbar \omega_{k}  \right)  \abs{ \bra{\psi_{f}^{\rm sys}} a \ket{\psi_{i}^{\rm sys}} }^{2}  \left( n_{\rm th}(\omega_{k})  + 1 \right),
    \label{eq:effectiveRateif}
\end{align}
where $n_{\rm th}(\omega_{k})$ represents the mean thermal occupation number for bath mode $k$ (mode energy $ \hbar \omega_{k}$). Finally, we take the continuum limit, define $\kappa=2\pi \hbar \rho(\omega_{k}) \abs{\lambda_{k}}^{2} $ with $\rho(\omega)$ denoting the bath density of states, and introduce $\omega_{jj'}^{\rm sys} = (E_{j}^{\rm sys} - E_{j'}^{\rm sys})/\hbar$, to obtain
\begin{align}
    \Gamma^{{\rm Purcell}, +}_{i \rightarrow f} &= \kappa_{\zeta} n_{\rm th}(\omega_{fi}^{\rm sys})   \abs{ \bra{\psi_{f}^{\rm sys}} a\dg \ket{\psi_{i}^{\rm sys}} }^{2} ,
    \label{eq:purcellPlusnl}
\end{align}
when $E_{f}^{\rm sys} >  E_{i}^{\rm sys}$, as well as a downward one
\begin{align}
    \Gamma^{{\rm Purcell}, -}_{i \rightarrow f} &= \kappa_{\zeta} \left(1 +  n_{\rm th}(\omega_{fi}^{\rm sys}) \right)   \abs{ \bra{\psi_{f}^{\rm sys}} a \ket{\psi_{i}^{\rm sys}} }^{2} ,
    \label{eq:purcellMinusnl}
\end{align}
when $E_{f}^{\rm sys} <  E_{i}^{\rm sys}$.
The final step is to note that over 
most of the relevant parameters discussed here, the qubit is in the dispersive 
regime with respect to the $\zeta$-mode. One can use this fact to label the 
eigenstates $\ket{\psi_{j}^{\rm sys}}$ with quantum numbers $l$ and $n$ 
corresponding to the number of qubit and $\zeta$-mode excitations respectively. 
As discussed in Sec~\ref{ssec:depolarization}, one way to do this is to look at 
a maximum overlap between the exact (numerically calculated) eigenstates and 
bare states where the coupling between the $\{\theta,\phi\}$ and $\zeta$ is set 
to zero. 
Another way is to approximate the eigenstates by treating the coupling $\sum_{l, l'}\left( g_{ll'}\ket{l}\!\bra{l'}a +{\rm h.c.}\right)$ from equation~\eqref{eq:Hqubitzeta} as a perturbation. 
In that case, we can express the dressed states $\ket{\overline{l,n}}$ in terms of bare eigenstates $\ket{l,n}$ as
\begin{equation}
    \ket{\overline{l,n}} = \ket{l,n} + \sum_{k} \alpha_{k}^{l,n} \ket{k,n-1} + \sum_{k} \beta_{k}^{l,n} \ket{k,n+1} + \dots, 
    \label{eq:purcDressedStates1}
\end{equation}
where we have taken 
\begin{equation}
    \alpha_{k}^{l,n}= \frac{g_{kl} \sqrt{n}}{E_{l,n}^{\rm sys} - E_{k,n-1}^{\rm sys}}  \,\,\,\,\,\,\,\, {\rm and} \,\,\,\,\,\,\,\, 
    \beta_{k}^{l,n}= \frac{g_{kl} \sqrt{n+1}}{E_{l,n}^{\rm sys} - E_{k,n+1}^{\rm sys}},
    \label{eq:purcAlphaBeta}
\end{equation}
with bare energies $E_{l,n}^{\rm sys} = E_{l}^{q} + n \hbar \Omega_{\zeta}$.
Defining $\ket{\psi_{i}^{\rm sys}}=\ket{\overline{l,n}}$, $\ket{\psi_{f}^{\rm sys}}=\ket{\overline{l',n'}}$, we thus find for the matrix element the leading-order expression
\begin{align}
    \abs{ \bra{\psi_{f}^{\rm sys}} a \ket{\psi_{i}^{\rm sys}} }^{2} &=
    \abs{\beta_{l'}^{l,n}\sqrt{n+1} \delta_{n,n'} + \left( \alpha_{l}^{l',n} \right)^{*} \sqrt{n} \delta_{n-1, n'-1} }^{2}  + \abs{ \alpha_{l'}^{l,n} \sqrt{n-1} \delta_{n-2, n'} + \left( \beta_{l}^{l',n-2} \right)^{*} \sqrt{n} \delta_{n-1, n'+1} }^{2} 
    \label{eq:purcExpa}
\end{align}
where we neglect terms beyond second order in $\alpha$ and $\beta$ from equations~\eqref{eq:purcAlphaBeta}. Substituting equation~\eqref{eq:purcExpa} and an analogous expression for $\abs{ \bra{\psi_{f}^{\rm sys}} a\dg \ket{\psi_{i}^{\rm sys}} }^{2}$ into \eqref{eq:effectiveRateif}, leads to the expressions \eqref{eq:purcellRatePlus} and \eqref{eq:purcellRateMinus}. In the last two steps we summed over the final $\zeta$-mode states $n'$, which conveniently resulted in the expression that is independent of $n$. In figure~\ref{fig:purcellfullPt} we show a comparison between the depolarization rates due to Purcell effect for PS1 (a), PS2 (b), and PS3 (c), calculated using both methods: the solid colored lines use numerical maximum-state-overlap method, while the black lines are from perturbation theory.

\section*{References}
\bibliographystyle{iopart-num}
\bibliography{library}

\end{document}